\documentclass[preprint2]{aastex}
\usepackage{amsmath, amsthm, amssymb}
\usepackage{natbib}
\shorttitle{Frequency Dependence of Pulse Width for Normal Pulsars}
\shortauthors{Chen J. L. AND Wang H. G. }

\begin{document}

\title{FREQUENCY DEPENDENCE OF PULSE WIDTH FOR 150 RADIO NORMAL PULSARS}
\author{J. L. Chen\altaffilmark{1,2}, H. G. Wang\altaffilmark{3,4}}
\altaffiltext{1}{Department of Physics \& Electronic Engineering, Yuncheng University, 044000,
             Yuncheng, Shanxi, China}
\altaffiltext{2}{Xinjiang Astronomical Observatory, 150, Science-1
Street, 830011, Urumqi, Xinjiang, China}
 \altaffiltext{3}{Center for Astrophysics, Guangzhou University,
  510006, Guangzhou, China; hgwang.gz@gmail.com}
 \altaffiltext{4}{ CSIRO Astronomy and Space Science, P.O. Box 76, Epping NSW 1710, Australia}

\begin{abstract}

The frequency dependence of pulse width is studied for 150 normal pulsars, mostly selected from the European Pulsar Network, for which the multifrequency 10\% pulse widths can be well fit with the Thorsett relationship $W_{10}=A\nu^{\mu}+W_{\rm 10, min}$. The relative fraction of pulse width change between 0.4~GHz and 4.85~GHz, $\eta=(W_{4.85}-W_{0.4})/W_{0.4}$, is calculated in terms of the best-fit relationship for each pulsar. It is found that 81 pulsars (54\%) have $\eta<-10$\% (group A), showing considerable profile narrowing at high frequencies, 40 pulsars (27\%) have $-$10\%$\leq\eta\leq 10$\% (group B), meaning a marginal change in pulse width, and 29 pulsars (19\%) have $\eta>10$\% (group C), showing a remarkable profile broadening at high frequencies. The fractions of the group-A and group-C pulsars suggest that the profile narrowing phenomenon at high frequencies is more common than the profile broadening phenomenon, but a large fraction of the group-B and group-C pulsars (a total of 46\%) is also revealed. The group-C pulsars, together with a portion of group-B pulsars with  a slight pulse broadening, can hardly be explained using the conventional radius-to-frequency mapping, which only applies to the profile narrowing phenomenon. Based on a recent version of the fan beam model, a type of broadband emission model, we propose that the diverse frequency dependence of pulse width is a consequence of different types of distribution of emission spectra across the emission region. The geometrical effect predicting a link between the emission beam shrinkage and the spectrum steepening is tested but disfavored.

\end{abstract}

%-----------------------------------------
\section{INTRODUCTION}           %% first-level sections will be auto-capitalized
\label{sect:intro}

It is well known that radio pulsars have diverse frequency dependence of average pulse profile. Some pulsars exhibit a stable pulse
morphology within a wide range of radio frequencies, while some others show remarkable variation in pulse shape and/or pulse width (e.g., Rankin 1983a; Hankin \& Rickett 1986; Lyne \& Manchester 1988, hereafter LM88; Johnston et al. 2008; Hankins \& Rankin 2010). Opposite trends have been observed among the pulsars with frequency-dependent profiles. For example, in a wide frequency range from tens of megahertz to a few gigahertz, the double components in the profile of PSR B1133$+$16 get closer as the frequency increases (e.g., Thorsett 1991; Phillips \& Wolszczan 1992; Hassall et al. 2012), while the double components of PSR B1919$+$21 get closer as the frequency decreases (Lyne et al. 1971; Hassall et al. 2012).

The frequency dependence of pulse width or component separation, a major property of profile morphology evolution, has been extensively studied. In many early studies it was suggested that the pulse component separation decreases with increasing frequency but breaks into two power laws for most pulsars (Craft \& Comella 1968; Lyne et al. 1971; Sieber et al. 1975; Rankin 1983b; Slee et al. 1987). By collecting a large number of published profiles for seven pulsars with double or multiple components, Thorsett (1991) studied the frequency dependence of peak separation between the outmost leading and trailing components. Unlike the early studies, they found that a simple power-law function with a constant term, $\Delta\theta=A\nu^{-\alpha}+\Delta\theta_{\rm min}$ (hereafter Thorsett relationship), is enough to fit the data for each pulsar.

Using a sample of 10 pulsars with conal components (including the 7 pulsars studied by Thorsett 1991), Mitra \& Rankin (2002, hereafter MR02) demonstrated that the Thorsett relationship is applicable to the beam radii derived from three kinds of pulse widths, i.e. $W_{10}$, the full width at the 10\% level of pulse peak, $W_{50}$, the full width at the 50\% level of pulse peak, and the peak separation. Based on the fit results, they suggested three kinds of beam-radius-to-frequency behaviors: a continuously decreasing trend of pulse width with increasing frequency (PSRs B0301$+$19, B0525$+$21, B1237$+$25, and B2045$-$16), a decreasing trend but approaching a constant pulse width at high frequency (PSRs B0329$+$54, B1133$+$16, and B2020$+$28), and a nearly constant trend of pulse width (PSRs B0834$+$06, B1604$-$00, and B1919$+$21).

The decreasing pulse width trend is usually interpreted with a scenario of narrowband emission, called radius-to-frequency mapping (RFM, Komesaroff et al. 1970; Cordes 1978), which assumes that high-frequency emission is generated at a low altitude and vice versa. Different emission models predict different indices for the RFM relationship, $r\propto \nu^\delta$. The index could be $-0.33$ in the inner vacuum gap model (Ruderman \& Sutherland 1975), $-0.45$ in the electron-bremsstrahlung model (Virtamo \& Jauho 1973), or $-0.14$ (or $-0.29$) in the curvature-plasma model (Beskin et al. 1988). In a model involving the propagation of waves in the pulsar magnetosphere (Barnard \& Arons 1986), the index may vary between $-0.5$ and 0 depending on the plasma density distribution and wave mode regimes (both narrowband and broadband scenarios were studied in this model). Dyks et al. (2010) proposed an alternative broadband interpretation in their stream-like fan beam model, which assumes that each thin and elongated plasma stream forms a split-fan beam via the curvature radiation. In this model, high-frequency pulse narrowing is not caused by the RFM, because the broadband emission is assumed to come from a narrow range of altitude; instead, it results from both the plasma density gradient at the outskirts of the stream and the intrinsic effect of the curvature beam.

Obviously, the phenomena of constant and increasing trends of pulse width with increasing frequency cannot be explained by the traditional RFM scenarios with negative indices. According to the empirical morphology classification for radio pulsars developed by Rankin (1983a), the profiles of PSRs B0834$+$06, B1604$-$00 and B1919$+$21 are dominated by the inner conal component (Rankin 1990); therefore, MR02 suggested that the inner conal component may show a different type of behavior from the outer conal component. In a more physical model invoking the inverse Compton scattering (ICS) of megahertz-frequency waves by relativistic particles to generate the radio emission, it is suggested that the outer cone follows the RFM-type frequency dependence, while the inner cone follows the opposite trend and the core component only shows a marginal pulse-width narrowing throughout the radio waveband (Qiao 1988, Qiao \& Lin 1998, Qiao et al. 2001). In the stream-like fan beam model (Dyks et al. 2010), the absence of pulse width variation is explained as a consequence of the fixed extent of the stream in the magnetic azimuth and the broadband nature of the emission.

Although this issue has been studied extensively, only dozens of pulsars have been involved in those studies, and pulsars with
the RFM-type frequency dependence have become the focus of those studies. The largest sample used to investigate the pulse width
change is a group of 87 pulsars in Kijak et al. (1998, hereafter K98), who compared the pulse widths $W_{10}$ at
1.4~GHz and 4.85~GHz. Their results show that 57\% of the pulsars exhibit a decrease in pulse width fraction\footnote{The pulse width fraction is defined as $(W_{4.85}-W_{1.4})/W_{1.4}$, where $W_{1.4}$ and $W_{4.85}$ are the pulse widths at 1.4~GHz and 4.85~GHz, respectively.} exceeding 5\%, 41\% of the pulsars show marginal pulse-width variation, namely, the absolute pulse width fraction being less than 5\%, and only 2\% of the pulsars show an increment of pulse width fraction exceeding 5\%. However, this is still a small portion of the 1079 pulsars in the profile database of the European Pulsar Network (EPN), where thousands of multifrequency profiles are archived.

This paper endeavors to perform a census of the frequency dependence of the pulse widths for normal pulsars by using the EPN database as well as some other small data sets in the literature. More than 170 normal pulsars with multifrequency profiles are selected following several criteria. The Thorsett relationship is  used to fit the relationship between the pulse width and the observing frequency. The paper is organized as follows. In Section 2, we describe our data reduction method. In Section 3, the results of the best-fit parameters and the grouping of the sample in terms of the relative fraction of pulse width change are presented. In section 4 we discusses possible explanations for the diverse frequency dependence of pulse width, but we mainly focuses on a new broadband interpretation. In Section 5, we test the geometric effect wherein the shrinkage of beam size may induce steepening of the spectrum. We present our conclusions and discussions in Section 6.

\section{DATA REDUCTION}

The EPN database is currently the largest multifrequency pulse profile database. Since the latest update of EPN in 2005, there have been few new multifrequency data sets published in literature. We note that Johnston et al. (2008) published pulse width data at five frequencies from 243~MHz to 3.1~GHz for 34 pulsars, of which three are not archived in the EPN. The 3.1~GHz profiles fill the gap between 2~GHz and $\sim$5~GHz for those pulsars archived in the EPN database. In addition to the data for this paper, we also collected the profile data from some other papers, that are complementary to the EPN. A total of 171 normal pulsars are selected from the EPN database and the complementary papers with the following criteria. (1) The signal-to-noise ratio of pulse profile must be high enough to ensure that the measured $W_{10}$ is not affected by the noise near the edge of pulse wings. (2) There are at least four available frequencies, and the highest frequency is at least three times the lowest one for each pulsar, in order to keep a relatively wide frequency range and enable a reliable fit for the pulse width-frequency relationship. (3) The average pulse profiles are not significantly affected by the interstellar scattering effect or poor sampling of phase bins. Those profiles with less than 200 sampling bins are excluded.\footnote{Most pulsars in the sample have 256 or more phase bins, except the following ones with the bin numbers between 200 and 256 at particular frequencies: PSR B0316$+$57 (408~MHz, 244 bins), B0540$+$23 (408~MHz, 209), B0740$-$28 (610~MHz, 244), B1556$-$44 (1560M, 243), B1735$-$32 (1400~MHz, 252), B1819$-$22 (410~MHz, 254), B1900$-$06 (1642~MHz, 223), B1930$+$22 (925~MHz, 250; 1408~MHz, 235), B1933$+$16 (610~MHz, 242; 1642~MHz, 229), B2027$+$37 (410~MHz, 216), B2053$+$36 (610~MHz, 244), B2324$+$60 (606~MHz, 205, 1600~MHz, 204).} The frequency range differs in different pulsars (see Tables 1-4), from the widest one within 400~MHz$-$32~GHz to the narrowest one within 400~MHz$-$1.4~GHz, depending on the quality of pulse profiles available. The references related to the used profiles of these pulsars are listed in Table 1, where the papers with profiles not archived in the EPN are marked with ``*", and the others are all archived in the EPN.

The most commonly published pulse width data are $W_{50}$ and $W_{10}$. For many pulsars with double or multiple components,
the heights of the outermost components are below the half maximum peak height at some frequencies. This will cause irregularity in the frequency evolution of $W_{50}$. In order to reduce
the influence of this effect, we prefer to use $W_{10}$, which is measured at the 10\% level of the maximum peak of pulse profile.
Our results show that selecting $W_{10}$ does provide acceptable fit in most cases.
When the $W_{10}$ data measured with the same method are available, we adopted the values from the literature
(see the papers marked with ``$^\dagger$" in Table 1, otherwise we measured the
width with the profiles downloaded from the EPN database or reproduced from the relevant papers. When the error of $W_{10}$ is not available from the literature, it is estimated by counting the contribution from the sampling time interval, the dispersion smearing and the scattering, of which the latter two are determined using the published dispersion measure, the observing frequency and the channel bandwidth in the relevant observations. The total error is the square root of the sum of the squares of three uncertainties.

We have discarded some $W_{10}$ data that are strongly affected by the evolution of the profile shape and hence violate the Thorsett relationship. In all of these cases one of the leading and trailing components is apparently below the 10\% peak intensity level at some frequencies while becoming prominent at the other frequencies; therefore, the $W_{10}$ data measured when the outrider is too weak are inconsistent with the other data.\footnote{They are PSR B$0355+54$ below 900~MHz (weak leading component, Gould \& Lyne 1998), PSR B1730$-$22 at 243~MHz (weak trailing component, Johnston et al. 2008), PSR B1822$-$09 at 400~MHz (weak leading component, Gould \& Lyne 1998), PSR B1857$-$26 at 243~MHz (weak leading component, Johnston et al. 2008).}

The $W_{10}$ data are fitted versus the observing frequencies $\nu$ (in units of GHz) with the Thorsett relationship:
\begin{equation}
W_{10}=A\nu^{\mu}+W_{\rm 10, 0}, \label{eq: thorsett
expression}
\end{equation}
where the best-fit coefficient $A$, asymptotic constant $W_{\rm 10, 0}$, and index $\mu$ are obtained with the weighted Levenberg$-$Marquardt nonlinear least-square fitting algorithm. $A$ and $W_{\rm 10, 0}$ are constrained to be non-negative values. The parameter errors at the 95\% confidence level are estimated by searching for the parameter space that satisfies $\chi^2\leq\chi_{\rm min}^2+\Delta\chi^2$ within the linearly spaced grids around the optimized values of these three parameters, where $\chi_{\rm min}^2$ is the minimized least square and $\Delta\chi^2$ is the chi-square increment corresponding to the 95\% confidence level for a degree of freedom $N-3$, where $N$ is the number of data points for each pulsar and 3 is the number of free model parameters (Press et al. 2007).

The probability $Q$ that $\chi^2$ exceeds $\chi_{\rm min}^2+\Delta\chi^2$ by chance is determined with $Q=1-P(N/2-1,\chi^2/2)$, where $P(a,x)$ is the incomplete gamma function defined as $P(a,x)=\int_0 ^x e^{\rm -t} t^{\rm a-1} {\rm d}t /\Gamma(a) (a>0)$ (Press et al. 2007). The following two considerations are taken into account to judge whether the fit is acceptable or not. (1) If $Q$ is extremely small (equivalently very large reduced $\chi^2$), and the data points are very scattered without clearly showing a monotonic trend, then the fit will be rejected. When $Q$ is very small (mainly influenced by the data points with small errors in a mildly scattered data set) but the best-fit curve can well represent the monotonic trend of pulse width variation, the results are still accepted. (2) In order to compare the difference in pulse width variation between pulsars, we determine the pulse widths at 0.4~GHz and 4.85~GHz using the best-fit relationship and then determine the fraction of pulse width change for each pulsar. However, we find that in some pulsars of which the best-fit relationship is obtained from a limited frequency range, e.g., lower than 3~GHz or higher than 0.6~GHz, the extrapolated pulse widths are incredibly too large, although the $Q$ values look acceptable. We exclude such cases with extrapolation problems. We eventually obtain a sample of 150 pulsars with acceptable results out of the total 171 pulsars. Among the 21 discarded pulsars, three cases have poor $Q$ values and non-monotonic trends, while the others have the extrapolation problem.

\section{RESULTS}

The best-fit model parameters and their 95\% confidence intervals are listed in Columns 3-5 in Tables 2-4, together with the other parameters related to the fit in Columns 2, 6-8, including the observing frequency range, the number of data points $N$, the minimum chi-square $\chi_{\rm min}^2$ and the probability $Q$. The best-fit $W_{10}-\nu$ relationship is presented for each pulsar in Figures 1-3. We should note that the confidence intervals are normally broad, because the three free parameters in the Thorsett relationship are highly correlated. For example, if $W_{10}$ is nearly constant at multiple frequencies, e.g., in PSRs B0105$+65$ and B$0450-18$ (see Figure 2 for the $W_{10}-\nu$ curves), the best-fit $A$ would be close to 0, then the relationship would be insensitive to $\mu$, which means that the uncertainty of $\mu$ would be very large.

\begin{figure*}
\centering
\resizebox{16cm}{17cm}{
\includegraphics{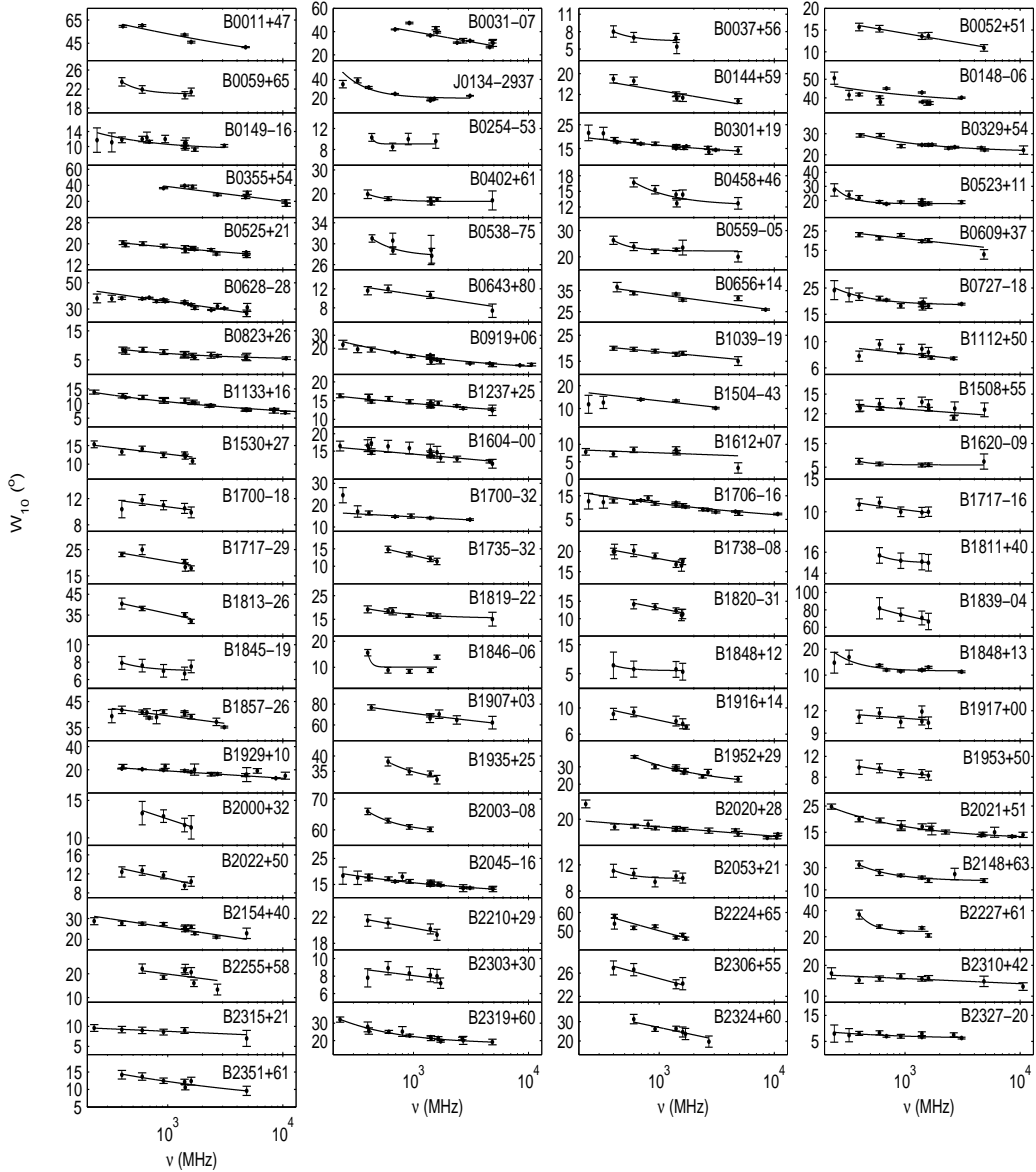}
}
\caption{10\% peak pulse width $W_{10}$ vs. the observing frequency $\nu$ for 81 group-A pulsars. The best-fit Thorsett relationship is presented as a solid curve for each pulsar. All of these pulsars have $\eta<-10$\%, where  $\eta=(W_{4.85}-W_{0.4})/W_{0.4}$ is the relative change of $W_{10}$ between 4.85~GHz and 0.4~GHz. See the text for details.}
   \label{Fig:groupa}
\end{figure*}

\begin{figure*}
\centering
\includegraphics[width=16cm]{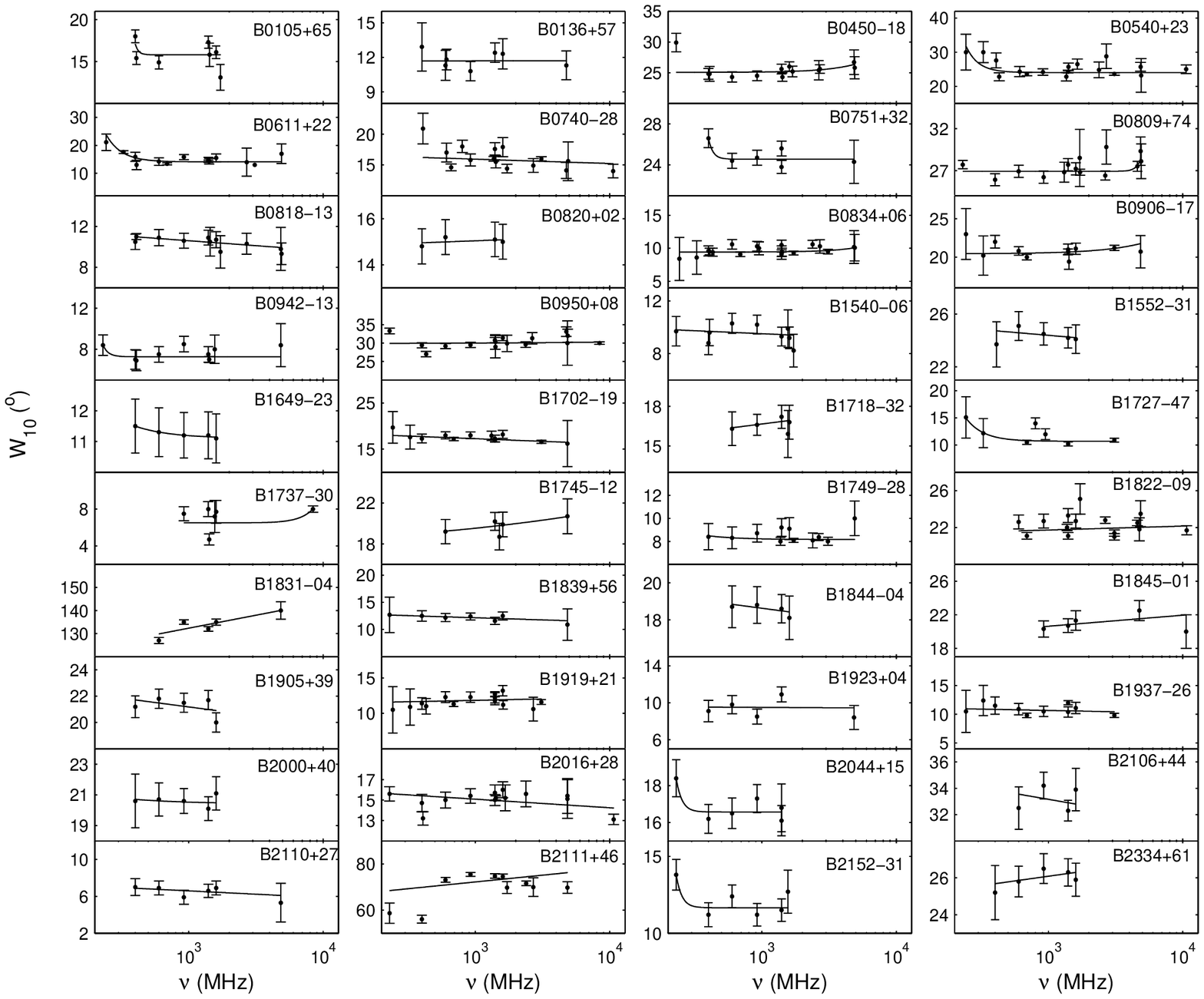}
\caption{$W_{10}-\nu$ diagrams for 40 group-B pulsars. The best-fit Thorsett relationship is presented as a solid curve for each pulsar. The criterion for this group is $|\eta|\leq 10$\%.}
   \label{Fig:groupb}
\end{figure*}

\begin{figure*}
\centering
\includegraphics[width=16cm]{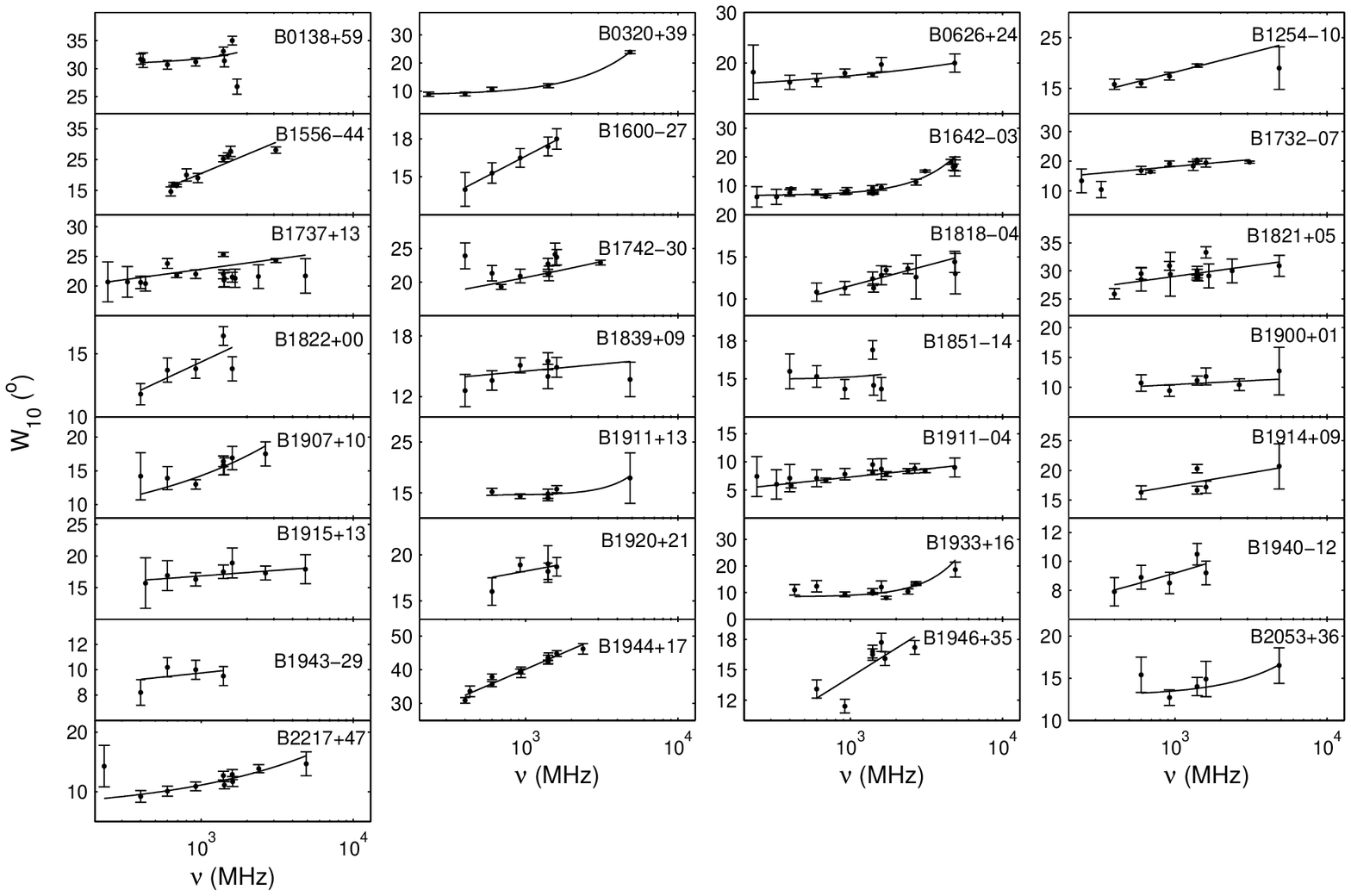}
\caption{$W_{10}-\nu$ diagrams for 29 group-C pulsars. The best-fit Thorsett relationship is presented as a solid curve for each pulsar. The criterion for this group is $\eta> 10$\%.}
   \label{Fig:groupc}
\end{figure*}

Below, we first present the statistical properties of the best-fit parameters and demonstrate that any single parameter is not suitable for classification, and then we show the global picture of the relative fraction of pulse width change determined using the best-fit relationship, on the basis of which the pulsars are divided into three groups as listed in Tables 2-4. Finally, we investigate whether different groups are associated with different physical parameters and morphology classes.

\subsection{Statistical Properties of the Best-fit Parameters}

$\mu$ is a parameter that can directly reflect the decreasing or increasing trend of pulse width evolution.
Among the 150 pulsars, 105 have $\mu<0$ and the other 45 have $\mu>0$, irrespective of their errors. However, it is inappropriate to use $\mu$ as a single parameter to classify the frequency dependence
of pulse width because of the following two reasons. First, the distribution of $\mu$ (Figure 4), mostly concentrated in $-1\leq\mu\leq1$, is continuous and peaked near 0. Considering the uncertainty of $\mu$ for those pulsars close to $\mu\sim0$, it is impossible to identify a clear boundary for negative and positive indices. Second, the Thorsett relationship, due to the presence of a third parameter $W_{10, 0}$, makes $\mu$ not always a straightforward and thorough parameter to describe the frequency dependence of pulse width, in other words, a large $|\mu|$ does not always represent a steep trend throughout the frequency range. For example, let us compare PSRs
B0942$-$13 and B0052$+$51, where $\mu=-10.9$ and $W_{10, 0}=7^\circ .3$ for the former and $\mu=-0.1$ and $W_{10, 0}=0^\circ.1$ for the latter. Obviously, the width change of PSR B0942$-$13 (see Figures 2) is less prominent than that of PSR B0052$+$51 for most of the frequency range above 0.4~GHz (Figure 1), even though the former has a much steeper $\mu$ than the latter, because a larger $W_{10, 0}$
makes the former relationship become much flatter at high frequencies.
%%------------ Fig 1:  cdf plot of W10,0
\begin{figure*}
\centering
\resizebox{12cm}{9cm}
{\includegraphics{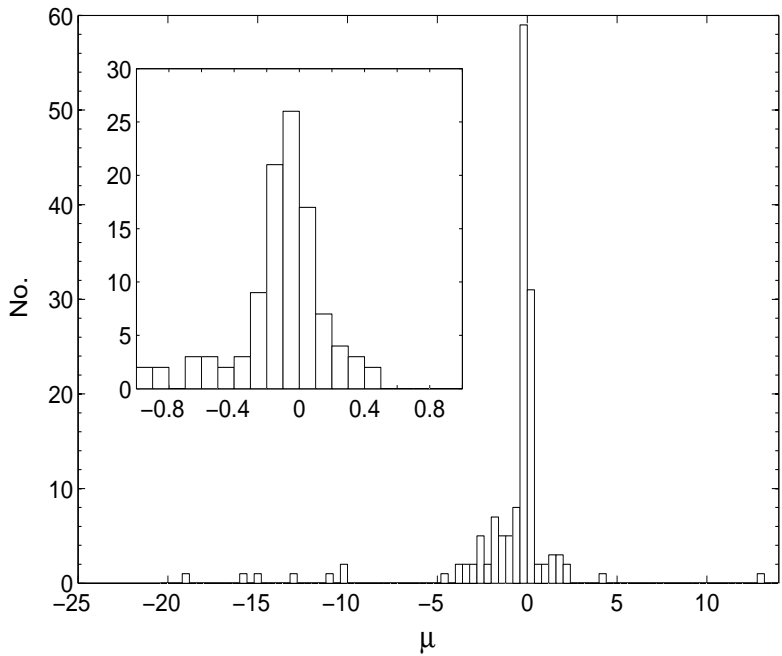}}
\caption{Histogram of $\mu$ for 150 pulsars with a bin size of 0.4. The inset is the snapshot for the distribution in $-1\leq\mu\leq1$ with a bin size of 0.1.}
\label{Fig:muhist}
\end{figure*}
%
%figure 4

 $W_{10, 0}$ is a quantity that the pulse width asymptotically approaches as the frequency increases, in the case $\mu<0$ or as the frequency decreases, in the case $\mu>0$. MR02 used the asymptotic term $\rho_0$ in another Thorsett relationship, $\rho=\rho_0+(\nu/\nu_0)^\mu$, to distinguish the 10 pulsars studied into 3 groups: the first group consisted of four pulsars with $\mu<0$ and a very small $\rho_0$, the second group consisted of three pulsars with $\mu<0$ and larger $\rho_0$ values than those in the first group, and the third group consisted of three pulsars with nearly a constant beam radius ($\mu$ is set as 0 in their fit). In other words, the former two groups both show a decreasing trend of pulse width, but the relationship of the first group is much closer to a pure power law. Since $W_{10, 0}$ and $W_{10}$ can be converted to the beam radii $\rho_0$ and $\rho$ if we know the inclination angle between the spin and magnetic axes and the impact angle between our line of sight (LOS) and the magnetic axis, it is worth examining whether MR02's finding also exists in our sample.

We plot the sequence diagrams of $W_{10,0}$ for a sub-sample of 105 pulsars with $\mu<0$ and a sub-sample of 45 pulsars with $\mu>0$, as shown in Figures 5(a) and (b), respectively, where $W_{10,0}$ and the 95\% confidence intervals (gray horizontal lines) are displayed one after another vertically. In panel (a), it seems that there is a ``gap'' between $W_{10, 0}\simeq 0^\circ.4$ and $3^\circ.5$, as shown by the two dashed lines, wherein only two data points are located. Although the confidence intervals of most pulsars are large and cover the gap region, the probability that such a gap is purely a coincidence seems to be very small. When only the data with $\mu<-0.1$ are used to avoid the uncertainty in the sign of $\mu$ for those pulsars with $\mu\sim0$, the gap still exists. However, in panel (b), there is no evidence for such a gap, even when the data with $\mu>0.05$ are used. Because a very small $W_{10,0}$ can also be related to a $\rho_{0}$ that is not as small, depending on the viewing geometry, it would be interesting to study whether this gap still holds when converting $W_{10,0}$ to $\rho_{0}$, which demands a sample of pulsars with well-constrained inclination and impact angles. Since in this paper we focus on the general trends of pulse width change, the gap problem of $W_{10, 0}$ will be investigated elsewhere.

 $A$ is highly correlated with the other two parameters, as can be seen from the $\mu-A$ and $W_{\rm 10, 0}-A$ diagrams in Figure 6. Generally, very small $A$ values are related to very large (positive) or very small (negative) $\mu$ and relatively large $W_{10, 0}$ values. This situation usually occurs when the pulse width undergoes a much steeper decreasing (or increasing) trend in a small fraction of frequency range near the low (or high) frequency end than in the other frequency range, with the result that a prominent negative or positive index is needed to model the abrupt change, and meanwhile a very small $A$ is needed to cancel its influence in the other frequency range. PSR B0254$-53$ and PSR B0523$+11$ are of this kind (see Figure 1) for the width-frequency curves). For those pulsars that can be modeled by a relationship close to a pure power law, i.e., $W_{10, 0}\sim0$, a relatively large $A$ and a normal index are needed, as can be seen from both panels in Figure 6.

%%------------ Fig :  W10,0
\begin{figure*}
\centering
\resizebox{10cm}{9cm}
{\includegraphics{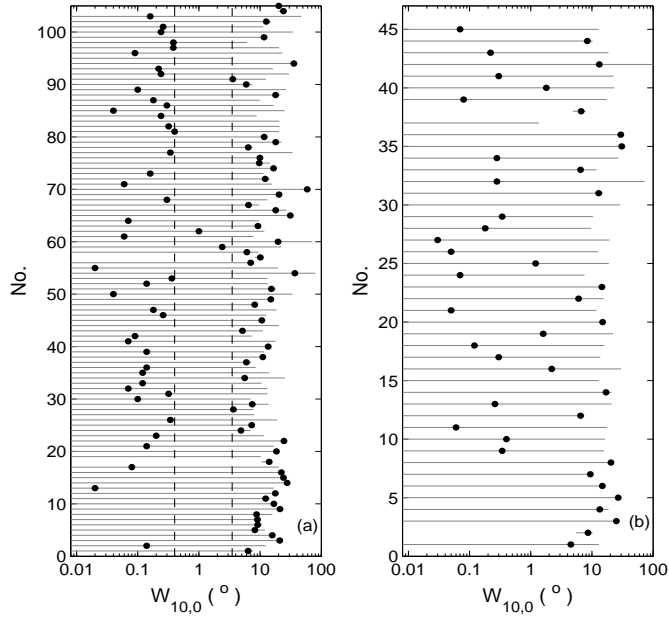}}
\caption{Sequence diagrams of $W_{10, 0}$ for the sub-samples of pulsars with (a) $\mu<0$ and (b) $\mu>0$. The ordinate is the sequence index of each pulsar in a sub sample. The horizontal gray lines are the error bars of $W_{10,0}$. }
\label{Fig:sd}
\end{figure*}
%

%%-- Fig :  A  (figure 5)
\begin{figure*}
\centering
\resizebox{8cm}{9cm}
{\includegraphics{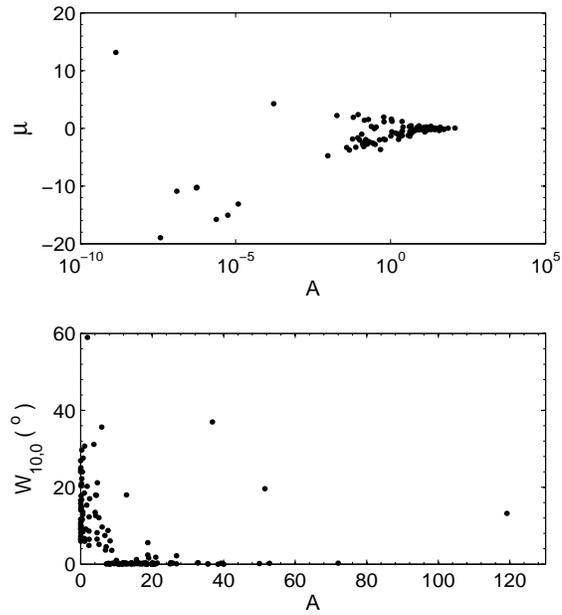}}
\caption{Diagrams for the best-fit parameters $\mu$, $A$ and $W_{\rm 10, 0}$ of the Thorsett relationship for 150 pulsars.}
\label{Fig:A}
\end{figure*}
%
%%-- (figure 6)
\subsection{Relative Fraction of Pulse Width Change}
The above results show that using any single parameter of $A$, $\mu$, or $W_{10,0}$ is not enough to identify different frequency-dependent behaviors of the pulse width. In order to assess the pulse width evolution, we selected three commonly used low, middle and high frequencies in the EPN database, that is, 0.4~GHz, 1.4~GHz, and 4.85~GHz, and calculated two fractions of pulse width change, $\eta=(W_{4.85}-W_{0.4})/W_{0.4}$ and $\eta^\prime=(W_{4.85}-W_{1.4})/W_{1.4}$, where $W_{0.4}$, $W_{1.4}$ and $W_{4.85}$ are the 10\% pulse widths determined using the best-fit Thorsett relationship at the subscripted frequencies. The advantages of using these fitted pulse widths rather than the observed data are as follows: (1)this method can reduce the bias caused by the random uncertainties in the observed data, especially when there are two different width values at a single frequency, and (2) this method can be applied to the pulsars without the observation at 4.85~GHz, and hence expand the sample. These fractions, to a large extent, can describe different cases of pulse width variation, no matter that the pulse width decreases sharply at low frequencies and then mildly at high frequencies (due to large $|\mu|$ and $W_{10,0}$), or varies steadily throughout the frequency range (due to $W_{10,0}\sim 0$). Another reason to select 1.4~GHz and 4.85~GHz is to compare our results with those given by K98, who used the same frequencies to calculate $\eta^{\prime}$ for a sample of 87 pulsars, of which 66 are also used in our sample. We are curious to see what is new when the sample is nearly doubled.
The results are listed in Columns 9 and 10 in Tables 2-4. The 95\% confidence intervals of $\eta$ are simultaneously determined when performing a grid search for the confidence intervals of the best-fit parameters, as presented in Column 9 of Tables 2-4.

The histograms of $\eta$ and $\eta^\prime$ are presented in Figure 7, which shows continuous distributions. The global feature is now very clear: many pulsars show considerable pulse width shrinkage, some pulsars have marginal variation, and the remaining pulsars exhibit notable width increment. This single-humped continuous distribution, similar to the continuous distribution of $\mu$, suggests that there are no clean boundaries for classifying the frequency dependence of pulse width into some completely different types. Nevertheless, as a practical or a phenomenological choice, it is still possible to select a reference value of $\eta$ to separate the pulsars into three groups, which show a remarkably decreasing trend (hereafter group A), a marginal width variation (hereafter group B) and a considerably increasing trend (hereafter group C), respectively. It should be
emphasized that such a criterion does not indicated an explicit physics boundary for different groups.

According to the distribution of $\eta$, we use 10\% as the criterion for the three groups, namely, group A with $\eta<-10$\%, group B with $-$10\%$\leq\eta\leq 10$\% and group C with $\eta> 10$\%. The boundaries of the groups are also shown by the dashed lines in Figure 7(a). We select 10\% rather than a smaller percentage because, on the one hand, this percentage is close to the uncertainty of $\eta$ (at the 1$\sigma$ level) for most pulsars in group B, on the other hand, the 10\% change of the typical pulse width of $20^\circ$ at 0.4 GHz is $2^\circ$, which is comparable to the typical uncertainty of $1^\circ-2^\circ$ of the observed $W_{10}$ data. Therefore, for group B, it means roughly that these pulsars have a marginal width variation within the errors. Based on this criterion, groups A, B ,and C have 81, 40 and 29 pulsars, occupying 54\%, 27\% , and 19\% of the sample, respectively. In group B, the number of pulsars with decreasing pulse width is about 50\% higher than that of pulsars with increasing pulse width irrespective of the error of $\eta$. In group C, a few pulsars have $\eta$ values larger than 100\%, which are not seen in group A. This is caused by the dramatically increasing trend of pulse width in these pulsars, as can be seen from the width-frequency curves in Figure 3 for PSRs B0320$+39$, B1642$-$03, B1946$+$35, B2217$+$47, etc. Note that all these large fractions are for the pulsars with data above 3~GHz (one at 3.1~GHz and the others above 4.7~GHz), so they are reliable.

Our results of $\eta^\prime$ largely modify the global picture obtained by K98. In Figure 7(b), the histogram of $\eta^\prime$ (black) is compared with the histogram of the same width fraction taken from K98 (red). A striking difference is that our results show a much higher percentage of pulsars with a pulse width broadening. Note that $|\eta^\prime|$ is always smaller than $|\eta|$, we use a criterion of 5\% to group the pulsars and compare them with K98's results. It is found that 42.0\% of the sample (63 pulsars) have $\eta^\prime<-5\%$, similar to 41\% in K98, 36.7\% (55 pulsars) have $-5$\%$\leq\eta^\prime\leq 5$\%, smaller than 57\% in K98, and 21.3\% (32 pulsars) have $\eta^\prime> 5$\%, considerably larger than 2\% in K98.

Based on a large sample, our results reveal a more comprehensive picture of the frequency evolution of the pulse width, particularly at a point where the pulsars with a weak pulse width variation and a prominent high-frequency profile broadening are nearly numerous those with a notably decreasing trend (groups B$+$C : group A $=$ 46\% : 54\%).

The group-C pulsars have been largely overlooked by many quantitative studies related to the pulse width evolution. Among these works, the first type is the study of the frequency dependence of pulse width or component separation (see reference in Table 5), the second type is the study of the RFM index (see reference in Table 6). In order to clarify this point, we collected 110 pulsars studied by the first type (Table 5) and the 45 pulsars studied by the second type (Table 6). In the first type of study, only seven pulsars are clearly identified with the increasing trend of pulse width or component separation, as marked with a symbol ``$\dagger$'' in Table 5.\footnote{The component separations of PSRs B0834$+$06 and B1919$+$21 were found to increase with increasing frequency by several independent studies, e.g., Lyne et al. (1971), Sieber et al. (1975), Rankin (1983b), Slee et al. (1987) and Hassall et al. (2012, only B1919$+$21). K98 pointed out that the pulse width at 4.85~GHz exceeds that at 1.4~GHz for PSRs B0402+61, B0450-18, B0626+24, B0906-17 and B1818-04. In fact, when examining a wider frequency range, we found that PSR B0402+61 belongs to group A, but PSRs B0450-18 and B0906-17 belong to group B.} In the second type of study, except for PSRs the B0834$+$06, B1604-00 and, B1919$+$21, which are regarded as pulsars without a frequency dependence of beam radius by MR02, all the others are identified as pulsars with negative RFM indices. In fact, according to our results, 36 of the 45 pulsars in Table 6 belong to group A, 12 pulsars belong to group B and the remaining 3 pulsars belong to group C. The main reason that we find different trends for some pulsars is that we use data at a wider frequency range. For instance, when the pulse width data at frequencies above 2.5~GHz are used for PSR B1604$-$00, we found a clear decreasing trend of pulse width (see Figure 1), but MR02 found a nearly constant trend when using the data below 2.5~GHz. In view of the above facts, the pulsars with the nearly constant and the increasing trends, due to their large fraction, should not be neglected when developing a general emission model for radio pulsars.

MR02 suggested that the behavior of the pulse profile shrinkage is normally related to the outer conal component, while the absence of pulse width variation is usually associated with the inner conal component. In order to test this scenario, we collect the morphological types given by Rankin (1990, 1993), as listed in Column ``R90'' in Tables 2-4, where the terms ``single'', ``double'', ``triple'', and ``multiple'' in the table note mean that the pulse profile consists of a single, double, triple, or multiple visible components. For many pulsars belonging to the ``conal single'', the ``conal double'', and the ``triple'' types, it was not well determined in the literature whether the conal components come from the outer cone or the inner cone. However, it is certain that the outermost outriders of ``multiple'' profiles are from the outer cone. The table shows that both the core and the conal components, including the outer cones in ``multiple'' profiles, can contribute to all three groups. We also collected the classification given by LM88, as listed in the last column in Tables 2-4, where they classified four basic types: the core-dominated (``core'' in the table), the cone-dominated (``cone''), cones with cores (``core-cone''), and partial cones (only a part of a cone is bright). There is no correlation between the profile type and the frequency development of pulse width under this classification system.

%
%%------------ Fig:  Histogram of distribution
\begin{figure*}
\centering
\resizebox{9cm}{9cm}
{\includegraphics{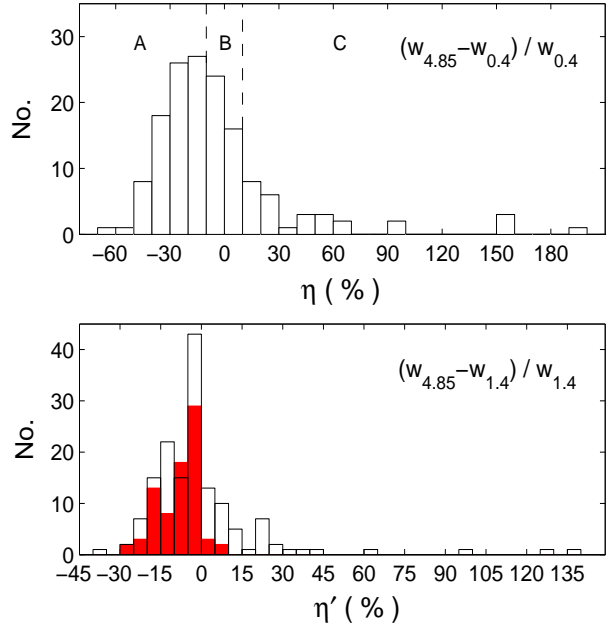}}
\caption{Histograms of the pulse-width change fraction $\eta$ (top) and $\eta^\prime$ (bottom), where $\eta\equiv(W_{4.85}-W_{0.4})/W_{0.4}$ and $\eta^\prime\equiv(W_{4.85}-W_{1.4})/W_{1.4}$ with the subscriptions standing for the frequency in units of GHz, at which the pulse width $W$ is determined using the best-fit Thorsett relationship. The red histogram is a reproduction of Figure 4 in K98, where the statistics were made for the fraction of the observed pulse-width change at 1.4~GHz and 4.85~GHz in terms of a sample of 87 pulsars. }
\label{Fig:eta}
\end{figure*}

%%%%%%%%%%%%%%%%%%figure 7
We also checked whether these groups show any difference in the parameters derived from $P$ and $\dot{P}$, e.g., the magnetic
field, the characteristic age, and the potential drop between the magnetic pole and the edge of the polar cap. However, there is no distinction between their distributions in the $P-\dot{P}$ diagram
(Figure 8), where the pulsars in groups A, B and C are plotted by the black dots, the ``+'' symbols, and squares, respectively.
Therefore, the diversity in the frequency development of pulse widths is unlikely to be related to a single physical parameter.
%
%%------------ Fig :  ppdot
\begin{figure*}
\centering
\resizebox{10cm}{7cm}
{\includegraphics{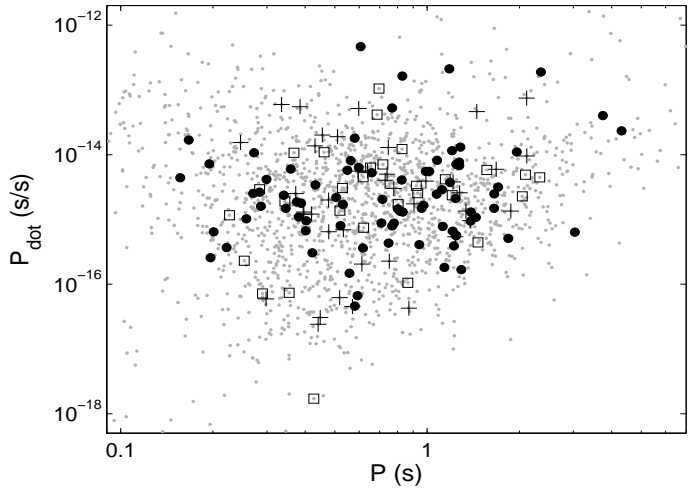}}
\caption{$P-\dot{P}$ diagram for the group-A (black dots), group-B (``+'') and group C (squares) pulsars. The gray dots represent the normal pulsars taken from the ATNF pulsar catalog (Manchester et al. 2005, http://www.atnf.csiro.au/research/pulsar/psrcat).}
\label{Fig:ppdot}
\end{figure*}
%%%%%%%%%%%%%%%%%figure 8
\subsection{Morphology-induced Pulse Width Change}

 Pulse width change is usually associated with the frequency development of pulse morphology, e.g., merging of components, profile bifurcation, or the relative intensity variation between the leading and trailing components. Many authors have noticed that in some pulsars, new outer components emerge in high-frequency profiles, whereas in some other pulsars, low-frequency outer components disappear at high frequencies. The former case may lead to an increasing trend of pulse width while the latter case may lead to a decreasing trend.

We have examined the sample for this effect and found that five pulsars in group A and eight pulsars in group C are apparently influenced by the morphology evolution, including (group A) PSRs J0134$-$2937, B0144$+$59, B0355$+$54 (above 0.9~GHz), B1717$-16$, B1738$-08$, and (group C) PSRs B1556$-$44, B1642$-$03, B1732$-$07, B1821$+$05, B1911$-$04, B1920$+$21, B1944$+$17 and B1946$+$35. For the five group-A pulsars, the decreasing pulse width is caused by the dramatic fading of the outer components at high frequencies. In contrast, for the nine group-C pulsars, the increasing pulse width is induced by the rise of the outer components. Their multifrequency profiles from the EPN database are shown in Figures. 9 and 10, respectively. Unlike some discarded pulsars for which the apparent morphology evolution leads to a non-monotonic frequency dependence of the pulse width, the 13 pulsars presented here follow the Thorsett relationship.

The above pulsars are the extreme cases of morphology evolution. Although the fraction of extreme cases in group C is larger than that in group A, we should stress that the pulse width change associated with a mild morphology evolution is also very common in group C. For instance, one can find several examples in group C showing a single profile broadening with increasing frequency, e.g., PSRs B1818$-$04, B1851$-$14, B1900$+$01, B1915$+$13, B2053$+$36 and ,B2217$+$47 (see the EPN database).

In many previous studies, the association between the merging of outer components in double, triple, and multiple profiles and the decreasing trend in pulse width is widely acknowledged. This is normally interpreted as the shrinkage of the emission cone at high frequency. However, the component merging is only one type of morphology evolution. For many pulsars, including the previously mentioned extreme and mild cases, it is more appropriate to describe their morphology evolution as a picture wherein different parts of the pulse profile follow different spectral behaviors. From our perspective, the frequency dependence of pulse width, together with the morphology evolution, can be interpreted as a consequence of the variation of the emission spectrum across the emission region in the pulsar magnetosphere, as will be presented in detail below.

\begin{figure*}
\resizebox{15.9cm}{4.5cm}
{
\includegraphics{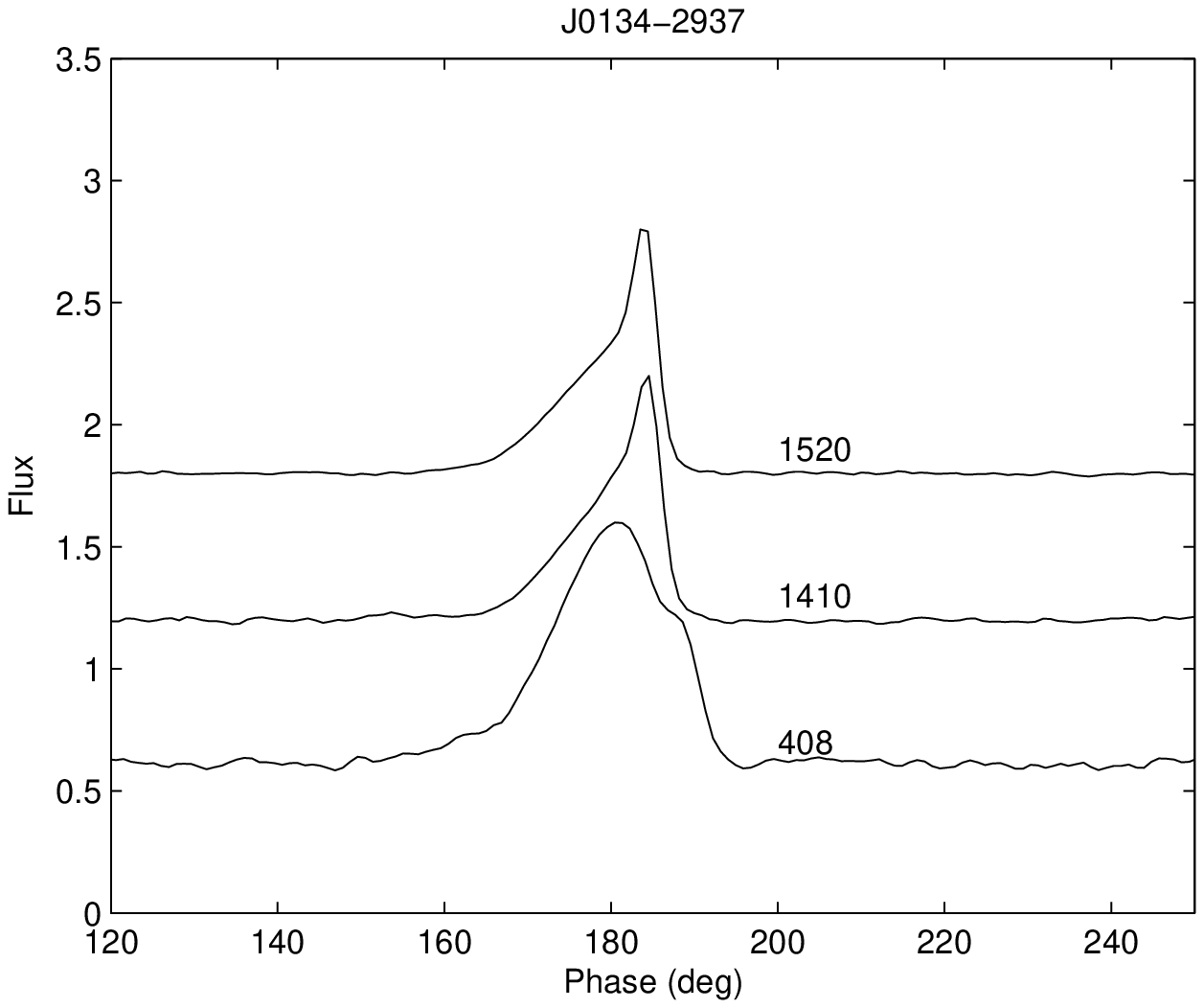}
\includegraphics{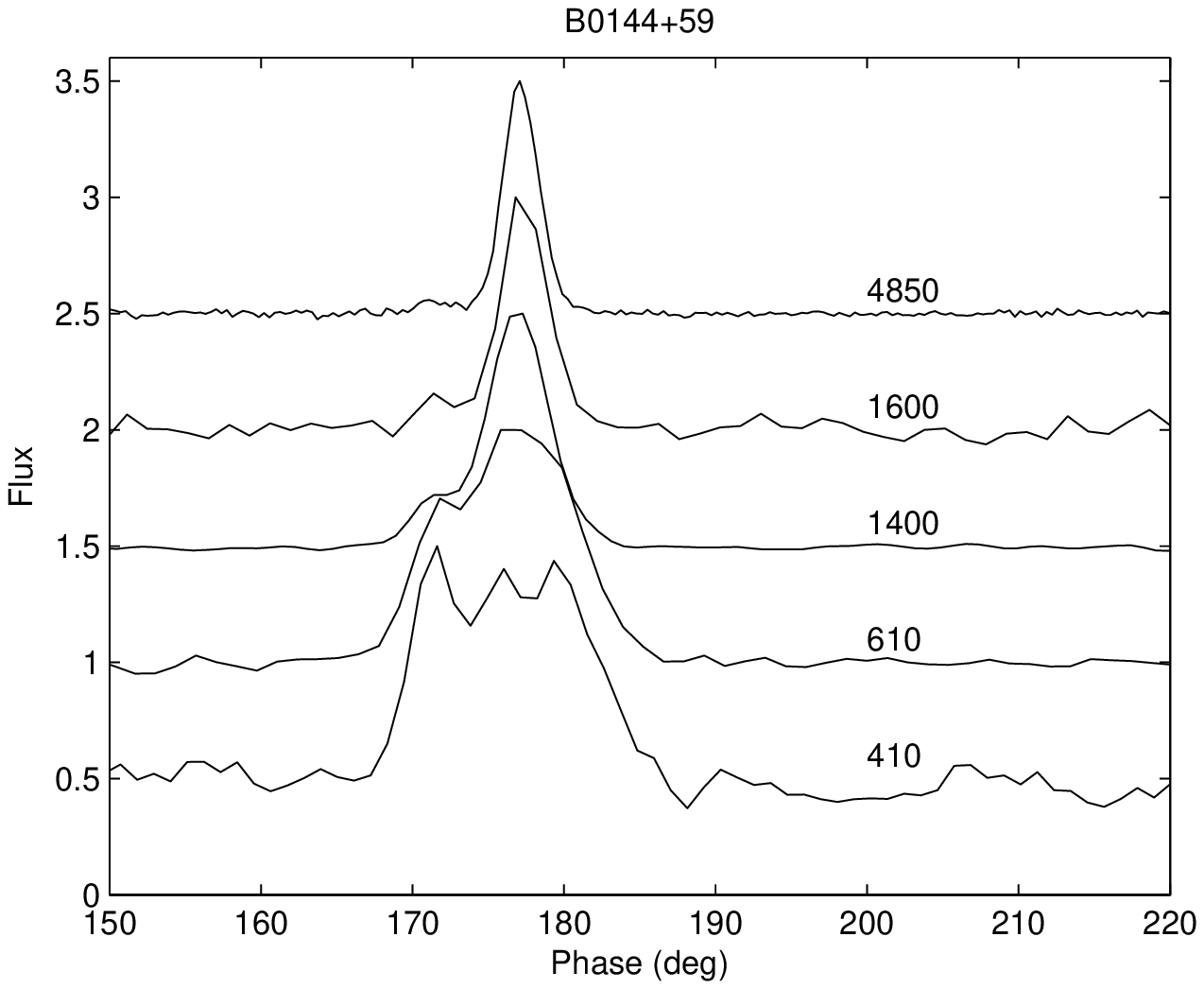}
\includegraphics{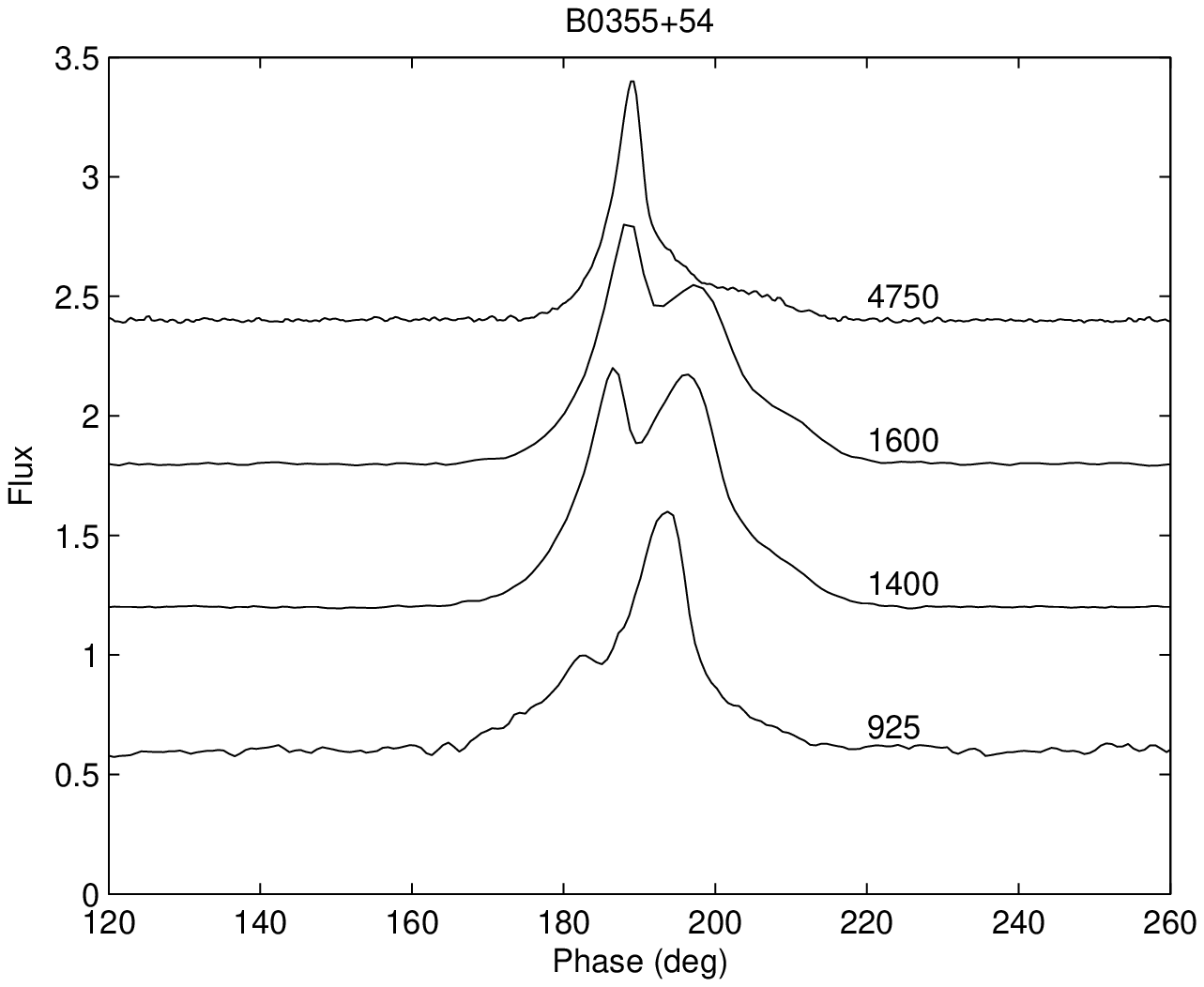}
}
\resizebox{10.6cm}{4.5cm}
{
\includegraphics{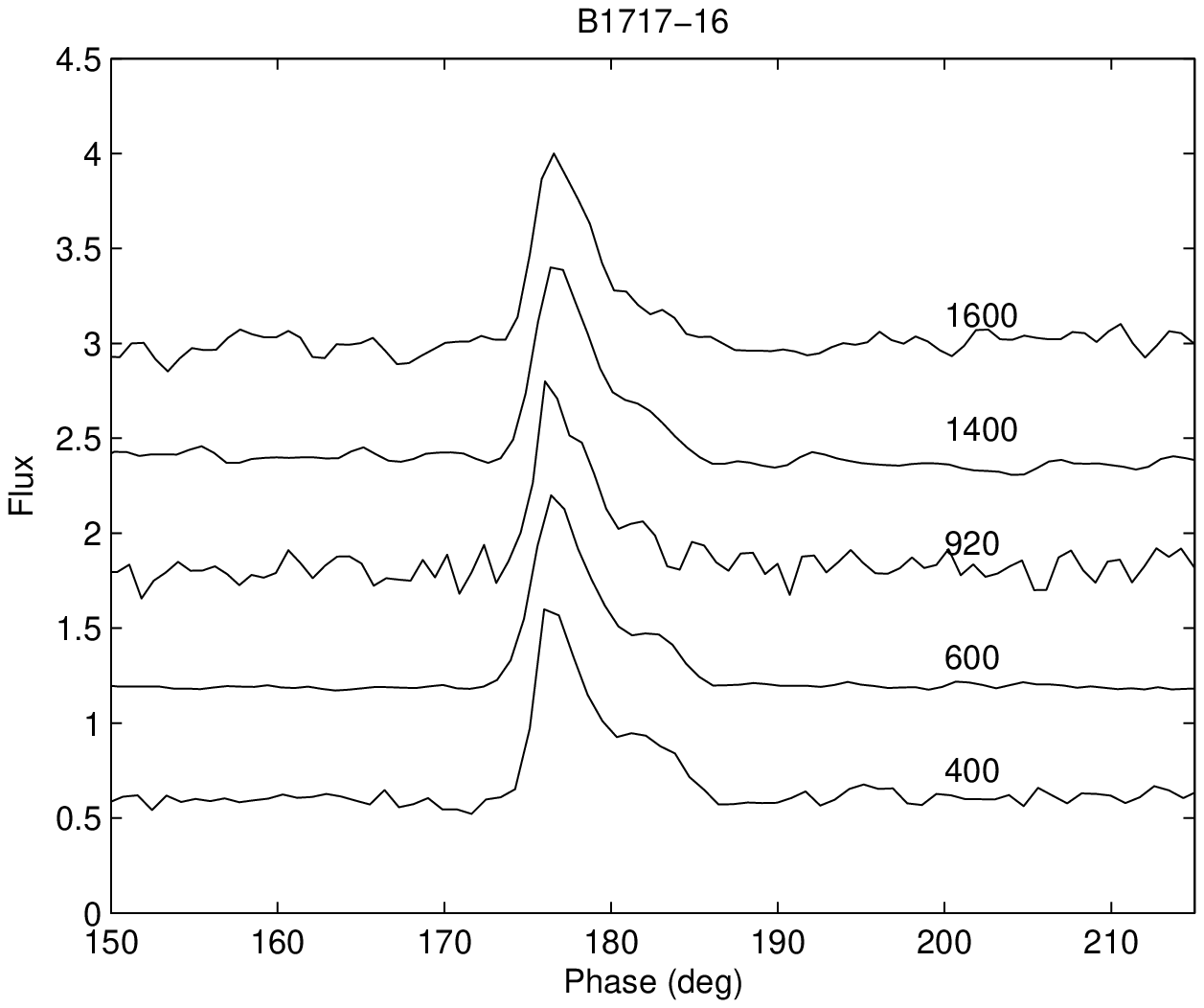}
\includegraphics{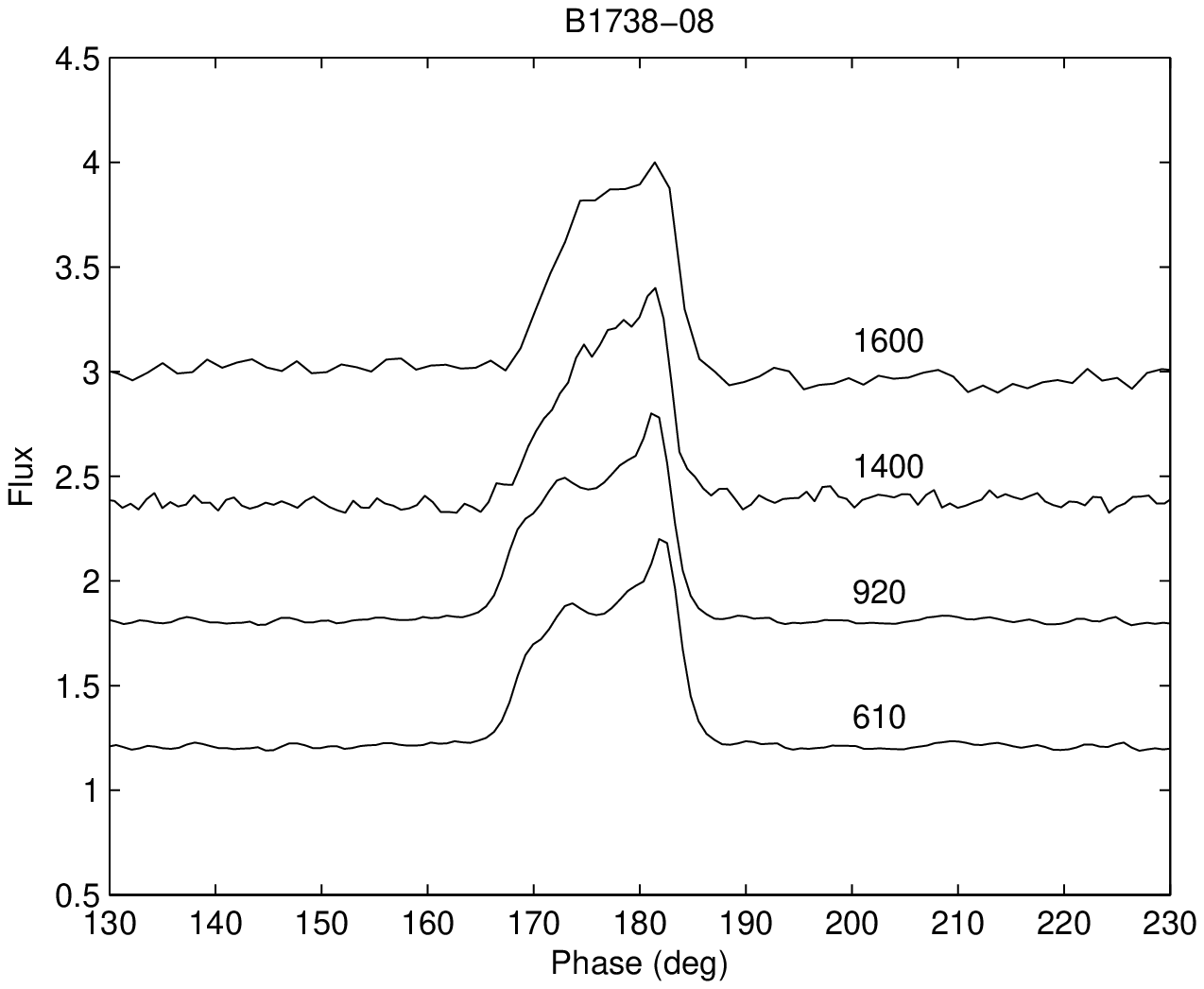}
}
\caption{Normalized multifrequency profiles of five pulsars in group A, for which the increasing trend of pulse width variation is apparently induced by the fading of the outermost components in the pulse profiles. The data are taken from the EPN database. The observing frequency is marked for each profile in units of MHz. The profiles are aligned in phase by eye.}
   \label{Fig:Aprof}
\end{figure*}
%
%%%%%%%%%%%%%%figure 9
%
\begin{figure*}
\resizebox{15.9cm}{4.5cm}
{
\includegraphics{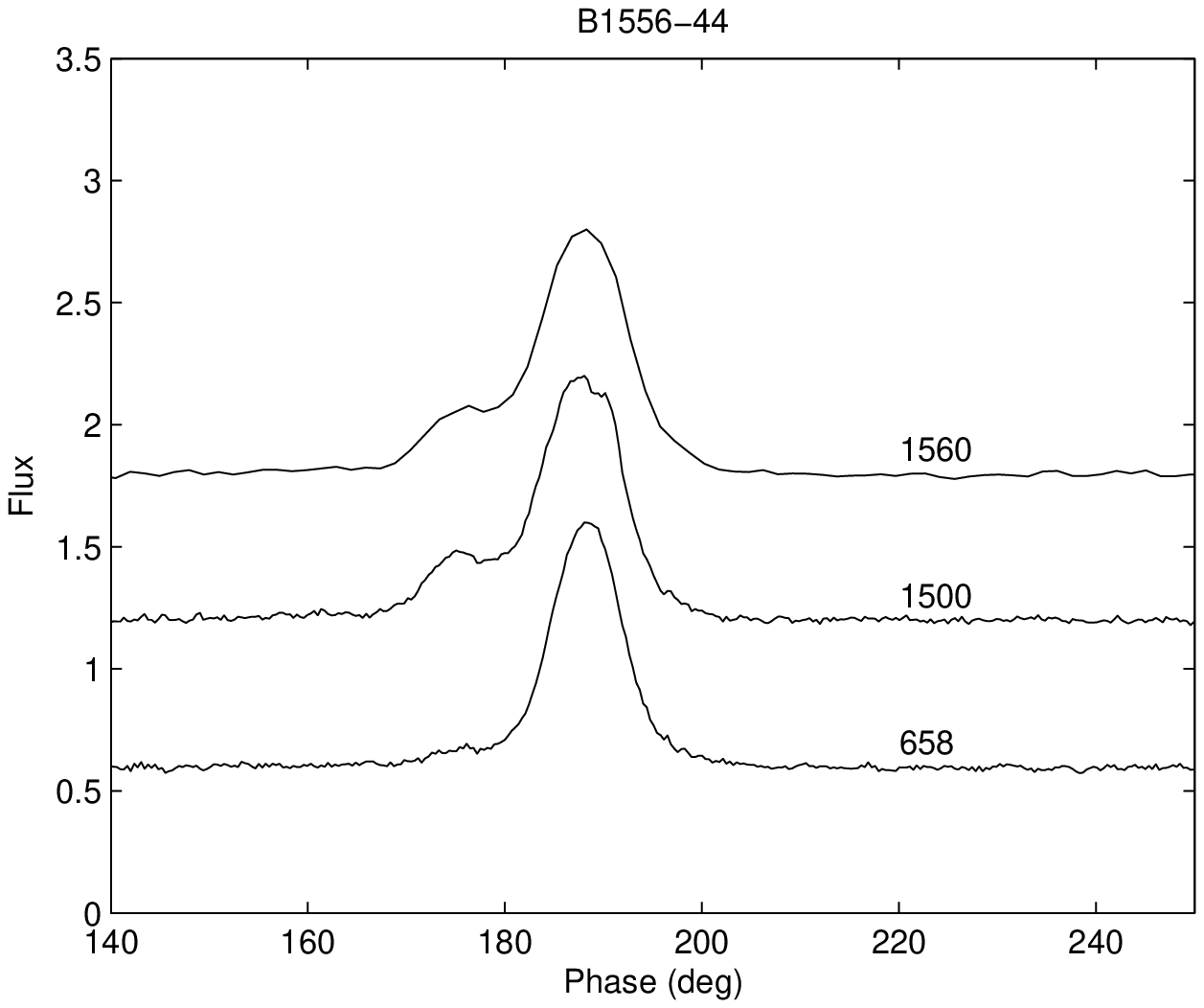}
\includegraphics{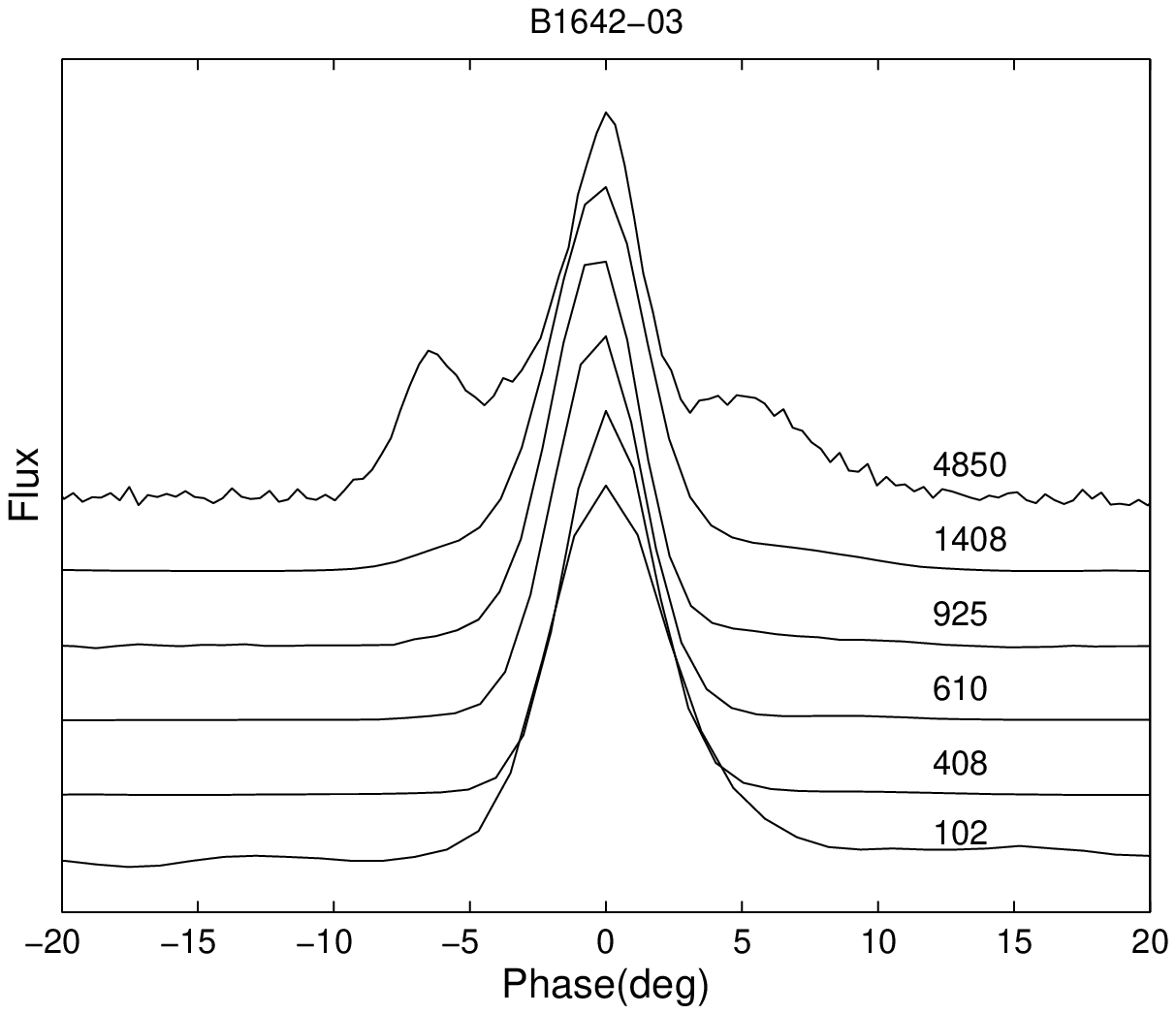}
\includegraphics{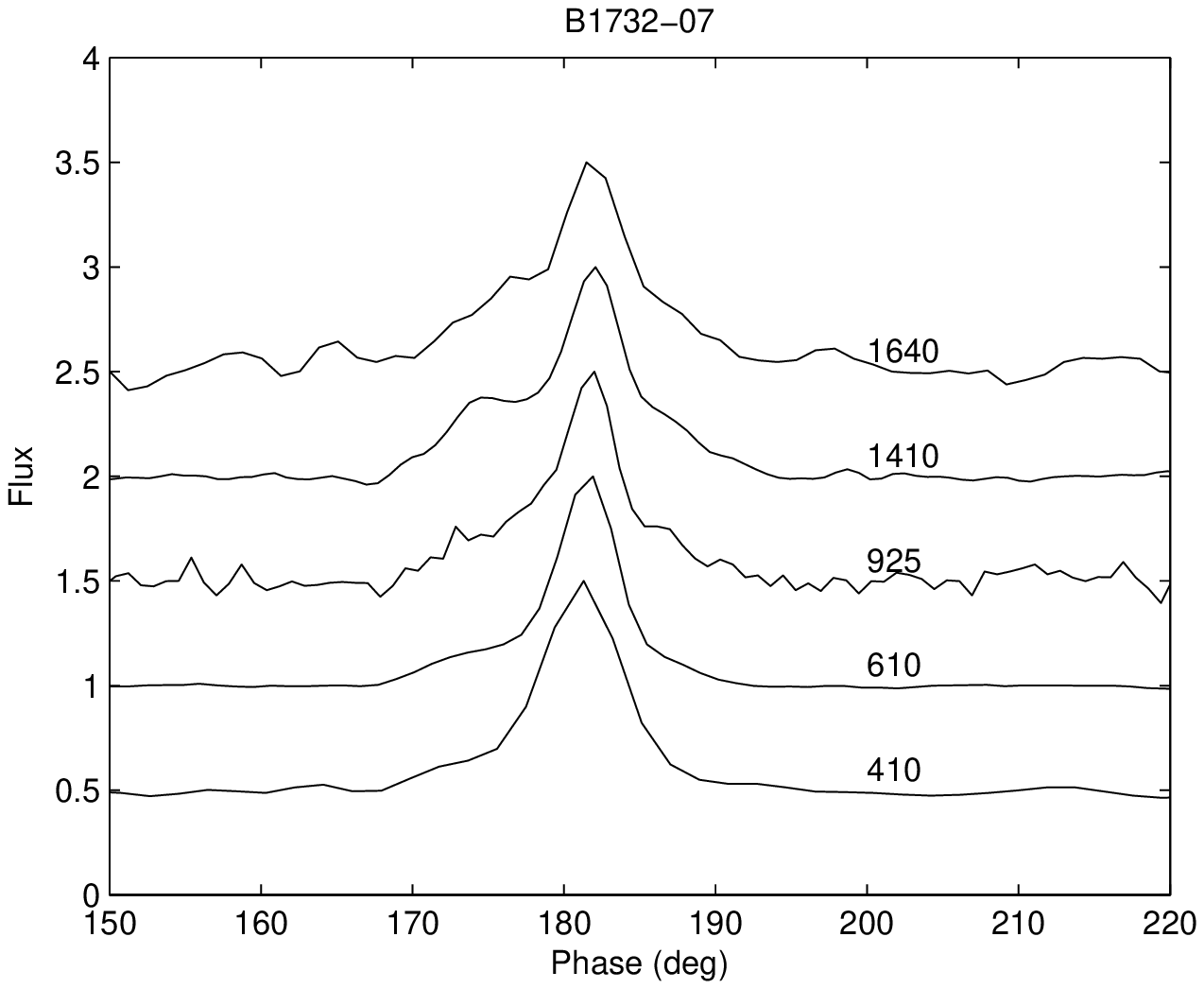}
}
\resizebox{15.9cm}{4.5cm}
{
\includegraphics{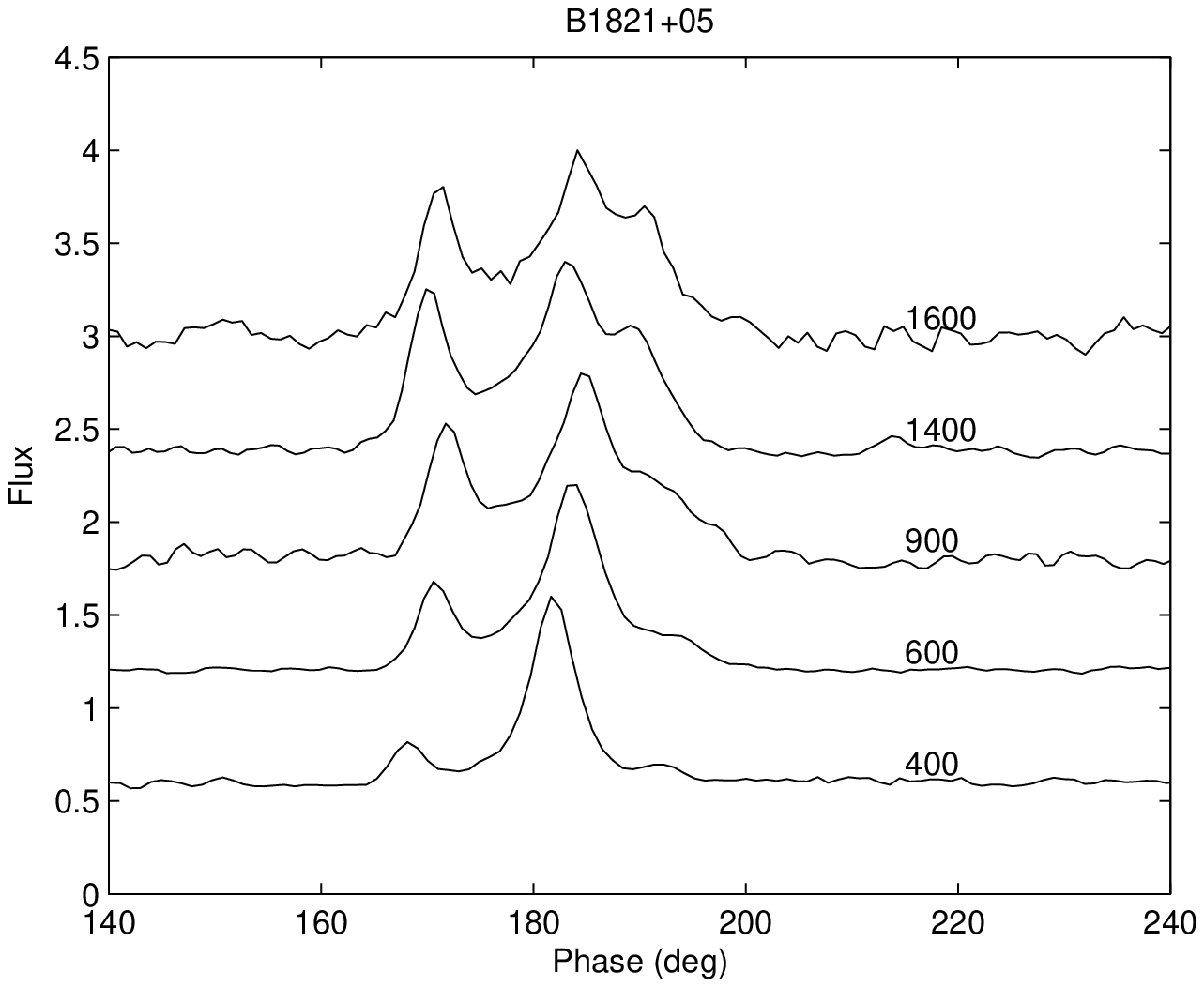}
\includegraphics{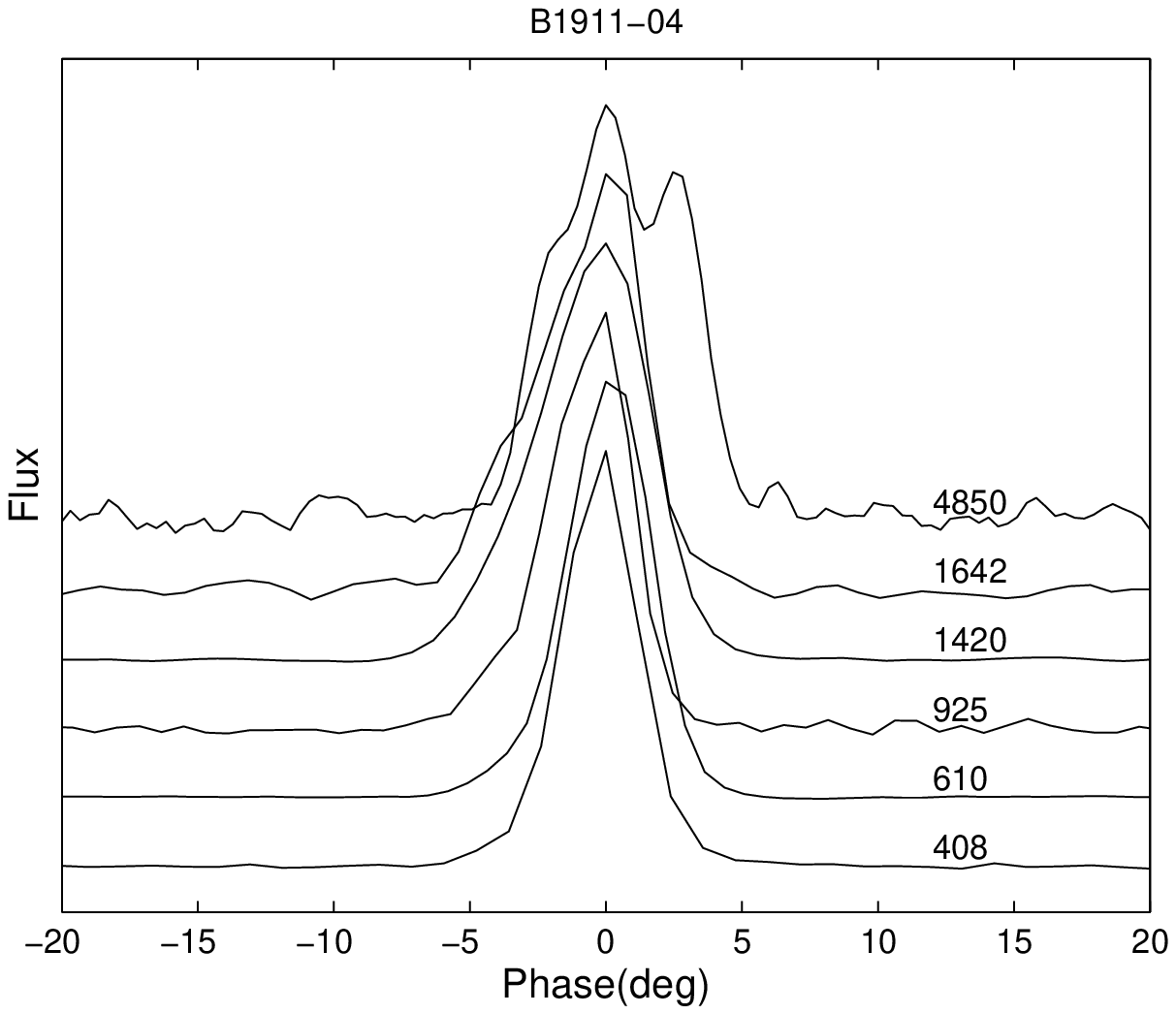}
\includegraphics{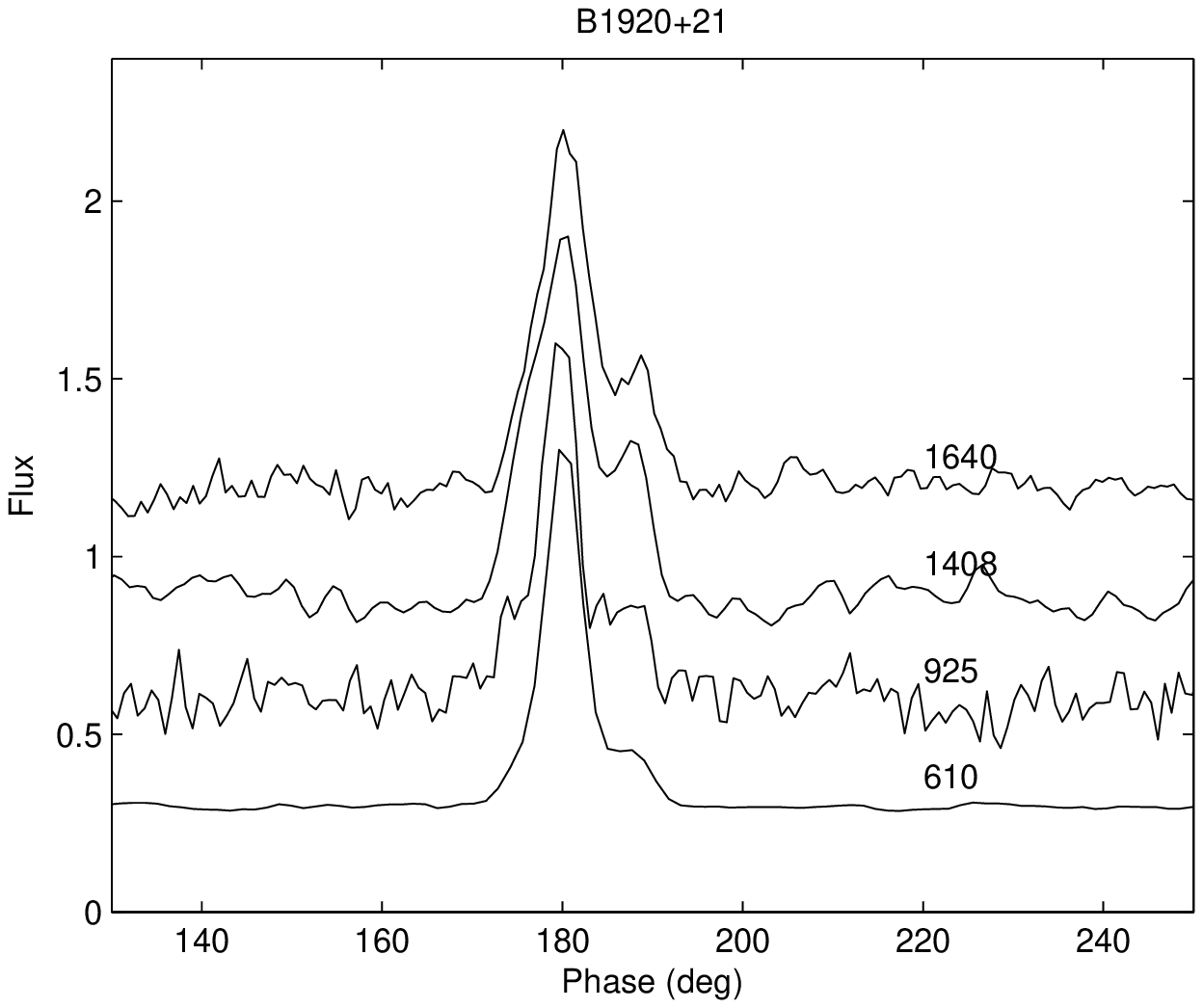}
}
\resizebox{10.6cm}{4.5cm}
{
\includegraphics{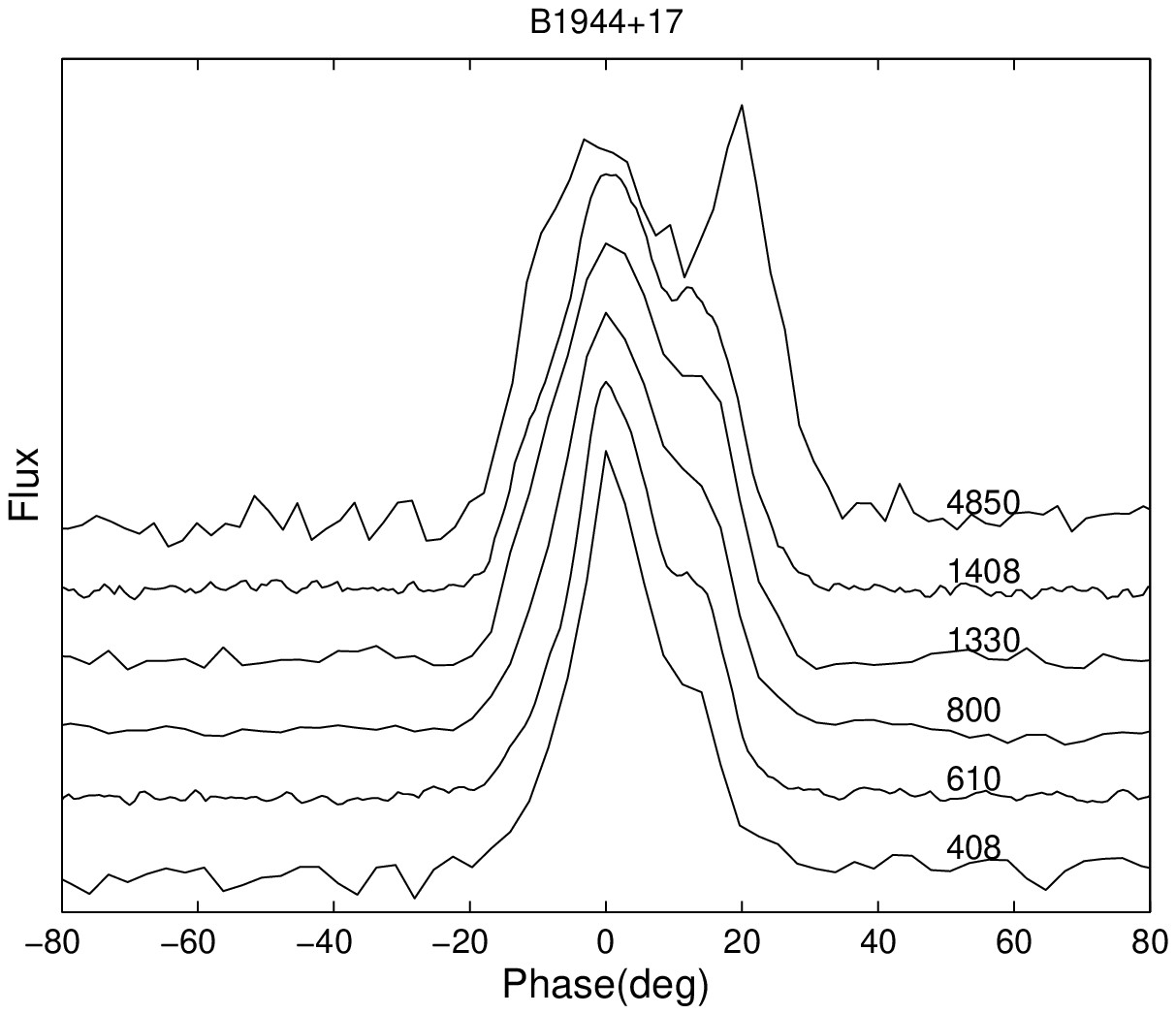}
\includegraphics{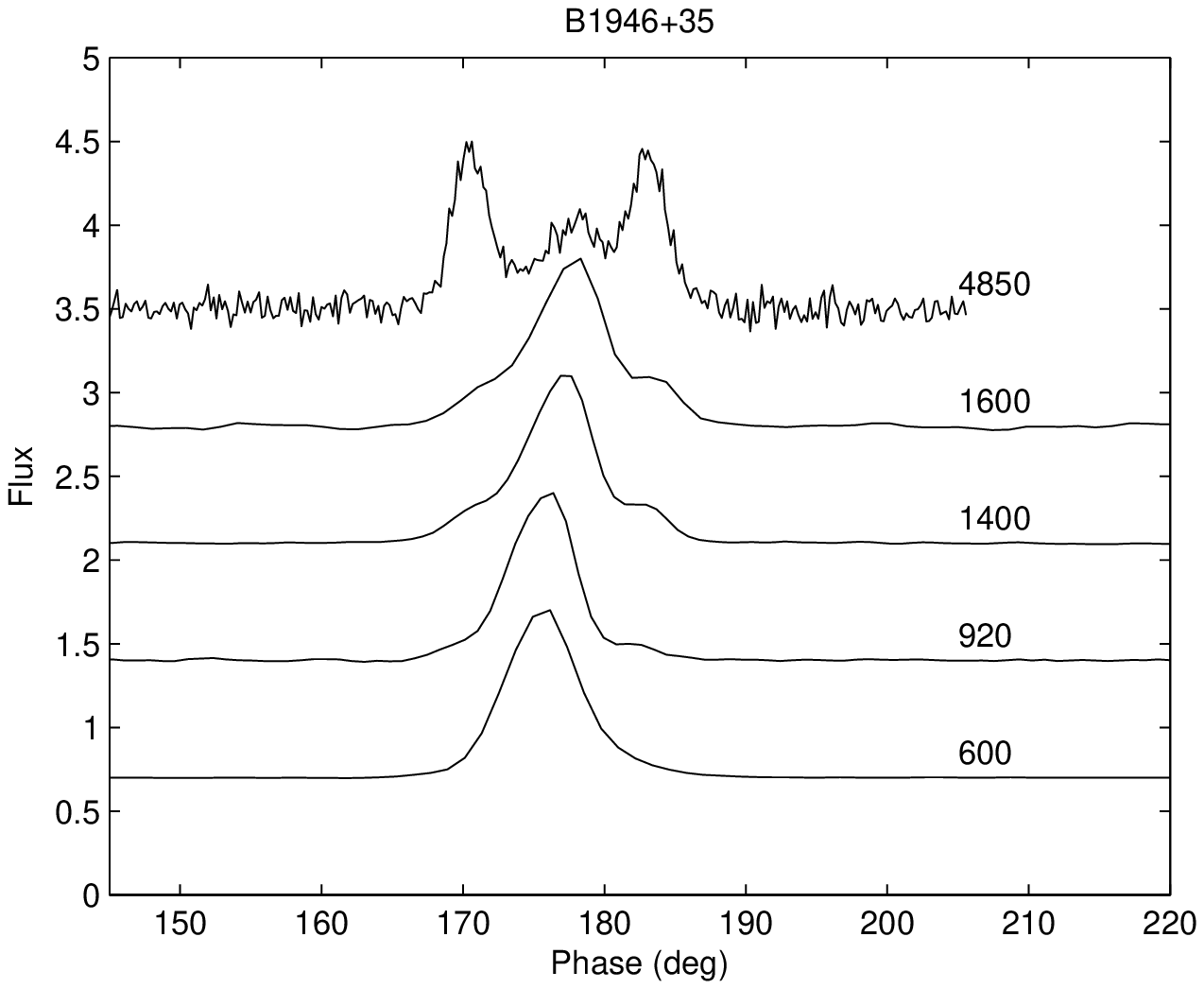}
}
\caption{Normalized multifrequency profiles of eight pulsars in group C, for which the increasing trend of pulse width variation is apparently induced by the rising of the outermost components at high frequency. The data are taken from the EPN database. The profiles are aligned in phase by eye.}
   \label{Fig:Cprof}
\end{figure*}
%
%figure 10

\section{INTERPRETATION of DIVERSE FREQUENCY DEPENDENCE of PULSE WIDTH}
We divide the emission models into narrowband and broadband models according to whether the emission from a single altitude is narrowband or broadband. In the following, we discuss the interpretations based on these two kinds of models separately. Narrowband emission mechanisms normally generate the conal emission beam.Due to the general shortcoming of current conal beam models when it comes to explaining the observed patchy beams of some precessional binary pulsars and a couple of statistical relationships as presented by Wang et al. (2014), we will concentrate on the broadband interpretation.

\subsection{Interpretation Based on Narrowband Models}
Incorporating the conventional RFM (hereafter CRFM), which assumes that the emission is narrowband at a given altitude and the emission frequency increases with decreasing altitude, the conal beam model can successfully explain the behavior of pulse profile narrowing at high frequencies. Using this model, in order to account for the behavior of quasi-constant pulse width, one has to modify the CRFM scenario by assuming that the emission at a wide range of frequencies is generated from a single altitude or a very narrow altitude range (non-RFM). As to the behavior of profile broadening, one must assume that the emission altitude increases with increasing frequency (hereafter anti-CRFM). A question then occurs: how can pulsars have different types of RFM?. A physical argument for the CRFM is that the plasma frequency of secondary particles decreases with increasing altitude because the number density of particles continues to decrease as the plasma flows out, and hence leads to the CRFM (Ruderman \& Sutherland 1975). However, this mechanism certainly excludes the possibility of non-RFM and anti-CRFM. Some other mechanisms mentioned in Section 1 cannot justify anti-CRFM either.

In an alternative narrowband model, the ICS model (Qiao 1988, Qiao \& Lin 1998, Qiao et al. 2001), emission is generated through the ICS process between the low-frequency waves, which are produced in the oscillatory sparking process in the inner vacuum gap, and the secondary relativistic particles. In this model, single-frequency emission can be generated at three distinct altitudes along a magnetic field line because of the combination of the altitude-dependent particle Lorentz factor and the incident angle between the particle velocity and the low-frequency wave vector, thus the whole magnetosphere forms a beam consisting of a core, an inner cone and an outer cone, which come from different altitudes. As the model predicts, the radius of the outer cone shrinks with increasing frequency, whereas the coral beam radius varies marginally with frequency, and the inner cone widens with frequency; therefore, it has more free space to account for various kinds of frequency dependence of pulse width. For example, when the outer cone is visible, one expects to see the profile narrowing with increasing frequency; when the outer cone is too weak and only the inner cone (and/or the core) is visible, one would see the profile broadening phenomenon.

A cone-shaped radio beam structure is a common feature of the current narrowband emission models. This is because when the secondary particles flow along a ring of open field lines, their emission at a certain frequency (generated at a particular altitude) will naturally form a circular beam or a somewhat deformed beam depending on the geometry of emission region. However, such conal beam models have encountered a couple of challenges, as will be summarized below. It is necessary to look for alternative interpretations to the pulse width frequency dependence.

\subsection{Interpretation Based on Broadband Models}

Two types of broadband models have been developed to explain the frequency dependence of the pulse width: the model invoking the wave propagation effects (Barnard \& Arons 1986), and the fan beam model invoking the broadband and coherent emission from particle streams along magnetic flux tubes (Michel 1987; Dyks et al. 2010; Dyks \& Rudak 2012; Dyks \& Rudak 2013, Wang et al. 2014). Michel (1987) proposed an early version of the fan beam model, where some narrow isolated flux tubes in non-dipolar magnetic fields generate wide individual fan beams via the coherent curvature radiation. On the one hand, the author adopted the CRFM while, on the other hand, he suggested that each flux tube (or beam component) may have its own spectrum. The second point is valuable, although the author did not develop a detailed explanation. In this paper, the general idea of our interpretation is very similar to that point. In an alternative fan beam model, Dyks et al. (2010) proposed that thin streams flowing along bunches of magnetic field lines generate split-fan beams via the coherent curvature radiation (see also Figure 1 in Dyks \& Rudak 2012 for a schematic). In this model, the low-frequency broadening of pulse width or component separation is attributed to the plasma gradient at the outer surface of the stream and the intrinsic $\nu^{-1/3}$ widening of the curvature beam, while the emission locations remain frequency independent. This model also allows for the existence of weak frequency dependence and constancy of pulse width due to the broadband nature of emission and the geometry of the streams and the cut of LOS.

Recently, Wang et al. (2014) developed a new version of the fan beam model, which assumes that the broadband and coherent emission from isolated flux tubes form wedge-like beams (as also suggested by Dyks et al. 2010). The model predicts that the emission intensity in the beam first increases when starting from the magnetic axis and then decrease after passing through a transition radius; meanwhile, the transverse sub-beam width increases gradually with increasing beam radius. The authors presented four pieces of observational evidence for this model, including (1) the patchy beam structures of two precessional binary pulsars, PSR J1141$-$6545 (Manchester et al. 2010) and PSR J1906$+$0746 (Desvignes et al. 2013); (2) the relationship between the pulse width and the impact angle for a sample of 64 pulsars; (3) the relationship between the modified intensity and the impact angle for the sample; and (4) the relationship between the pulsar distance and the impact angle. These observational features, as demonstrated in that paper, are difficult to explain using current conal beam models.

Based on this fan beam model, we present an alternative interpretation of the diverse frequency development of pulse width. The original idea was partially proposed by Chen et al. (2007), who studied the phase-resolved spectrum (hereafter the absolute PHRS) of PSR B1133+16 and suggested that the spectral variation across the emission region is responsible to the frequency dependence of the pulse width. In Section 4.3, the absolute PHRS, the spectrum for absolute phase intervals, is compared with the normalized PHRS, i.e., the spectra for normalized phase intervals. We clarify that they are applicable to the broadband and narrowband emission, respectively. We will show that the absolute PHRS is a natural interpretation for the frequency dependence of pulse width under the broadband assumption.

\subsubsection{The Basic Idea}

In the case of broadband emission, it is very likely that the emission spectrum is not homogeneous everywhere in a flux tube. For example, supposing a single-peak profile coming from a flux tube, if the spectra in the leading and trailing edge are steeper than that in the central part of the flux tube, the pulse profile wings will have steeper spectra with respect to the central pulse phase. Consequently, the intensity in the wings will decrease faster than the central phase, and therefore the pulse width will decrease with frequency. In contrast, if both edges have flatter spectra than the central part of the flux tube, then one will see the opposite frequency dependence of pulse width. Therefore, in this case, the different frequency-dependent behaviors of pulse width can be attributed to the different kinds of distribution of emission spectra across the flux tube or the whole emission region if there are multiple flux tubes. Let us explain this in a bit more detail.
\begin{itemize}
  \item For group-A pulsars, the profile narrowing is caused by the steepening spectrum toward the edge of the profile wing, which may be physically induced by the steepening emission spectrum from the inner part to the rim of the emission region.
  \item For group-B pulsars, the nearly constant pulse width reflects the absence of spectral variation in the profile wings, which may be caused by marginal spectral variation in the outer part of the emission region.
  \item For group-C pulsars, the profile broadening reflects a flattening spectrum toward the edge of the profile wing, which may be induced by a flattening emission spectrum from the inner part to the rim of the emission region.
\end{itemize}

One can see the essential difference between our interpretation and the narrowband interpretations. In the narrowband interpretations, the pulse width evolution with frequency is caused by the beam size evolution, which results from the frequency dependence of the emission altitude. In our interpretation, there is no frequency dependence of altitude due to the broadband assumption, and the pulse width change is merely a byproduct of inhomogeneous emission spectra across the emission region.

\subsubsection{Simulations of the Thorsett Relationship Cases}

In order to quantitatively show how the spectral variation impacts the frequency evolution of the pulse width, we simulated two cases of spectral index variation with the pulse phase, that is, the PHRS, one with a steepening spectrum toward the edge of the pulse profile
(Figure 11(a)) and the other with the opposite trend (Figure 11(b)). Note that here we do not start the simulations by assuming the spectral index distribution in flux tubes to avoid the complexity of mapping phase intervals into a three-dimensional magnetosphere. Instead, we directly assume a certain type of PHRS. This is sufficient to illustrate the general idea, because the spectra indices in individual phase intervals are related to the spectra in different parts of the magnetosphere.

For simplicity, the spectral index $\alpha$ across the pulse phase (PHRS) is assumed to be symmetrical and piecewise, and the initial pulse profile at the lowest frequency 100~MHz is assumed to be double-peaked and centered at the phase $\Phi=0$, as shown in the right panels of Figures 11(a) and (b). Given the initial pulse profile at 100~MHz, the profiles at the other nine frequencies are simulated following the PHRS before the pulse widths $W_{10}$ are determined. The frequency development of the pulse width is then plotted in the $W_{10}-\nu$ diagrams, as shown by the left
panels of Figures 9(a) and (b), where the best-fit $W_{10}-\nu$ relationships are also plotted. The normalized average
pulse profiles at 100~MHz (solid), 500~MHz (dashed) and 4100~MHz (dash-dotted) are presented in top right of each panel of
Figures 11(a) and (b).
\begin{figure*}
\centering
\resizebox{16cm}{6cm}
{
\includegraphics{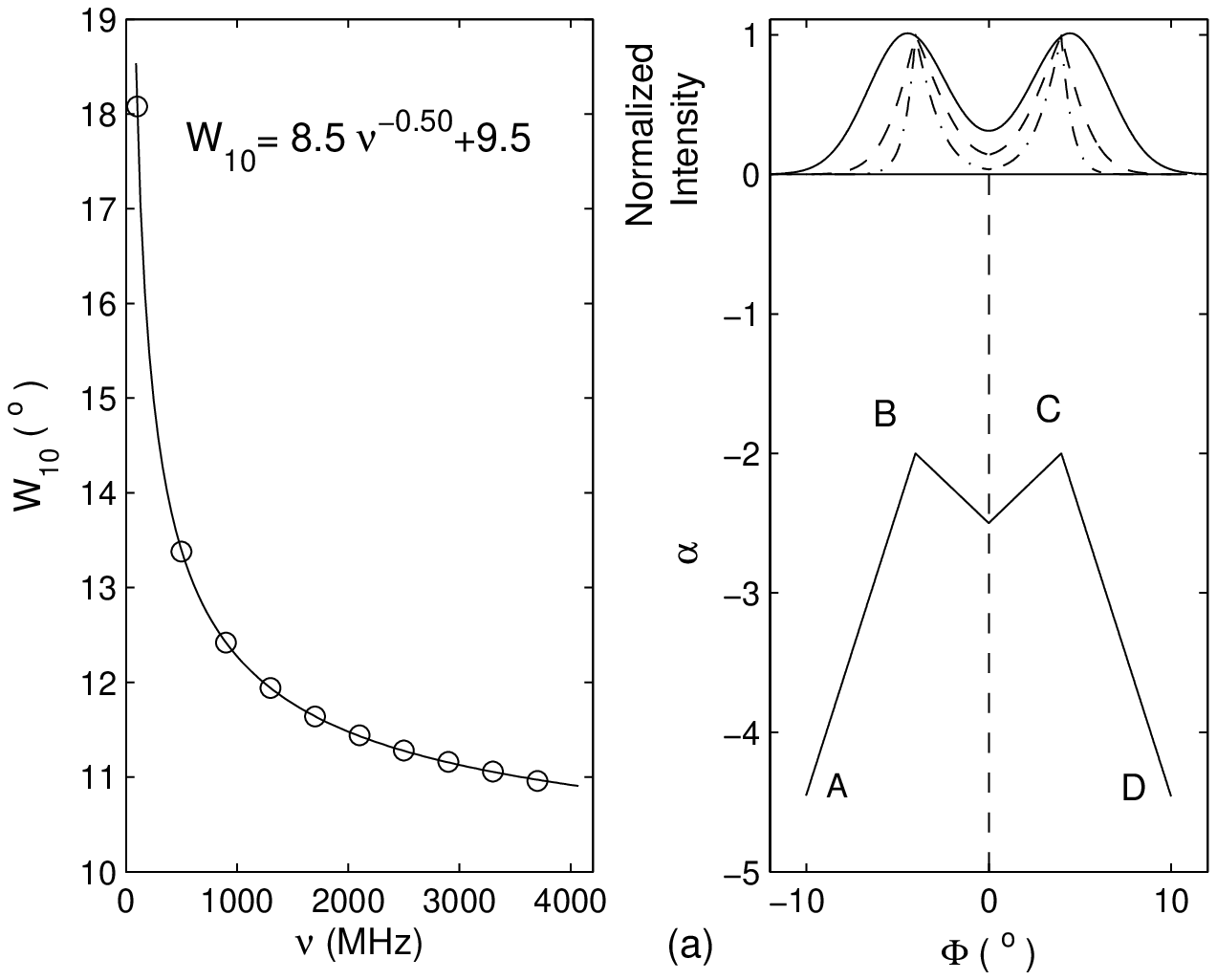}
\includegraphics{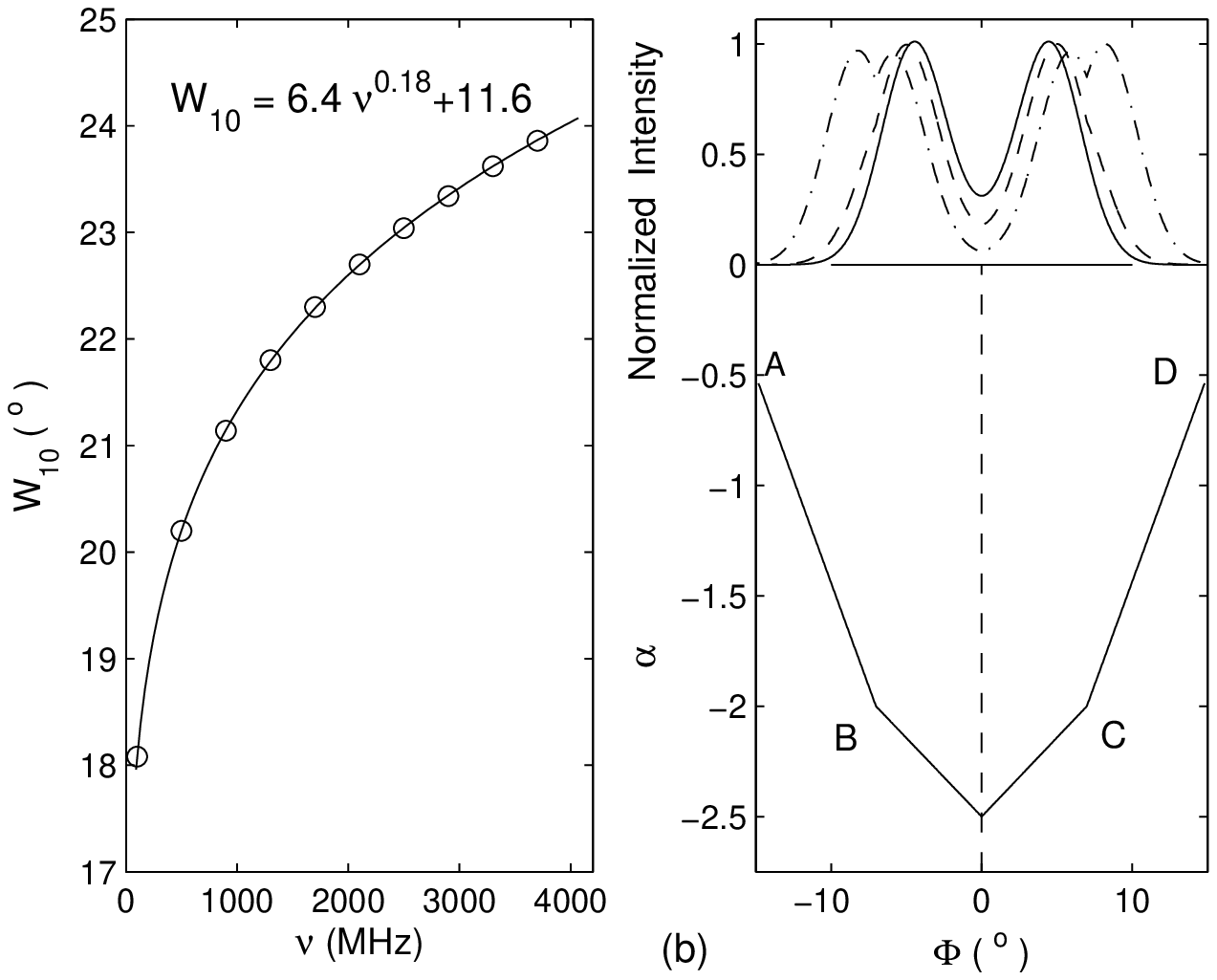}
}
\caption{Simulations for the decreasing (a) and increasing (b) frequency dependence of pulse width. In each right panel of plots (a) and (b), we show the normalized pulse profiles at 100~MHz (solid), 500~MHz (dashed), and 4100~MHz (dash-dotted, top) and the curve of phase-resolved spectral index (PHRS, bottom). The PHRS is assumed to steepen outward in the outer wings of the pulse profile in plot (a), and to be the opposite in plot (b). The double-peaked initial profiles at 100MHz are formed by superposing three Gaussian components, of which the centers are located at the longitudes $\Phi=-4.5, 0$ and $4.5$, with the same half widths of $2^\circ.5$ and different amplitudes of $1, 0.1$ and $1$, respectively. The high-frequency profiles are determined using the assumed PHRS curves. In the left panel of each plot, we plot the 10\% pulse width at 10 frequencies from 100~MHz to 3700~MHz (open circles) and the best-fit Thorsett relationship (solid curve). }
\label{Fig:prs_sim}
\end{figure*}
%%%%%%%%%%%%%%%%%%figure 11
The simulations clearly show that different kinds of PHRS can lead to different types of pulse width evolution, which can follow the Thorsett relationship very well. In fact, the presumed PHRS in Figure 11(a) is very similar to those of PSR B1133$+$16 (Chen et al. 2007), PSR B0525$+$21 and PSR B2020$+$28 (Chen et al. 2011), which are derived from phase-aligned multifrequency profiles published in the literature (Hankins et al. 1991; Kramer et al. 1997; Kuzmin et al. 1998). The method for deriving the PHRS is to divide the profiles into a handful of phase intervals and then calculate the relative spectral index for each interval with respect to a reference interval, which is usually near a pulse peak. Since all the three pulsars belong to group A, the derived PHRS can be regarded as a support for this interpretation.

An interesting feature is that the break points in PHRS can lead to some complex morphology evolution, e.g., bifurcation, which can
be seen in Figure 11(b) in the leading and trailing components at 4100~MHz, where the bifurcation is obviously related
to the break points B and C in the spectral index curve.

\subsubsection{Simulations of the Non-Thorsett Cases}

In our simulations, we found that the pulse width variation normally follows the Thorsett relationship when the PHRS is not highly asymmetric in the leading and trailing profile wings, but the PHRS can deviate from the Thorsett relationship when it is highly asymmetric. This can be seen in Figure 12, which displays how a double profile and its 10\% pulse width evolve with frequency in different cases of PHRS. We found similar behaviors of pulse width variation when applying the same sequence of PHRSs to single and multiple profiles, as long as the profile evolution does not lead to abrupt disappearance or emergence of outriders. This feature, i.e., the pulse width is generally in agreement with the Thorsett relationship, is consistent with the fact that most of the pulsars in our sample can be fitted by a Thorsett relationship. The influence of highly asymmetric PHRS may partially account for the poor quality of fitting (very low $Q$ value) for some pulsars.

Some cases of drastic deviation are also found in our simulations. Figure 13 gives two examples for complex forms of PHRS and the resultant non-monotonic pulse width variation. In case (a), the pulse width first increases and then decreases with frequency because of the asymmetry of PHRS in the leading and trailing wings. In case (b), on the contrary, the pulse width first decreases and then increases, accompanied by the emergence of the leading component and the disappearance of the trailing component at high frequency, which is caused by the different PHRS in the two parts.

In fact, we do find some pulsars with the non-monotonic $W_{10}-\nu$ relationship. One is PSR J1844$+$1454 (B1842$+$14), which shows an increasing trend below 1.5~GHz and then a slightly decreasing trend up to 3.1~GHz, as shown by Figure 14. This turnover behavior is accompanied with an interesting morphology evolution similar to our simulations in Figure 13(a): the leading wing of the single-humped profile at 243~MHz becomes stronger as frequency increases while the trailing wing becomes weaker, forming a gradually separating double-component profile above 600~MHz, and eventually the leading component becomes the dominant one at 3.1~GHz (see Figure 5 in Johnston et al. 2008). Another pulsar is PSR B0809$+$74, for which the pulse width decreases from 14.9~MHz to around 400~MHz and then increases slowly above $400$~MHz (the latter part has been shown in Figure 2). The pulse profile first shows a merging trend of double components below 400~MHz and then a bifurcating trend up to 7.85~GHz (see Figure 13 in Hassall et al. 2012). This morphology evolution is similar to that in Figure 13(b). From our perspective, these examples indicate that some sorts of asymmetric and complex spectral distribution across the emission region may result in the complex frequency development of pulse width.

\begin{figure*}
\centering
\resizebox{16cm}{10cm}
{
\includegraphics{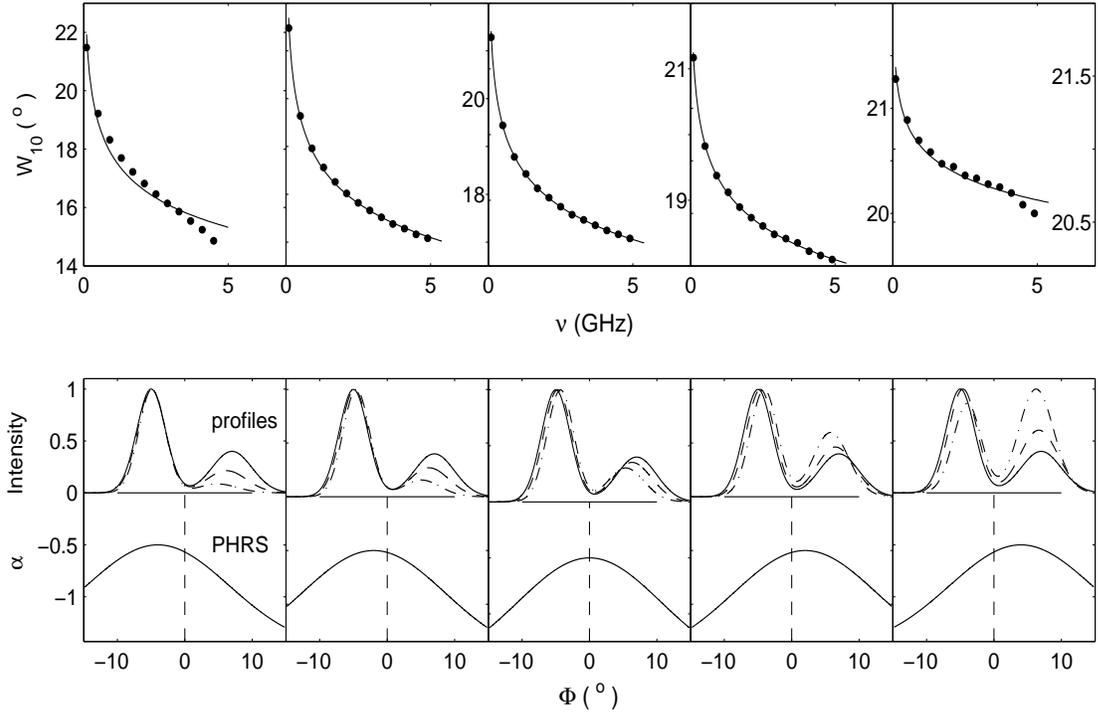}
}
\caption{Simulation for deviation from the Thorsett relationship induced by the asymmetry in the PHRS and profiles. The initial profile at 0.1~GHz is the same for all the five cases (solid curves in the bottom panels) and, consists of two Gaussians with the trailing one being weaker than the leading one. In the top panel of each case, the $W_{10}$ pulse width measured from the profiles are plotted as dots, accompanied with the best-fit curve of Thorsett relationship. In the first and fifth cases, the highly asymmetric PHRS in the leading and trailing profile wings leads the $W-\nu$ relationship deviate from the Thorsett relationship. See the caption of Figure 11 for details.}
\label{Fig:stack}
\end{figure*}
%%%%%%%%%figure 12
%
\begin{figure*}
\centering
\resizebox{16cm}{6cm}
{
\includegraphics{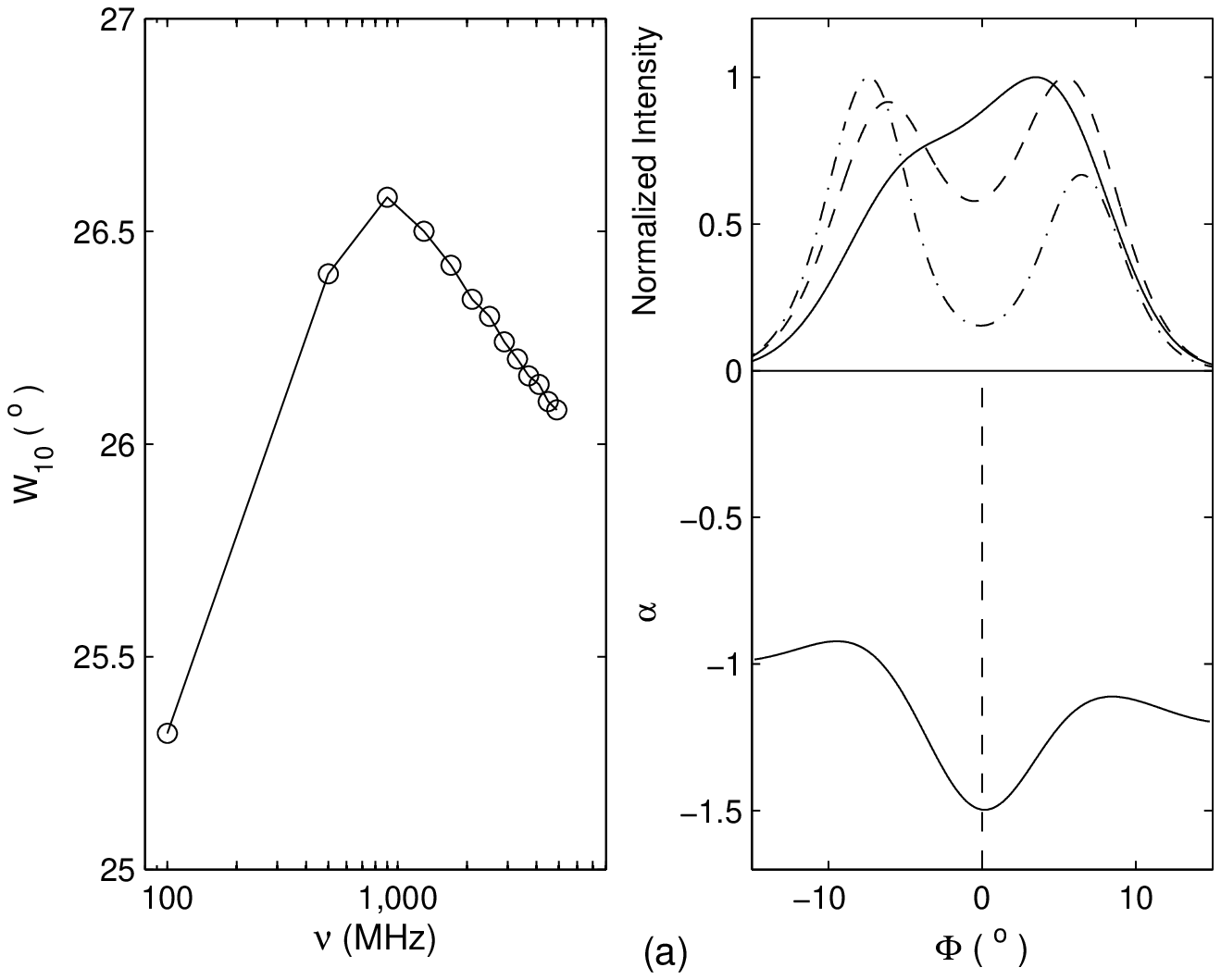}
\includegraphics{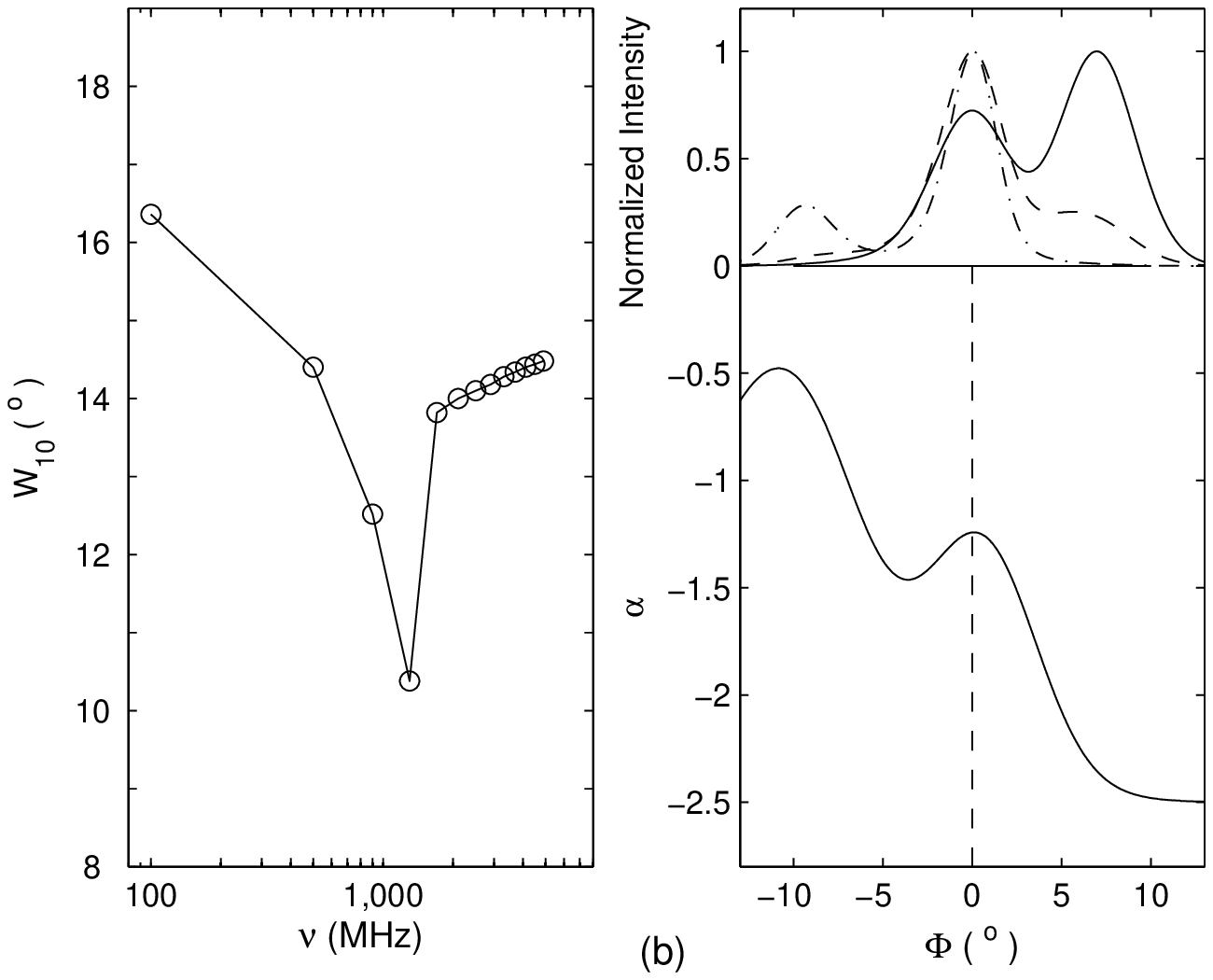}
}
\caption{Simulation for the non-monotonic frequency dependence of pulse width. The twofold width-frequency relationship and the profile evolution are the consequence of the assumed asymmetric PHRS (the bottom curve in each right panel). See the caption of Figure 11 for more details, but different initial pulse profiles at 100~MHz are used here. }
\label{Fig:complex}
\end{figure*}
%%%%%%%%%%%%%%%figure 13
%
\begin{figure*}
\centering
\resizebox{10cm}{7cm}
{
\includegraphics{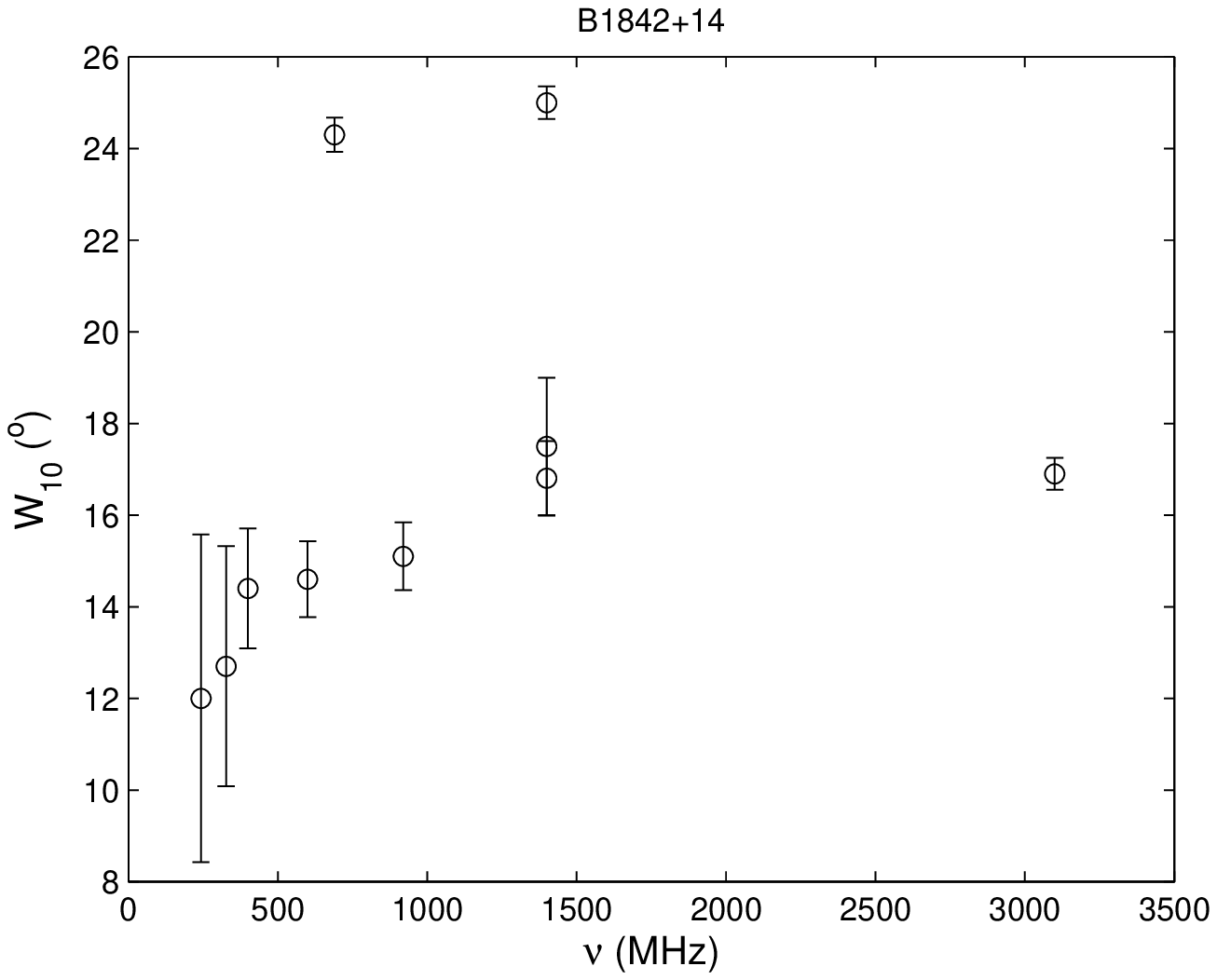}
}
\caption{Diagram of $W_{10}$ vs. observing frequency for PSR B1842$+$14, which shows non-monotonic frequency dependence of pulse width. }
\label{Fig:1842}
\end{figure*}
%figure 14
\subsection{Difference Between the Absolute PHRS and the Normalized PHRS}

The above interpretation is different from previous ideas cocerning the component spectrum that attribute the pulse width and morphology evolution to the spectral difference in the pulse components, e.g., conal and core components, or outer and inner (with respect to the beam or profile center) components (Rankin 1983a; LM88; Kramer et al. 1994). In order to derive the component spectrum, some normalization techniques have been developed, which first normalize the multifrequency beam radii and then calculate the spectral index for each normalized beam-radius interval or component (LM88; Kramer et al. 1994; Kramer 1994). These studies found that the spectra of the outer (or conal) components are generally flatter than those of the inner (or core) components. Since most of the pulsars studied are of group A (see Table 6), it is possible to compare their interpretation with ours. First, we suggest that the spectrum in the outermost parts of the pulse profile, i.e., the profile wings, needs to be steeper than that in the inner part for group-A pulsars, which is contrary to their interpretation. Second, the normalization of the beam radius included a prior assumption that the variation of the pulse width is due to the frequency-dependent evolution of the beam radius. This assumption is a narrowband interpretation; therefore, it is different from our broadband interpretation. Below, we will illustrate the difference.

Let us suppose a narrowband scenario, in which a low-frequency cone (LFC) and a high-frequency cone (HFC) are generated at different altitudes in the pulsar magnetosphere following the CRFM, as shown in panel (a) in Figure 15. In panel (b), the gray ring and the ring confined by two thick circles are the projections of the LFC and HFC onto the celestial sphere centered on the pulsar, represented by the longitude-colatitude ($\Phi-\Theta$) plane, where $\Theta$ is counted from the pulsar spin axis. For a static dipole, the magnetic field lines are projected as straight lines radiating from the stellar center (``+''). As the LOS sweeps across the cones, the profiles at low frequency $\nu_{\rm L}$ and high frequency $\nu_{\rm H}$ are formed, as plotted versus the absolute pulse phase in panel (c).

Following the method used by LM88, the narrower high-frequency profile is first stretched to have the same width as the low-frequency profile before computing the normalized PHRS (see panel (d)).\footnote{This is equivalent to the beam radius normalization, but here we do not normalize the pulse widths to unity for the purpose of comparison with panel (c).} Then, the normalized phase interval within the two dashed lines in panel (d) corresponds to the absolute interval between the dashed lines ``1'' and ``1$^\prime$'' at $\nu_{\rm L}$ and a narrower phase interval between ``2'' and ``2$^\prime$'' at $\nu_{\rm H}$ in panel (c). These intervals can be traced back to different positions in the cones, as denoted by the black dots ``1'' and ``1$^\prime$'' in the LFC and the gray dots ``2'' and ``2$^\prime$'' in the HFC in panels (b) and (a). Obviously, they come from different magnetospheric locations, and hence the spectrum of the normalized phase interval actually compares the multifrequency intensities from different locations.

An alternative normalized-PHRS method is to derive the component spectra by invoking Gaussian decomposition techniques (Kramer et al. 1994; Kramer 1994). Normally a fixed number of Gaussian components is used to fit the multifrequency profiles, so that the flux densities of the same component are used to compute the component spectrum. Since the phase interval of the same component usually changes with frequency, this method is similar to the normalization method used by LM88; therefore, the component spectrum also reflect the multifrequency intensities from different locations.

The above analysis shows that the normalized PHRS is only valid for the case of narrowband emission. If the emission is broadband, then each magnetospheric location will have its own spectrum, and the spectrum of an absolute phase interval is needed to reproduce the real spectrum for a single location.

\begin{figure*}
%\centering
\resizebox{15cm}{11cm}
{
\includegraphics{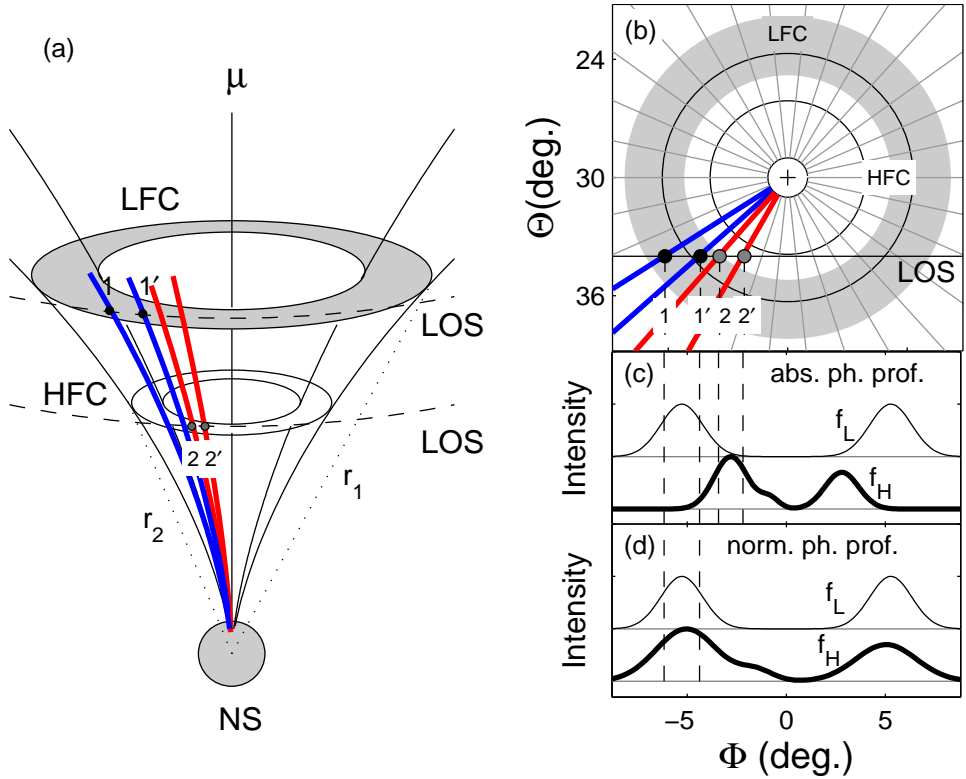}
}
\caption{Schematic diagram showing how the normalized phase-resolved spectrum is associated with radius-to-frequency mapping (RFM). (a) Following the conventional RFM, the LFC and the HFC, symmetrical around the magnetic axis $\mu$, are assumed to be generated at different altitudes, $r_1$ and $r_2$, respectively. Since the line of sight (LOS) forms a cone surrounding the pulsar's spin axis, which is behind $\mu$ but not shown here, we will see different parts of the cones when beam size shrinks at high frequency, as shown by the trajectories of LOS at the LFC and the HFC. Two pairs of points, (1, 1$^\prime$) and (2, 2$^\prime$), located at the blue and red field lines, respectively, correspond to two separated phase intervals as shown in the next two panels. (b) Projections of the LFC (gray annular) and the HFC (the annular confined by two outer thick circles) onto the longitude-colatitude ($\Phi-\Theta$) plane, which stands for the celestial sphere centered on the neutron star, where $\Theta$ is counted from the spin axis. The symbol ``+'' and the radial lines represent the projections of the magnetic pole and the magnetic field lines of a pure dipole. The innermost circle around ``+'' stands for the boundary of polar cap. (c) Profiles at low frequency $\nu_{\rm L}$ and high frequency $\nu_{\rm H}$. The ``abs. ph. prof.'' means the profile in the absolute pulse phase. (d) Profiles plotted in the ``normalized'' pulse phase, where the high-frequency profile (dark line) is stretched to have the same pulse width as the low-frequency profile (light line). The section of the dark profile within the absolute phase interval (2, 2$^\prime$) in panel (c) is equivalent to the section of the stretched dark profile within the interval (1, 1$^\prime$) in panel (d).}
\label{Fig:rfm}
\end{figure*}
%%%%%%%%figure 15

\section{STATISTICAL TEST of A GEOMETRICAL EFFECT on SPECTRUM}

 Kramer et al. (1994) and Sieber (1997) suggested that the observed radio spectrum of a pulsar may be steeper than the intrinsic one if the emission beam continues to shrink as frequency increases. Employing the Gaussian decomposition technique, Kramer et al. (1994) found that the spectra of the core and inner conal components are generally steeper than those of the outer conal components, confirming the similar conclusion made by Rankin (1983a) and LM88. They argued that the difference may be caused by the geometric effect of the shrinkage of nested emission cones rather than different emission mechanisms as proposed by Rankin (1983a). The reason for this is that the core or the inner cone, which is visible at low frequencies, may move out of the LOS at high frequencies as the beam radius shrinks, but the outer cone may still be wide enough to be grazed by the LOS, thus exerting an extra influence of spectral steepening on the core or inner cone spectrum.

Assuming that both the angular sizes of both the cones and the core decrease with frequency, Sieber (1997) demonstrated that an intrinsic spectrum with a power-law index $-$1.5 could be observed as a spectrum with an index of $<-$2.0 averaged over the whole profile. If this geometrical effect is prevalent in group-A pulsars, then their spectra are expected to be statistically steeper than those of group-B and group-C pulsars. This effect can be tested by comparing the spectral indices provided by Maron et al. (2000) for the three groups.

Maron et al. (2000) derived the spectral indices for 281 pulsars. Apart from fifteen pulsars that show double-power-law spectra, all others pulsars exhibit single-power-law spectra, mostly from 0.4 to 1.4~GHz and for some pulsars up to 4.9~GHz or even 10.6~GHz. We collected the single-power-law spectral indices for 132 pulsars in our sample, which are listed in Table 7.

The numbers of pulsars with spectral index data in groups A, B and C are 68, 36 and 28, respectively. The mean spectral index $\alpha_{\rm m}$ and its standard deviation $\sigma_\alpha$ are given in Table 7 for each group. In order to test whether the difference between the mean spectral indices of these groups is statistically significant, we perform the Student's $t$-test for each pair of groups, i.e., (A, B), (A, C), and (B, C), and obtain the probability $p$ of observing a $t$ value as extreme of the distribution of the statistics $$t=\frac{\alpha_{\rm m,i}-\alpha_{\rm m,j}}{\sqrt{\sigma_{\alpha,{\rm i}}^2/n_{\rm i}+\sigma_{\alpha,{\rm j}}^2/n_{\rm j}}}$$ under the null hypothesis that the groups ``i'' and ``j'' are independent samples from normal distributions with equal mean spectral indices, where $n$ is the number of data points of each group. The 95\% confidence interval $c$ for the difference of the true means of a pair of groups is also obtained and listed in Table 7. The 4.5\% and 5.0\% $p$ values for (A, B) and (A, C) suggest that the probability that the group-A pulsars have the same mean spectral index as the other two groups is low, but the mean index of the group-A pulsars, $-1.71$, is smaller than the typical index of $\sim -1.9$ for groups B and C, which is contrary to the prediction. When a sub-sample of 20 pulsars with $\eta\leq-30$\% are selected, the mean spectral index is $-1.52$, which is even flatter than that of all group-A pulsars. Therefore, we conclude that the current data do not support the pure geometrical effect. The weak tendency shown by the results, i.e., that the group-A pulsars statistically have a flatter spectrum, still requires further data for verification.

\section{CONCLUSIONS and DISCUSSIONS}

The frequency dependence of pulse width is studied for a sample of 150 normal pulsars, of which the multifrequency 10\% pulse width can be well fitted with the Thorsett relationship $W_{10}=A\nu^{\mu}+W_{\rm 10, 0}$ (Thorsett 1991). In terms of the best-fit relationship, we calculate the relative fraction of pulse width change between 0.4~GHz and 4.85~GHz, $\eta=(W_{4.85}-W_{0.4})/W_{0.4}$. The following major points are found from the results.

 1. Most of the best-fit $\mu$ values are continuously distributed within $-1$ and 1 and peaked at a negative value close to 0. $\mu$ is not sufficient to be used as a single parameter to classify different kinds of pulse width evolution.

 2. In the 105 pulsars with negative $\mu$, nearly half of the pulsars have a very small asymptotic width $W_{10, 0}$, and the other half have a much larger $W_{10, 0}$ ($>3^\circ.5$). There is likely a gap between $0^\circ.4$ and $3^\circ.5$.

 3. Using $|\eta|=10$\% as a criterion, the sample is divided into three groups: 81 pulsars (54\%) in group A with $\eta<-10$\%, showing a considerable profile narrowing at high frequency, 40 pulsars (27\%) in group B with $-$10\%$\leq\eta\leq 10$\%, exhibiting a marginal change in pulse width, and 29 pulsars (19\%) in group C with $\eta>10$\%, showing a remarkable profile broadening at high frequency.

The results suggest that the profile narrowing phenomenon at high frequency (group A) is more common than the profile broadening phenomenon (group C). This may partially explain why the profile narrowing phenomenon has become the focus of previous studies. However, the fraction of  group-C pulsars is found to be considerably larger than the fraction noted in previous studies. The group-C pulsars, together with a portion of group-B pulsars that exhibit a slight pulse broadening at high frequencies, pose a challenge to the conventional scenario that the radio beam size shrinks with increasing frequency, which can only account for the phenomenon of pulse width narrowing at high frequency. On the basis of the fan beam model (Wang et al. 2014) assuming a broadband nature for the radio emission, we propose an alternative interpretation for the diverse pulse width evolution, which suggests that the pulse width change is a consequence of different kinds of distribution of emission spectrum across the emission region. For the group-A (or group-C) pulsars, the profile narrowing (or broadening) is caused by a steepening (or flattering) spectrum toward the leading and trailing edges of the emission region. For the group-B pulsars, the nearly constant pulse width is due to weak spectral variation or the absence of spectral variation in the outer parts of emission region. The geometrical effect that the emission beam shrinkage may lead to a steepening of the spectrum is tested but disfavored.

We should point out that there may be a slight bias in our sample, because the observing frequency range is relatively narrower for a small fraction of pulsars compared with the others. In group A, 28 pulsars have a frequency range below 2~GHz, accounting for a percentage of 34.6\%. The numbers are similar in group B and group C, 13 (32.5\%) and 7 (24.1\%), respectively. Some of these pulsars have quite large uncertainties in $\eta$, e.g., PSRs B0820$+$02 and B1727$-$47, because of the relatively narrow frequency range and the uncertainty of pulse width. Therefore, future observations at high frequencies are vital to verify our results for these pulsars. However, even without these pulsars, the proportions of the three groups in the remained sample will still be close to the present values.

Regarding the underlying physics that lead to different kinds of spectral variation across the emission region, Chen et al. (2007) have discussed three possibilities: the emission spectrum may evolve with altitude as non-monoenergetic secondary particles flow along a flux tube; the emission spectrum may also vary with azimuth and colatitude in the pulsar magnetosphere. The exact reason is not clear yet, but at least the following mechanisms are possible causes of inhomogeneity in the particle energy spectrum and emission spectrum distribution, if we assume that the emission spectrum is tightly or at least partially related to the energy spectrum of secondary relativistic particles.
\begin{itemize}
  \item The energy spectrum of the secondary particle can be modified by the resonant ICS near the polar cap when they collide with the thermal X-ray photons from the stellar surface (Lyubarskii \& Petrova 2000). The difference in magnetic field strength and geometry within different flux tubes may cause inhomogeneity across the emission region.
  \item The possible asymmetry in the field line structure across the polar gap may result in an inhomogeneous distribution of the particle energy spectrum during the pair production process (e.g., Harding \& Muslimov 2011).
  \item When the non-neutral secondary plasma flows along a flux tube, the free-flow charge density and the Goldreich-Julian charge density (Goldreich \& Julian 1969) do not follow the same function of the altitude, and therefore a weak parallel electric field will be induced and modify the particle energy spectrum. This effect, which relies on the geometry of magnetic field, may cause an inhomogeneous spectral distribution depending on the altitude, azimuth, and colatitude.
  \item Propagation effects, e.g., the refraction effect, which depends on the secondary plasma distribution in the pulsar magnetosphere, may cause redistribution of the emission spectrum (Petrova 2002).
\end{itemize}

The following studies will help us to improve our understanding of the spectral distribution in the pulsar magnetosphere and the underlying mechanism. (1) Three-dimensional simulations for the discharge process in pulsar polar gaps may be helpful for finding out what kinds of energy spectral distribution of the secondary particles may be formed. (2) If the phase-aligned multifrequency intensity map of the radio beam can be derived for some precessional binary pulsars in the future, then we will be able to measure the spectra across the beam and infer how the spectral index depends on the altitude, azimuth, and colatitude. (3) Collecting a sample of pulsars with phase-aligned multifrequency profiles and known viewing geometry may enable us to map the spectral distribution in the magnetosphere. (4) Deep observations will be helpful for determining whether or not there is weak emission out of the pulse window that is limited by the current observing sensitivity, particularly at high frequencies for the group-A pulsars. In fact, the weak level emission components have been discovered for more millisecond pulsars as the signal-to-noise ratio significantly increases  (Yan et al. 2011, Dai et al. 2014 in preparation). Weak precursors or postcursors were also discovered for a few normal pulsars with higher observing sensitivities (Mitra \&
Rankin 2011). If the weak and extended emission is found at high (or low) frequencies for group-A (or group-C) pulsars, then it would favor the broadband interpretation, otherwise it would favor the narrowband interpretation.

\vspace{0.2cm} %
\noindent {\it Acknowledgments}:

We appreciate the anonymous referee for valuable suggestions and comments. We acknowledge Y.B. Li, Q. Guo, L. Wang, and D. Zhang for their help in collecting part of the data. This work was funded by the Chinese Scholarship Council (201308440093), NSFC key project 11178001, Yuncheng University (YQ-2011033) and WLFC (XBBS201422).

\clearpage

\begin{table}
\caption{References for the pulse width data used in this paper.}
 \label{Tab:freq}
\[
\begin{tabular}{ll}
\hline
${\rm Refercence}$ & ${\rm Frequency~(MHz)}$ \\
\hline
{\rm Arzoumanian et al. 1994}		&	800, 1640	\\
{\rm Downs  1979*} &	2388	\\
{\rm Gould \& Lyne  1998$^\dagger$}		&	230, 400, 600, 920, 1400, 1600	\\
{\rm Hankins \& Rickett  1986*}	&	430, 600, 932, 1358, 1387,  1414,  1665, 2380\\
{\rm Hankins et al.,  1991*}	&	430, 2370, 4870	\\
{\rm Hoensbroech \& Xilouris 1997}		&	1710, 4850, 10450	\\
{\rm Hoensbroech  et  al. 1998}	&	4850	\\
{\rm Johnston et al. 2006*$^\dagger$}		&	8400	\\
{\rm Johnston et al. 2008*$^\dagger$}		&	243, 327, 690, 1400, 3100	\\
{\rm Karastergiou et  al. 2005*$^\dagger$}		&	3100	\\
{\rm Karastergiou \& Johnston 2006*$^\dagger$}		&	1375, 3100	\\
{\rm Kijak et al. 1998}		&	4850	\\
{\rm Kramer et al. 1997}		&	1410, 2250, 4750, 8500, 10550	\\
{\rm Kuzmin et al. 1986*}		&	4600, 10700	\\
{\rm Lyne et al. 1971*}		&	1612, 2650	\\
{\rm Lyne \& Manchester 1988*}	&	409,415,611,1420\\
{\rm Manchester et al. 1998$^\dagger$}		&	430, 660, 1502	\\
{\rm  Morris et al. 1981*}		&	1720, 2650, 8700, 14800	\\
{\rm van Ommen et al. 1997*$^\dagger$}		&	800, 950	\\
{\rm Phillips \& Wolszczan 1992*}	&	430, 1418, 2380, 4800	\\
{\rm Popov \& Soglasnov 1987*$^\dagger$}		&	927, 6000	\\
{\rm Qiao et al. 1995}	&	660, 1440	\\
{\rm Seiradakis  et al. 1995}		&	1310, 1420, 1615, 4750, 10550	\\
{\rm Sieber et al. 1975*}&	2700, 4900, 10700	\\
{\rm Stinebring  et al. 1984a*}		&	1404	\\
{\rm Stinebring et al. 1984b*}		&800	\\
{\rm Wu et al. 1993} &1560	\\
\hline
\end{tabular}
\]
\tablenotetext{\space}{*The profile data not archived in the EPN. }
\tablenotetext{\space}{$^\dagger$ The profile data that we adopted from literature directly. }
\end{table}

%%%%%%%%%%%%%%%%%%%%%%%%%%%%%%%%%%%%%%
% table 2  group A

%\begin{sideways}

\begin{deluxetable}{lllllllllllll}
%%%\begin{sidewaystable}{llllllllllll}
\tabletypesize{\scriptsize}
\rotate
\tablecaption{Parameters for 81 group-A pulsars \label{Tab:groupA}}
\tablewidth{0pt}
\tablehead{ \colhead{PSR B}&\colhead{Freq. (GHz)} & \colhead{A} &\colhead{$\mu$}&
\colhead{$W_{\rm 10,0} (^\circ)$ }    &\colhead{N}
& \colhead{$\chi^{2}_{\rm min}$} & \colhead{$Q$} &\colhead{$\eta$} &\colhead{$\eta^\prime$}&\colhead{$\alpha$}&
\colhead{R90} & \colhead{LM88}  }
\startdata
0011$+$47		&		0.408$-$4.75		&		5.28E$+$01	($^{+0.8}_{-30.4}$)		&		 -1.60E$-$01	($^{+0.01}_{-0.24}$)		 &		2.40E$-$01	 ($^{+29.39}_{-0.24}$)		&		5		 &	37.90 	&		5.75E$-$09		&	-0.33 	 ($^{+0.02}_{-0.02}$)	 &	-0.18 	&		-1.3 		 &		 $-$ 		 &		$-$ 		\\
0031$-$07		&		0.69$-$4.9		&		3.96E$+$01	($^{+0.8}_{-16.0}$)		&		 -2.10E$-$01	($^{+0.02}_{-0.17}$)		 &		 2.20E$-$01	 ($^{+15.96}_{-0.22}$)		&		11		 &	186.00 	&		0.00E$+$00		&	-0.41 	 ($^{+0.03}_{-0.03}$)	 &	 -0.23 	&		-1.4 		 &		 CNS		 &		CN		\\
0037$+$56		&		0.4$-$1.42		&		2.01E$-$01	($^{+7.40}_{-0.20}$)		&		 -2.29E$+$00	 ($^{+12.29}_{-8.71}$)		 &		6.36E$+$00	 ($^{+1.26}_{-6.36}$)		&		4		 &	1.10 	&		2.95E$-$01		 &	-0.20 	($^{+0.43}_{-0.42}$)	 &	-0.01 	&		-1.8 		 &		 $-$ 		 &		$-$ 		\\
0052$+$51		&		0.4$-$4.85		&		1.40E$+$01	($^{+0.8}_{-12.6}$)		&		 -1.40E$-$01	($^{+0.07}_{-1.08}$)		 &		 1.40E$-$01	 ($^{+12.22}_{-0.14}$)		&		5		 &	0.52 	&		7.73E$-$01		&	-0.29 	 ($^{+0.13}_{-0.12}$)	 &	 -0.16 	&		-0.7 		 &		 $-$ 		 &		$-$ 		\\
0059$+$65		&		0.4$-$1.6		&		2.56E$-$01	($^{+21.94}_{-0.25}$)		&		 -2.52E$+$00	($^{+2.5}_{-4.06}$)		 &		 2.09E$+$01	 ($^{+1.14}_{-20.94}$)		&		4		&	 0.44 	&		5.12E$-$01		&	-0.11 	 ($^{+0.09}_{-0.17}$)	 &	 -5.0E-3 	&		-1.6 		 &		 $-$ 		 &		$-$ 		\\
J0134$-$2937		&		0.243$-$3.1		&		1.80E$+$00	($^{+1.20}_{-0.80}$)		&		 -1.93E$+$00	 ($^{+0.48}_{-0.52}$)		 &		2.02E$+$01	 ($^{+0.97}_{-1.09}$)		&		8		 &	156.00 	&		0.00E$+$00		 &	 -0.34 	($^{+0.07}_{-0.06}$)	 &	-0.04 	&		$-$ 		 &		 $-$ 		 &		$-$ 		\\
0144$+$59		&		0.4$-$4.85		&		4.60E$+$00	($^{+10.40}_{-3.80}$)		&		 -9.20E$-$01	($^{+0.8}_{-1.08}$)		 &		 8.20E$+$00	 ($^{+3.83}_{-8.20}$)		&		6		&	 8.65 	&		3.43E$-$02		&	-0.51 	 ($^{+0.26}_{-0.18}$)	 &	 -0.20 	&		-1.0 		&		 $-$ 		 &		$-$ 		\\
0148$-$06		&		0.243$-$3.1		&		5.89E$+$00	($^{+36.11}_{-4.69}$)		&		 -4.18E$-$01	($^{+0.39}_{-1.06}$)		 &		3.57E$+$01	 ($^{+4.39}_{-35.66}$)		&		11		 &	222.00 	&		0.00E$+$00		&	-0.13 	 ($^{+0.05}_{-0.05}$)	 &	-0.05 	&		-2.7 		 &		 CD		 &		CN		\\
0149$-$16		&		0.243$-$3.1		&		1.45E$+$00	($^{+9.95}_{-1.45}$)		&		 -8.10E$-$01	 ($^{+10.81}_{-9.19}$)		 &		9.14E$+$00	 ($^{+1.65}_{-9.14}$)		&		13		 &	13.70 	&		1.86E$-$01		 &	-0.22 	($^{+0.24}_{-0.18}$)	 &	-0.07 	&		-2.1 		 &		 $-$ 		 &		CN		\\
0254$-$53		&		0.434$-$1.56		&		2.37E$-$06	($^{+10.20}_{-2.37{\rm E}-6}$)		&		 -1.58E$+$01	 ($^{+78.78}_{-63.22}$)		&		 9.06E$+$00	 ($^{+1.07}_{-9.06}$)		&		4		 &	1.70 	&		2.23E$-$01		 &	 -0.33 	($^{+0.15}_{-0.14}$)	&	-1.3E-9 	&		$-$ 		 &		 $-$ 		&		PCN		\\
0301$+$19		&		0.243$-$4.85		&		7.64E$+$00	($^{+9.36}_{-7.40}$)		&		 -2.39E$-$01	 ($^{+0.18}_{-2.5}$)		 &		8.74E$+$00	 ($^{+7.02}_{-8.74}$)		&		16		 &	14.10 	&		3.67E$-$01		&	 -0.23 	 ($^{+0.12}_{-0.11}$)	 &	-0.12 	&		$-$ 		 &		 CD		 &		CN		\\
0329$+$54		&		0.408$-$10.7		&		4.67E$+$00	($^{+21.93}_{-3.27}$)		&		 -6.78E$-$01	 ($^{+0.61}_{-1.15}$)		 &		2.12E$+$01	 ($^{+2.55}_{-21.16}$)		&		11		 &	17.30 	&		2.75E$-$02		 &	 -0.24 	($^{+0.09}_{-0.07}$)	 &	-0.09 	&		-2.2 		 &		 T		 &		CRCN		\\
0355$+$54		&		0.92$-$10.7		&		3.84E$+$01	($^{+0.8}_{-3.6}$)		&		 -2.40E$-$01	($^{+0.02}_{-0.04}$)		 &		 0.00E$+$00	 ($^{+3.66}_{0}$)		&		8		&	 51.00 	&		8.82E$-$10		&	-0.45 	($^{+0.02}_{-0.04}$)	 &	 -0.26 	 &		$-$ 		&		 CRS		 &		PCN		\\
0402$+$61		&		0.4$-$4.85		&		2.81E$-$01	($^{+17.92}_{-0.28}$)		&		 -2.62E$+$00	 ($^{+12.62}_{-10.38}$)		 &		1.68E$+$01	 ($^{+1.24}_{-16.76}$)		&		6		 &	4.85 	&		1.83E$-$01		 &	 -0.16 	($^{+0.64}_{-0.23}$)	 &	-0.01 	&		-1.4 		 &		 $-$ 		 &		CRCN		\\
0458$+$46		&		0.6$-$4.85		&		2.37E$+$00	($^{+13.23}_{-2.36}$)		&		 -1.25E$+$00	($^{+1.21}_{-8.75}$)		 &		1.23E$+$01	 ($^{+2.33}_{-12.30}$)		&		6		 &	4.44 	&		2.20E$-$01		&	-0.36 	 ($^{+0.27}_{-0.59}$)	 &	-0.09 	&		-1.3 		 &		 $-$ 		 &		CN		\\
0523$+$11		&		0.243$-$3.1		&		1.37E$-$01	($^{+18.46}_{-0.14}$)		&		 -3.15E$+$00	 ($^{+16.15}_{-12.85}$)		 &		1.78E$+$01	 ($^{+0.98}_{-17.77}$)		&		12		 &	17.40 	&		4.25E$-$02		 &	 -0.12 	($^{+1.97}_{-0.16}$)	 &	-2.6E-3 	&		-2.0 		 &		 M		 &		CN		\\
0525$+$21		&		0.408$-$4.85		&		1.86E$+$01	($^{+0.8}_{-17.4}$)		&		 -9.00E$-$02	($^{+0.04}_{-1.38}$)		 &		2.00E$-$02	 ($^{+16.76}_{-0.02}$)		&		14		 &	15.50 	&		1.60E$-$01		&	-0.20 	 ($^{+0.09}_{-0.09}$)	 &	-0.11 	&		-1.5 		 &		 CD		 &		CN		\\
0538$-$75		&		0.436$-$1.44		&		6.75E$-$01	($^{+29.13}_{-0.66}$)		&		 -1.95E$+$00	 ($^{+1.95}_{-5.38}$)		 &		2.76E$+$01	 ($^{+2.51}_{-27.58}$)		&		5		 &	1.59 	&		4.53E$-$01		 &	 -0.13 	($^{+0.12}_{-0.26}$)	 &	-0.01 	&		$-$ 		 &		 $-$ 		 &		CRCN		\\
0559$-$05		&		0.4$-$4.85		&		3.19E$-$01	($^{+23.88}_{-0.32}$)		&		 -2.79E$+$00	 ($^{+13.79}_{-11.21}$)		 &		2.22E$+$01	 ($^{+1.48}_{-22.23}$)		&		6		 &	2.22 	&		5.28E$-$01		 &	 -0.16 	($^{+0.20}_{-0.22}$)	 &	-0.01 	&		-1.7 		 &		 $-$ 		 &		CN		\\
0609$+$37		&		0.4$-$4.85		&		2.10E$+$01	($^{+1.0}_{-20.2}$)		&		 -1.30E$-$01	($^{+0.09}_{-1.54}$)		 &		 8.00E$-$02	 ($^{+20.24}_{-0.08}$)		&		6		 &	9.83 	&		2.01E$-$02		&	-0.28 	 ($^{+0.19}_{-0.14}$)	 &	 -0.15 	&		-1.5 		 &		 $-$ 		 &		$-$ 		\\
0628$-$28		&		0.243$-$4.9		&		3.56E$+$01	($^{+0.4}_{-23.2}$)		&		 -1.60E$-$01	($^{+0.03}_{-0.30}$)		 &		 9.00E$-$02	 ($^{+22.85}_{-0.09}$)		&		17		 &	69.00 	&		2.98E$-$09		&	-0.33 	 ($^{+0.06}_{-0.04}$)	 &	 -0.18 	&		-1.9 		 &		 CNS		 &		CN		\\
0643$+$80		&		0.4$-$4.85		&		1.08E$+$01	($^{+0.8}_{-10.2}$)		&		 -1.40E$-$01	($^{+0.10}_{-1.41}$)		 &		 0.00E$+$00	 ($^{+10.27}_{0}$)		&		4		&	 1.54 	&		2.14E$-$01		&	-0.30 	($^{+0.20}_{-0.17}$)	 &	 -0.16 	 &		-1.9 		&		 CNS		 &		$-$ 		\\
0656$+$14		&		0.43$-$8.4		&		3.28E$+$01	($^{+0.8}_{-20.0}$)		&		 -1.10E$-$01	($^{+0.02}_{-0.23}$)		 &		 3.80E$-$01	 ($^{+19.88}_{-0.38}$)		&		6		 &	24.90 	&		1.61E$-$05		&	-0.24 	 ($^{+0.04}_{-0.04}$)	 &	 -0.13 	&		-0.5 		 &		 T		 &		CN		\\
0727$-$18		&		0.243$-$3.1		&		1.02E$+$00	($^{+19.38}_{-1.01}$)		&		 -1.32E$+$00	($^{+1.32}_{-3.76}$)		 &		1.85E$+$01	 ($^{+1.31}_{-18.5}$)		&		12		 &	8.27 	&		5.19E$-$01		&	-0.15 	 ($^{+0.15}_{-0.12}$)	 &	-0.03 	&		-1.6 		 &		 $-$ 		 &		CN		\\
0823$+$26		&		0.408$-$10.7		&		2.35E$+$00	($^{+6.05}_{-2.35}$)		&		 -5.10E$-$01	 ($^{+10.51}_{-9.49}$)		 &		4.86E$+$00	 ($^{+2.22}_{-4.86}$)		&		14		 &	2.73 	&		9.94E$-$01		 &	 -0.31 	($^{+0.34}_{-0.22}$)	 &	-0.14 	&		$-$ 		 &		 CRS		 &		CRCN		\\
0919$+$06		&		0.243$-$10.6		&		1.43E$+$01	($^{+1.3}_{-6.1}$)		&		 -3.85E$-$01	($^{+0.09}_{-0.38}$)		 &		3.83E$-$01	 ($^{+5.70}_{-0.38}$)		&		16		 &	54.20 	&		5.57E$-$07		&	-0.61 	 ($^{+0.08}_{-0.07}$)	 &	-0.37 	&		-1.8 		 &		 T		 &		PCN		\\
1039$-$19		&		0.4$-$4.85		&		1.82E$+$01	($^{+1.2}_{-18.1}$)		&		 -1.00E$-$01	($^{+0.09}_{-1.90}$)		 &		 3.40E$-$01	 ($^{+18.47}_{-0.34}$)		&		6		 &	0.65 	&		8.85E$-$01		&	-0.22 	 ($^{+0.19}_{-0.14}$)	 &	 -0.12 	&		-1.5 		 &		 M		 &		CN		\\
1112$+$50		&		0.4$-$2.65		&		8.20E$+$00	($^{+0.60}_{-8.20}$)		&		 -1.00E$-$01	($^{+2.10}_{-1.90}$)		 &		0.00E$+$00	 ($^{+8.01}_{0}$)		&		8		&	 8.11 	&		1.50E$-$01		&	-0.22 	($^{+0.32}_{-0.19}$)	 &	 -0.12 	&		-1.7 		&		 T	 	&		 $-$ 		\\
1133$+$16		&		0.23$-$32		&		6.96E$+$00	($^{+4.04}_{-4.56}$)		&		 -2.57E$-$01	($^{+0.13}_{-0.49}$)		 &		3.65E$+$00	 ($^{+4.14}_{-3.65}$)		&		18		 &	25.00 	&		5.04E$-$02		&	-0.33 	 ($^{+0.07}_{-0.06}$)	 &	-0.17 	&		-1.9 		 &		 CD		 &		CN		\\
1237$+$25		&		0.23$-$4.85		&		6.80E$+$00	($^{+8.00}_{-6.56}$)		&		 -1.67E$-$01	($^{+0.13}_{-1.65}$)		 &		7.44E$+$00	 ($^{+6.22}_{-7.44}$)		&		15		 &	11.60 	&		4.86E$-$01		&	-0.18 	 ($^{+0.10}_{-0.08}$)	 &	-0.09 	&		$-$ 		 &		 M		 &		CRCN		\\
1504$-$43		&		0.243$-$3.1		&		1.32E$+$01	($^{+0.4}_{-6.8}$)		&		 -1.90E$-$01	($^{+0.05}_{-0.21}$)		 &		 1.00E$-$01	 ($^{+6.84}_{-0.10}$)		&		5		 &	10.40 	&		5.47E$-$03		&	-0.38 	 ($^{+0.08}_{-0.07}$)	 &	 -0.21 	&		$-$ 		 &		 $-$ 		 &		$-$ 		\\
1508$+$55		&		0.4$-$4.9		&		1.24E$+$01	($^{+0.8}_{-12.4}$)		&		 -5.00E$-$02	($^{+2.05}_{-1.95}$)		 &		 3.20E$-$01	 ($^{+12.72}_{-0.32}$)		&		11		 &	14.10 	&		7.93E$-$02		&	-0.11 	 ($^{+0.21}_{-0.13}$)	 &	 -0.06 	&		-2.2 		 &		 T		 &		CRCN		\\
1530$+$27		&		0.23$-$1.64		&		1.26E$+$01	($^{+0.8}_{-12.6}$)		&		 -1.20E$-$01	($^{+0.10}_{-1.88}$)		 &		 7.00E$-$02	 ($^{+13.01}_{-0.07}$)		&		7		 &	6.06 	&		1.95E$-$01		&	-0.26 	 ($^{+0.24}_{-0.15}$)	 &	 -0.14 	&		-1.4 		 &		 $-$ 		 &		PCN		\\
1604$-$00		&		0.23$-$4.85		&		1.40E$+$01	($^{+1.0}_{-14.0}$)		&		 -9.00E$-$02	($^{+0.07}_{-1.91}$)		 &		 1.20E$-$01	 ($^{+13.86}_{-0.12}$)		&		16		 &	11.00 	&		6.14E$-$01		&	-0.20 	 ($^{+0.16}_{-0.14}$)	 &	 -0.11 	&		-1.5 		 &		 T		 &		PCN		\\
1612$+$07		&		0.23$-$4.85		&		7.40E$+$00	($^{+1.00}_{-7.40}$)		&		 -6.00E$-$02	($^{+2.06}_{-1.94}$)		 &		1.40E$-$01	 ($^{+8.35}_{-0.14}$)		&		6		 &	9.73 	&		2.10E$-$02		&	-0.14 	 ($^{+0.37}_{-0.27}$)	 &	-0.07 	&		-2.6 		 &		 CNS		 &		PCN		\\
1620$-$09		&		0.4$-$4.85		&		7.57E$-$02	($^{+7.12}_{-7.57{\rm E}-2}$)		&		 -3.28E$+$00	 ($^{+16.28}_{-12.72}$)		&		 5.98E$+$00	 ($^{+1.18}_{-5.98}$)		&		5		 &	0.22 	&		8.98E$-$01		 &	 -0.20 	($^{+1.55}_{-0.41}$)	&	-4.1E-3 	&		-1.7 		 &		 $-$ 		 &		$-$ 		\\
1700$-$18		&		0.4$-$1.6		&		1.06E$+$01	($^{+1.0}_{-10.6}$)		&		 -9.00E$-$02	($^{+2.09}_{-1.91}$)		 &		 1.40E$-$01	 ($^{+11.44}_{-0.14}$)		&		5		 &	1.59 	&		4.52E$-$01		&	-0.20 	 ($^{+1.13}_{-0.29}$)	 &	 -0.10 	&		-1.9 		 &		 $-$ 		 &		$-$ 		\\
1700$-$32		&		0.243$-$3.1		&		6.00E$-$01	($^{+14.60}_{-0.48}$)		&		 -1.87E$+$00	($^{+1.85}_{-0.13}$)		 &		1.36E$+$01	 ($^{+0.93}_{-13.6}$)		&		7		 &	7.35 	&		1.19E$-$01		&	-0.20 	 ($^{+0.15}_{-0.13}$)	 &	-0.02 	&		-3.1 		 &		 T		 &		CRCN		\\
1706$-$16		&		0.243$-$10.7		&		1.16E$+$01	($^{+0.8}_{-7.6}$)		&		 -2.20E$-$01	($^{+0.07}_{-0.53}$)		 &		9.00E$-$02	 ($^{+7.17}_{-0.09}$)		&		17		 &	32.30 	&		3.68E$-$03		&	-0.42 	 ($^{+0.10}_{-0.10}$)	 &	-0.24 	&		-1.5 		 &		 CRS		 &		CN		\\
1717$-$16		&		0.4$-$1.6		&		5.15E$+$00	($^{+5.85}_{-5.15}$)		&		 -1.98E$-$01	($^{+2.20}_{-1.80}$)		 &		5.15E$+$00	 ($^{+5.91}_{-5.15}$)		&		5		 &	0.86 	&		6.51E$-$01		&	-0.21 	 ($^{+0.26}_{-0.26}$)	 &	-0.11 	&		-2.2 		 &		 $-$ 		 &		$-$ 		\\
1717$-$29		&		0.4$-$1.6		&		2.06E$+$01	($^{+1.0}_{-20.2}$)		&		 -1.50E$-$01	($^{+0.08}_{-1.85}$)		 &		 0.00E$+$00	 ($^{+20.2}_{0}$)		&		5		&	 5.98 	&		5.02E$-$02		&	-0.31 	($^{+0.20}_{-0.13}$)	 &	 -0.17 	 &		-2.2 		&		 $-$ 		 &		CN		\\
1735$-$32		&		0.6$-$1.6		&		1.28E$+$01	($^{+1.0}_{-12.2}$)		&		 -2.60E$-$01	($^{+0.18}_{-1.74}$)		 &		 2.60E$-$01	 ($^{+12.33}_{-0.26}$)		&		4		 &	0.05 	&		8.28E$-$01		&	-0.47 	 ($^{+0.30}_{-0.18}$)	 &	 -0.27 	&		-0.9 		 &		 $-$ 		 &		$-$ 		\\
1738$-$08		&		0.4$-$1.6		&		1.80E$+$01	($^{+1.2}_{-17.9}$)		&		 -1.20E$-$01	($^{+0.11}_{-1.88}$)		 &		 1.80E$-$01	 ($^{+18.43}_{-0.18}$)		&		7		 &	2.12 	&		7.14E$-$01		&	-0.26 	 ($^{+0.23}_{-0.17}$)	 &	 -0.14 	&		-2.2 		 &		 M		 &		CN		\\
1811$+$40		&		0.6$-$1.6		&		1.77E$-$01	($^{+15.82}_{-1.77{\rm E}-1}$)		&		 -2.72E$+$00	 ($^{+13.72}_{-11.28}$)		&		 1.50E$+$01	 ($^{+0.96}_{-14.99}$)		&		4		 &	0.01 	&		9.42E$-$01		 &	 -0.13 	($^{+0.79}_{-0.80}$)	&	-4.5E-3 	&		-1.8 		 &		 $-$ 		 &		CN		\\
1813$-$26		&		0.4$-$1.6		&		3.56E$+$01	($^{+0.8}_{-34.4}$)		&		 -1.40E$-$01	($^{+0.06}_{-1.86}$)		 &		 4.00E$-$02	 ($^{+34.03}_{-0.04}$)		&		4		 &	3.24 	&		7.19E$-$02		&	-0.30 	 ($^{+0.14}_{-0.11}$)	 &	 -0.16 	&		-1.4 		 &		 $-$ 		 &		CN		\\
1819$-$22		&		0.4$-$4.85		&		1.72E$+$00	($^{+16.48}_{-1.72}$)		&		 -9.56E$-$01	 ($^{+10.96}_{-9.04}$)		 &		1.53E$+$01	 ($^{+2.74}_{-15.34}$)		&		7		 &	1.32 	&		8.61E$-$01		 &	-0.19 	($^{+0.43}_{-0.27}$)	 &	-0.05 	&		-1.7 		 &		 $-$ 		 &		CN		\\
1820$-$31		&		0.6$-$1.6		&		1.28E$+$01	($^{+1.2}_{-12.8}$)		&		 -2.20E$-$01	($^{+2.22}_{-1.78}$)		 &		 1.40E$-$01	 ($^{+13.19}_{-0.14}$)		&		5		 &	0.50 	&		7.79E$-$01		&	-0.42 	 ($^{+0.99}_{-0.28}$)	 &	 -0.24 	&		-2.1 		 &		 $-$ 		 &		CN		\\
1839$-$04		&		0.6$-$1.6		&		3.68E$+$01	($^{+46.0}_{-36.8}$)		&		 -3.75E$-$01	 ($^{+10.37}_{-9.63}$)		 &		3.70E$+$01	 ($^{+42.99}_{-36.99}$)		&		4		 &	0.05 	&		8.32E$-$01		 &	-0.36 	($^{+0.60}_{-0.61}$)	 &	-0.17 	&		-1.6 		 &		 $-$ 		 &		$-$ 		\\
1845$-$19		&		0.4$-$1.6		&		1.58E$-$01	($^{+7.84}_{-1.58{\rm E}-1}$)		&		 -1.95E$+$00	 ($^{+11.95}_{-8.05}$)		&		 7.00E$+$00	 ($^{+1.06}_{-7.00}$)		&		5		 &	0.77 	&		6.81E$-$01		 &	 -0.12 	($^{+0.38}_{-0.39}$)	&	-0.01 	&		-2.0 		 &		 $-$ 		 &		CN		\\
1846$-$06		&		0.4$-$1.6		&		5.58E$-$06	($^{+0.26}_{-5.58{\rm E}-6}$)		&		 -1.51E$+$01	 ($^{+28.93}_{7.71}$)		&		 1.01E$+$01	 ($^{+0.90}_{-1.89}$)		&		5		 &	30.30 	&		1.86E$-$07		 &	 -0.35 	($^{+0.12}_{-0.11}$)	&	-3.5E-9 	&		-2.2 		 &		 $-$ 		 &		$-$ 		\\
1848$+$12		&		0.4$-$1.6		&		1.29E$-$01	($^{+9.27}_{-0.13}$)		&		 -2.89E$+$00	 ($^{+14.89}_{-11.11}$)		 &		6.07E$+$00	 ($^{+3.32}_{-6.07}$)		&		4		 &	0.04 	&		8.43E$-$01		 &	 -0.23 	($^{+0.71}_{-0.71}$)	 &	-0.01 	&		-1.9 		 &		 $-$ 		 &		$-$ 		\\
1848$+$13		&		0.243$-$3.1		&		4.38E$-$01	($^{+11.96}_{-0.43}$)		&		 -1.98E$+$00	($^{+1.98}_{-3.03}$)		 &		1.16E$+$01	 ($^{+0.64}_{-11.61}$)		&		9		 &	27.90 	&		1.05E$-$04		&	-0.19 	 ($^{+0.18}_{-0.13}$)	 &	-0.02 	&		-1.4 		 &		 $-$ 		 &		$-$ 		\\
1857$-$26		&		0.327$-$3.1		&		3.92E$+$01	($^{+0.8}_{-34.0}$)		&		 -6.00E$-$02	($^{+0.02}_{-0.4}$)		 &		 2.40E$-$01	 ($^{+33.95}_{-0.24}$)		&		12		&	 98.50 	&		0.00E$+$00		&	-0.14 	 ($^{+0.05}_{-0.06}$)	&	 -0.07 	&		-2.1 		&		 M		 &		CRCN		\\
1907$+$03		&		0.43$-$4.85		&		5.15E$+$01	($^{+22.9}_{-51.2}$)		&		 -1.23E$-$01	($^{+0.11}_{-1.88}$)		 &		1.96E$+$01	 ($^{+51.39}_{-19.62}$)		&		6		 &	0.88 	&		8.32E$-$01		&	-0.20 	 ($^{+0.17}_{-0.14}$)	 &	-0.10 	&		-1.8 		 &		 T		 &		CRCN		\\
1916$+$14		&		0.4$-$1.72		&		8.00E$+$00	($^{+0.60}_{-7.86}$)		&		 -2.00E$-$01	($^{+0.15}_{-1.80}$)		 &		6.00E$-$02	 ($^{+7.72}_{-0.06}$)		&		5		 &	1.34 	&		5.12E$-$01		&	-0.39 	 ($^{+0.30}_{-0.17}$)	 &	-0.22 	&		-0.3 		 &		 T		 &		CN		\\
1917$+$00		&		0.4$-$1.6		&		1.00E$+$01	($^{+1.6}_{-10.0}$)		&		 -5.00E$-$02	($^{+2.05}_{-1.95}$)		 &		 1.00E$+$00	 ($^{+10.73}_{-1.00}$)		&		6		 &	3.49 	&		3.22E$-$01		&	-0.11 	 ($^{+1.52}_{-0.28}$)	 &	 -0.05 	&		-2.2 		 &		 T		 &		CRCN		\\
1929$+$10		&		0.4$-$10.5		&		1.86E$+$01	($^{+0.6}_{-9.8}$)		&		 -1.60E$-$01	($^{+0.03}_{-0.21}$)		 &		 7.00E$-$02	 ($^{+9.57}_{-0.07}$)		&		15		 &	20.00 	&		6.74E$-$02		&	-0.33 	 ($^{+0.05}_{-0.05}$)	 &	 -0.18 	&		-1.6 		 &		 T		 &		CN		\\
1935$+$25		&		0.6$-$1.6		&		3.72E$+$00	($^{+32.28}_{-3.70}$)		&		 -1.23E$+$00	($^{+1.18}_{-8.77}$)		 &		3.12E$+$01	 ($^{+3.89}_{-31.16}$)		&		4		 &	0.90 	&		3.44E$-$01		&	-0.26 	 ($^{+0.16}_{-0.66}$)	 &	-0.06 	&		-0.7 		 &		 $-$ 		 &		$-$ 		\\
1952$+$29		&		0.61$-$4.85		&		1.28E$+$01	($^{+19.6}_{-10.4}$)		&		 -6.33E$-$01	($^{+0.47}_{-2.09}$)		 &		1.80E$+$01	 ($^{+8.86}_{-18.01}$)		&		10		 &	7.92 	&		3.41E$-$01		&	-0.44 	 ($^{+0.12}_{-0.11}$)	 &	-0.20 	&		$-$ 		 &		 T		 &		CRCN		\\
1953$+$50		&		0.4$-$1.6		&		2.38E$+$00	($^{+7.42}_{-2.38}$)		&		 -4.51E$-$01	 ($^{+10.45}_{-9.55}$)		 &		6.47E$+$00	 ($^{+3.24}_{-6.47}$)		&		5		 &	0.27 	&		8.79E$-$01		 &	-0.24 	($^{+0.35}_{-0.35}$)	 &	-0.10 	&		-1.6 		 &		 CRS		 &		$-$ 		\\
2000$+$32		&		0.6$-$1.6		&		1.22E$+$01	($^{+1.2}_{-12.2}$)		&		 -1.80E$-$01	($^{+2.18}_{-1.82}$)		 &		 3.00E$-$01	 ($^{+12.85}_{-0.30}$)		&		4		 &	0.11 	&		7.37E$-$01		&	-0.35 	 ($^{+1.20}_{-0.31}$)	 &	 -0.20 	&		-1.1 		 &		 $-$ 		 &		$-$ 		\\
2003$-$08		&		0.4$-$1.4		&		1.89E$+$00	($^{+59.71}_{-1.81}$)		&		 -1.44E$+$00	($^{+1.39}_{-3.16}$)		 &		5.90E$+$01	 ($^{+2.37}_{-58.98}$)		&		4		 &	0.07 	&		8.42E$-$01		&	-0.10 	 ($^{+0.04}_{-0.10}$)	 &	-0.02 	&		-1.4 		 &		 T		 &		CN		\\
2020$+$28		&		0.23$-$10.7		&		4.20E$+$00	($^{+14.00}_{-3.40}$)		&		 -5.60E$-$01	($^{+0.51}_{-1.37}$)		 &		1.26E$+$01	 ($^{+2.33}_{-12.6}$)		&		14		 &	37.80 	&		8.33E$-$05		&	-0.27 	 ($^{+0.14}_{-0.11}$)	 &	-0.11 	&		$-$ 		 &		 T		 &		CN		\\
2021$+$51		&		0.23$-$10.6		&		5.11E$+$00	($^{+13.69}_{-3.51}$)		&		 -6.05E$-$01	($^{+0.49}_{-0.75}$)		 &		1.21E$+$01	 ($^{+2.37}_{-12.08}$)		&		14		 &	4.11 	&		9.67E$-$01		&	-0.33 	 ($^{+0.09}_{-0.07}$)	 &	-0.14 	&		$-$ 		 &		 CNS		 &		PCN		\\
2022$+$50		&		0.4$-$1.6		&		1.08E$+$01	($^{+1.0}_{-10.8}$)		&		 -2.00E$-$01	($^{+0.18}_{-1.80}$)		 &		 1.60E$-$01	 ($^{+11.22}_{-0.16}$)		&		5		 &	2.21 	&		3.31E$-$01		&	-0.39 	 ($^{+0.36}_{-0.21}$)	 &	 -0.22 	&		-0.8 		 &		 $-$ 		 &		$-$ 		\\
2045$-$16		&		0.243$-$4.9		&		6.01E$+$00	($^{+10.39}_{-5.21}$)		&		 -3.22E$-$01	($^{+0.26}_{-1.53}$)		 &		9.67E$+$00	 ($^{+4.79}_{-9.67}$)		&		19		 &	9.53 	&		9.00E$-$01		&	-0.25 	 ($^{+0.12}_{-0.10}$)	 &	-0.12 	&		-2.1 		 &		 T		 &		CRCN		\\
2053$+$21		&		0.4$-$1.6		&		1.25E$-$01	($^{+10.88}_{-0.12}$)		&		 -2.54E$+$00	 ($^{+12.54}_{-10.46}$)		 &		9.91E$+$00	 ($^{+1.08}_{-9.91}$)		&		5		 &	1.21 	&		5.54E$-$01		 &	 -0.11 	($^{+0.33}_{-0.33}$)	 &	-0.01 	&		-0.8 		 &		 $-$ 		 &		$-$ 		\\
2148$+$63		&		0.4$-$4.85		&		4.21E$+$00	($^{+20.39}_{-4.20}$)		&		 -1.35E$+$00	($^{+1.27}_{-6.73}$)		 &		1.80E$+$01	 ($^{+4.43}_{-17.96}$)		&		7		 &	1.98 	&		7.39E$-$01		&	-0.43 	 ($^{+0.26}_{-0.17}$)	 &	-0.11 	&		-1.8 		 &		 CNS		 &		CN		\\
2154$+$40		&		0.23$-$4.85		&		2.52E$+$01	($^{+1.2}_{-20.2}$)		&		 -1.40E$-$01	($^{+0.06}_{-0.45}$)		 &		 4.00E$-$01	 ($^{+19.95}_{-0.40}$)		&		11		 &	24.50 	&		1.86E$-$03		&	-0.29 	 ($^{+0.11}_{-0.08}$)	 &	 -0.16 	&		-1.6 		 &		 CD		 &		CN		\\
2210$+$29		&		0.4$-$1.6		&		2.00E$+$01	($^{+1.0}_{-20.0}$)		&		 -7.00E$-$02	($^{+2.07}_{-1.93}$)		 &		 3.20E$-$01	 ($^{+20.49}_{-0.32}$)		&		4		 &	0.43 	&		5.14E$-$01		&	-0.16 	 ($^{+0.16}_{-0.13}$)	 &	 -0.08 	&		-1.5 		 &		 $-$ 		 &		$-$ 		\\
2224$+$65		&		0.408$-$1.7		&		5.00E$+$01	($^{+0.8}_{-47.6}$)		&		 -1.40E$-$01	($^{+0.03}_{-1.61}$)		 &		 1.60E$-$01	 ($^{+46.71}_{-0.16}$)		&		7		 &	19.60 	&		6.03E$-$04		&	-0.29 	 ($^{+0.11}_{-0.07}$)	 &	 -0.16 	&		-1.9 		 &		 T		 &		PCN		\\
2227$+$61		&		0.4$-$1.6		&		4.70E$-$01	($^{+4.93}_{-0.46}$)		&		 -3.66E$+$00	($^{+2.82}_{-4.55}$)		 &		2.40E$+$01	 ($^{+1.65}_{-4.49}$)		&		5		 &	15.50 	&		4.40E$-$04		&	-0.36 	 ($^{+0.18}_{-0.12}$)	 &	-0.01 	&		-2.6 		 &		 $-$ 		 &		$-$ 		\\
2255$+$58		&		0.6$-$2.7		&		1.98E$+$01	($^{+1.6}_{-19.8}$)		&		 -1.30E$-$01	($^{+2.13}_{-1.87}$)		 &		 0.00E$+$00	 ($^{+20.55}_{0}$)		&		7		&	 18.40 	&		1.04E$-$03		&	-0.28 	($^{+0.70}_{-0.32}$)	 &	 -0.15 	 &		-2.1 		&		 CRS		 &		$-$ 		\\
2303$+$30		&		0.4$-$1.72		&		7.80E$+$00	($^{+1.00}_{-7.80}$)		&		 -9.00E$-$02	($^{+2.09}_{-1.91}$)		 &		2.40E$-$01	 ($^{+8.52}_{-0.24}$)		&		6		 &	2.14 	&		5.45E$-$01		&	-0.20 	 ($^{+1.48}_{-0.32}$)	 &	-0.10 	&		-2.3 		 &		 CNS		 &		CR		\\
2306$+$55		&		0.4$-$1.6		&		2.50E$+$01	($^{+0.8}_{-24.8}$)		&		 -9.00E$-$02	($^{+0.07}_{-1.91}$)		 &		 4.00E$-$02	 ($^{+25.16}_{-0.04}$)		&		4		 &	0.28 	&		5.99E$-$01		&	-0.20 	 ($^{+0.16}_{-0.13}$)	 &	 -0.11 	&		-1.9 		 &		 CD		 &		CN		\\
2310$+$42		&		0.23$-$10.6		&		1.54E$+$01	($^{+1.4}_{-15.4}$)		&		 -5.00E$-$02	($^{+2.05}_{-1.95}$)		 &		 3.00E$-$01	 ($^{+16.24}_{-0.30}$)		&		8		 &	2.94 	&		7.09E$-$01		&	-0.12 	 ($^{+0.21}_{-0.17}$)	 &	 -0.06 	&		-1.9 		 &		 CNS		 &		CN		\\
2315$+$21		&		0.23$-$4.85		&		8.60E$+$00	($^{+1.20}_{-8.60}$)		&		 -6.00E$-$02	($^{+2.06}_{-1.94}$)		 &		1.80E$-$01	 ($^{+9.67}_{-0.18}$)		&		6		 &	0.64 	&		8.88E$-$01		&	-0.14 	 ($^{+0.53}_{-0.28}$)	 &	-0.07 	&		-2.1 		 &		 CNS		 &		CR		\\
2319$+$60		&		0.23$-$4.85		&		4.42E$+$00	($^{+19.38}_{-3.82}$)		&		 -8.01E$-$01	($^{+0.69}_{-1.36}$)		 &		1.79E$+$01	 ($^{+3.51}_{-17.89}$)		&		13		 &	14.70 	&		1.46E$-$01		&	-0.29 	 ($^{+0.16}_{-0.15}$)	 &	-0.10 	&		$-$ 		 &		 T		 &		CRCN		\\
2324$+$60		&		0.6$-$2.7		&		2.68E$+$01	($^{+2.2}_{-26.8}$)		&		 -2.10E$-$01	($^{+2.21}_{-1.79}$)		 &		 1.00E$-$01	 ($^{+26.61}_{-0.10}$)		&		6		 &	2.93 	&		4.02E$-$01		&	-0.41 	 ($^{+0.43}_{-0.25}$)	 &	 -0.23 	&		$-$ 		 &		 $-$ 		 &		CN		\\
2327$-$20		&		0.243$-$3.1		&		1.10E$+$00	($^{+6.50}_{-1.10}$)		&		 -5.95E$-$01	 ($^{+10.59}_{-9.41}$)		 &		5.95E$+$00	 ($^{+1.53}_{-5.95}$)		&		10		 &	7.07 	&		4.25E$-$01		 &	-0.19 	($^{+0.68}_{-0.25}$)	 &	-0.07 	&		-2.0 		 &		 T		 &		PCN		\\
2351$+$61		&		0.4$-$4.85		&		8.72E$+$00	($^{+4.68}_{-8.71}$)		&		 -2.35E$-$01	($^{+0.23}_{-6.53}$)		 &		3.58E$+$00	 ($^{+8.92}_{-3.58}$)		&		7		 &	3.60 	&		4.65E$-$01		&	-0.33 	 ($^{+0.32}_{-0.23}$)	 &	-0.18 	&		-1.1 		 &		 $-$ 		 &		$-$ 		\\

\enddata
\tablenotetext{\space}{The abbreviations in the last two columns mean the following profile types: in the classification from Rankin (1990), CNS - conal single, CRS - core single, CD - conal double, T - triple, M - Multiple; in the classification from Lyne \& Manchester (1988), CN - cone, PCN - partial cone, CRCN - both core and cone.   }
\end{deluxetable}

%%%%%%%TABLEa
% table 3  group B
%\usepackage{rotating}
%\begin{sideways}
\begin{deluxetable}{lllllllllllll}
%\tabletypesize{\scriptsize}
\tabletypesize{\tiny}
\rotate
\tablewidth{0pt}
 \tablecaption{Parameters for 40 group-B pulsars. \label{Tab:groupB}}
\tablehead{ \colhead{PSR B}&\colhead{Freq. (GHz)} & \colhead{A} &\colhead{$\mu$}&
\colhead{$W_{\rm 10,0} (^\circ)$ }    &\colhead{N}
& \colhead{$\chi^{2}_{\rm min}$} & \colhead{$Q$} &\colhead{$\eta$} &\colhead{$\eta^\prime$}&\colhead{$\alpha$}&
\colhead{R90} & \colhead{LM88}  }
\startdata

0105$+$65		&		0.4$-$1.72		&		3.75E$-$08	($^{+17}_{-3.75{\rm E}-8}$)		&		 -1.90E$+$01	 ($^{+94.95}_{-76.05}$)		 &		1.58E$+$01	 ($^{+1.23}_{-15.81}$)		&		7		 &	14.50 	&		1.16E$-$02		 &		 -7.63E$-$02	 ($^{+0.15}_{-0.14}$)		&		 -4.03E$-$12		&		 -1.9 		 &		 CRS		&		CN		 \\
0136$+$57		&		0.4$-$4.75		&		7.16E$+$00	($^{+5.64}_{-7.16}$)		&		 6.20E$-$04	($^{+2.00}_{-2.00}$)		 &		4.53E$+$00	 ($^{+8.41}_{-4.53}$)		&		7		 &	2.55 	&		6.31E$-$01		&		9.48E$-$04	 ($^{+0.49}_{-0.37}$)		&		 4.72E$-$04		&		$-$ 		 &		 CRS		&		CN		\\
0450$-$18		&		0.23$-$4.9		&		6.27E$-$02	($^{+26.14}_{-6.27{\rm E}-2}$)		&		 1.91E$+$00	 ($^{+8.09}_{-11.91}$)		&		2.50E$+$01	 ($^{+1.33}_{-25.05}$)		&		13		 &	7.24 	&		1.27E$-$01		 &		 5.07E$-$02	($^{+0.23}_{-0.19}$)		&		 4.62E$-$02		&		-2.0 		 &		 T		&		CRCN		 \\
0540$+$23		&		0.243$-$10.6		&		9.45E$-$03	($^{+25.19}_{9.45{\rm E}-3}$)		&		 -4.74E$+$00	 ($^{+23.74}_{-19.26}$)		&		 2.40E$+$01	 ($^{+0.98}_{-24.03}$)		&		17		 &	19.00 	&		1.68E$-$01		 &		 -2.94E$-$02	($^{+0.17}_{-0.14}$)		&		 -7.95E$-$05		 &		$-$ 		&		 $-$ 		&		PCN		 \\
0611$+$22		&		0.243$-$4.9		&		4.71E$-$02	($^{+3.75}_{-3.71{\rm E}-1}$)		&		 -3.77E$+$00	 ($^{+3.38}_{-1.63}$)		&		 1.41E$+$01	 ($^{+1.02}_{-3.56}$)		&		14		 &	16.30 	&		1.34E$-$01		 &		 -9.53E$-$02	($^{+0.07}_{-0.15}$)		&		 -9.32E$-$04		 &		 -2.1 		&		 CRS		&		CN		 \\
0740$-$28		&		0.408$-$10.6		&		1.58E$+$01	($^{+1.2}_{-15.8}$)		&		 -2.00E$-$02	($^{+2.02}_{-1.98}$)		 &		1.40E$-$01	 ($^{+16.47}_{-0.14}$)		&		15		 &	27.90 	&		5.69E$-$03		&		-4.83E$-$02	 ($^{+0.22}_{-0.26}$)		&		 -2.43E$-$02		&		 -2.0 		 &		 CRS		&		PCN		\\
0751$+$32		&		0.4$-$4.85		&		1.22E$-$05	($^{+25.4}_{-1.22{\rm E}-5}$)		&		 -1.31E$+$01	 ($^{+65.12}_{-52.88}$)		&		 2.46E$+$01	 ($^{+0.92}_{-24.56}$)		&		6		 &	3.79 	&		3.10E$-$01		 &		 -7.66E$-$02	($^{+0.21}_{-0.13}$)		&		 -6.00E$-$09		 &		 -1.5 		&		 CD		&		$-$ 		 \\
0809$+$74		&		0.23$-$4.9		&		1.37E$-$09	($^{+27.6}_{-1.37{\rm E}-9}$)		&		 1.32E$+$01	 ($^{+52.84}_{-66.16}$)		&		2.69E$+$01	 ($^{+0.87}_{-26.9}$)		&		14		 &	11.40 	&		4.96E$-$01		 &		 5.39E$-$02	($^{+0.09}_{-0.13}$)		&		 5.39E$-$02		&		-1.4 		 &		 CNS		&		PCN		 \\
0818$-$13		&		0.4$-$4.9		&		1.04E$+$01	($^{+1.0}_{-10.4}$)		&		 -4.00E$-$02	($^{+2.04}_{-1.96}$)		 &		 2.00E$-$01	 ($^{+11.3}_{-0.20}$)		&		11		 &	1.59 	&		9.91E$-$01		&		-9.33E$-$02	 ($^{+0.30}_{-0.22}$)		 &		 -4.76E$-$02		&		 -2.3 		 &		 CNS		&		CN		\\
0820$+$02		&		0.4$-$1.6		&		2.40E$-$01	($^{+15.56}_{-0.24}$)		&		 3.30E$-$01	($^{+1.67}_{-2.33}$)		 &		1.48E$+$01	 ($^{+0.96}_{-14.8}$)		&		4		 &	0.13 	&		7.18E$-$01		&		1.51E$-$02	 ($^{+1.08}_{-0.19}$)		&		 9.02E$-$03		&		-2.4 		 &		 CNS		&		$-$ 		\\
0834$+$06		&		0.243$-$4.9		&		1.86E$-$02	($^{+9.98}_{-1.86{\rm E}-2}$)		&		 2.23E$+$00	 ($^{+8.77}_{-12.23}$)		&		9.44E$+$00	 ($^{+0.73}_{-9.44}$)		&		19		 &	12.70 	&		6.98E$-$01		 &		 6.67E$-$02	($^{+0.76}_{-0.28}$)		&		 6.25E$-$02		&		-2.7 		 &		 CD		&		CN		\\
0906$-$17		&		0.243$-$4.85		&		1.43E$-$01	($^{+21.06}_{-0.14}$)		&		 1.41E$+$00	 ($^{+8.59}_{-11.41}$)		 &		2.04E$+$01	 ($^{+0.91}_{-20.43}$)		&		11		 &	8.77 	&		3.63E$-$01		 &		6.28E$-$02	 ($^{+0.33}_{-0.17}$)		&		 5.30E$-$02		&		-1.4 		 &		 $-$ 		&		PCN		 \\
0942$-$13		&		0.23$-$4.85		&		1.27E$-$07	($^{+8.0}_{-1.27{\rm E}-7}$)		&		 -1.09E$+$01	 ($^{+54.89}_{-43.11}$)		&		 7.26E$+$00	 ($^{+0.80}_{-7.26}$)		&		9		 &	4.66 	&		6.21E$-$01		 &		 -3.76E$-$04	($^{+1.15}_{-0.41}$)		&		 -4.47E$-$10		 &		 -3.0 		&		 CRS		&		CR		 \\
0950$+$08		&		0.23$-$8.4		&		3.42E$-$01	($^{+30.46}_{-0.34}$)		&		 2.38E$-$01	 ($^{+9.76}_{-10.24}$)		 &		2.97E$+$01	 ($^{+1.37}_{-29.67}$)		&		16		 &	38.20 	&		2.63E$-$06		 &		7.43E$-$03	 ($^{+0.06}_{-0.07}$)		&		 4.23E$-$03		&		-2.2 		 &		 CNS		&		CN		 \\
1540$-$06		&		0.23$-$1.72		&		9.40E$+$00	($^{+1.20}_{-9.40}$)		&		 -2.00E$-$02	($^{+2.02}_{-1.98}$)		 &		1.20E$-$01	 ($^{+10.48}_{-0.12}$)		&		9		 &	3.87 	&		6.94E$-$01		&		-4.81E$-$02	 ($^{+2.19}_{-0.33}$)		&		 -2.42E$-$02		&		 -2.0 		 &		 CNS		&		$-$ 		\\
1552$-$31		&		0.41$-$1.6		&		1.88E$+$01	($^{+6.6}_{-18.8}$)		&		 -2.00E$-$02	($^{+2.02}_{-1.98}$)		 &		 5.60E$+$00	 ($^{+19.86}_{-5.6}$)		&		5		 &	0.61 	&		7.38E$-$01		&		-3.77E$-$02	 ($^{+1.06}_{-0.22}$)		 &		 -1.89E$-$02		&		 -1.6 		 &		 $-$ 		&		CN		\\
1649$-$23		&		0.4$-$1.6		&		8.79E$-$02	($^{+11.91}_{-8.79{\rm E}-2}$)		&		 -1.64E$+$00	 ($^{+11.64}_{-8.36}$)		&		 1.11E$+$01	 ($^{+1.00}_{-11.1}$)		&		5		 &	0.01 	&		9.97E$-$01		 &		 -3.38E$-$02	($^{+0.31}_{-0.31}$)		&		 -3.95E$-$03		 &		 -1.7 		&		 $-$ 		&		CR		 \\
1702$-$19		&		0.243$-$4.85		&		1.72E$+$01	($^{+0.8}_{-17.2}$)		&		 -3.00E$-$02	($^{+2.03}_{-1.97}$)		 &		7.00E$-$02	 ($^{+17.72}_{-0.07}$)		&		12		 &	5.06 	&		8.29E$-$01		&		-7.18E$-$02	 ($^{+0.22}_{-0.14}$)		&		 -3.64E$-$02		&		 -1.3 		 &		 T		&		CN		\\
1718$-$32		&		0.6$-$1.6		&		1.66E$+$01	($^{+1.2}_{-16.6}$)		&		 3.00E$-$02	($^{+1.97}_{-2.03}$)		 &		 6.00E$-$02	 ($^{+17.80}_{-0.06}$)		&		5		 &	0.49 	&		7.81E$-$01		&		7.74E$-$02	 ($^{+2.51}_{-0.43}$)		 &		 3.78E$-$02		&		-2.3 		 &		 $-$ 		&		CN		\\
1727$-$47		&		0.243$-$3.1		&		3.85E$-$02	($^{+11.36}_{-3.85{\rm E}-1}$)		&		 -3.33E$+$00	 ($^{+16.33}_{-13.67}$)		&		 1.07E$+$01	 ($^{+0.65}_{-10.7}$)		&		7		 &	15.10 	&		4.70E$-$03		 &		 -7.06E$-$02	($^{+1.04}_{-0.21}$)		&		 -1.15E$-$03		 &		 $-$ 		&		 T		&		CN		 \\
1737$-$30		&		0.92$-$8.4		&		1.70E$-$04	($^{+7.40}_{-1.70{\rm E}-4}$)		&		 4.27E$+$00	 ($^{+1.60}_{-4.27}$)		&		6.50E$+$00	 ($^{+1.03}_{-6.50}$)		&		6		 &	15.30 	&		1.65E$-$03		 &		 2.21E$-$02	($^{+0.66}_{-0.02}$)		&		 2.20E$-$02		&		-1.3 		 &		 $-$ 		&		$-$ 		 \\
1745$-$12		&		0.6$-$4.85		&		2.52E$+$00	($^{+18.48}_{-2.52}$)		&		 2.35E$-$01	 ($^{+9.76}_{-10.24}$)		 &		1.70E$+$01	 ($^{+3.97}_{-17.01}$)		&		5		 &	1.00 	&		6.10E$-$01		 &		8.52E$-$02	 ($^{+0.37}_{-0.94}$)		&		 4.69E$-$02		&		-2.1 		 &		 M		&		$-$ 		 \\
1749$-$28		&		0.4$-$4.9		&		5.94E$-$02	($^{+8.94}_{-5.94{\rm E}-2}$)		&		 -1.84E$+$00	 ($^{+11.84}_{-8.16}$)		&		 8.16E$+$00	 ($^{+0.48}_{-8.16}$)		&		11		 &	5.92 	&		6.64E$-$01		 &		 -3.76E$-$02	($^{+0.83}_{-0.33}$)		&		 -3.50E$-$03		 &		 $-$ 		&		 CRS		&		CRCN		 \\
1822$+$00		&		0.6$-$10.7		&		2.14E$+$01	($^{+1.0}_{-21.4}$)		&		 1.00E$-$02	($^{+1.99}_{-2.01}$)		 &		 3.00E$-$01	 ($^{+22.10}_{-0.30}$)		&		15		 &	40.10 	&		6.95E$-$05		&		2.49E$-$02	 ($^{+0.08}_{-0.15}$)		 &		 1.23E$-$02		&		-2.4 		 &		 $-$ 		&		$-$ 		\\
1831$-$04		&		0.6$-$4.85		&		1.19E$+$02	($^{+14}_{-118}$)		&		 4.00E$-$02	($^{+1.45}_{-0.03}$)		 &		 1.32E$+$01	 ($^{+118.48}_{-13.2}$)		&		5		 &	19.40 	&		5.98E$-$05		&		9.41E$-$02	 ($^{+0.07}_{-0.06}$)		 &		 4.59E$-$02		&		-1.3 		 &		 M		&		CRCN		\\
1839$+$56		&		0.23$-$4.85		&		1.18E$+$01	($^{+1.2}_{-11.8}$)		&		 -3.00E$-$02	($^{+2.03}_{-1.97}$)		 &		 3.60E$-$01	 ($^{+12.80}_{-0.36}$)		&		7		 &	1.11 	&		8.92E$-$01		&		-7.00E$-$02	 ($^{+0.69}_{-0.30}$)		 &		 -3.55E$-$02		&		 -1.5 		 &		 $-$ 		&		CN		\\
1844$-$04		&		0.6$-$1.6		&		1.86E$+$01	($^{+1.0}_{-18.6}$)		&		 -2.00E$-$02	($^{+2.02}_{-1.98}$)		 &		 2.00E$-$02	 ($^{+19.50}_{-0.02}$)		&		4		 &	0.14 	&		7.13E$-$01		&		-4.86E$-$02	 ($^{+1.35}_{-0.29}$)		 &		 -2.45E$-$02		&		 -2.2 		 &		 CRS		&		$-$ 		\\
1845$-$01		&		0.92$-$10.6		&		1.90E$+$01	($^{+3.0}_{-19.0}$)		&		 3.00E$-$02	($^{+1.97}_{-2.03}$)		 &		 1.60E$+$00	 ($^{+20.46}_{-1.60}$)		&		5		 &	1.87 	&		3.92E$-$01		&		7.15E$-$02	 ($^{+0.22}_{-0.40}$)		 &		 3.51E$-$02		&		-1.6 		 &		 CT		&		CRCN		\\
1905$+$39		&		0.4$-$1.6		&		1.88E$+$01	($^{+3.2}_{-18.8}$)		&		 -3.00E$-$02	($^{+2.03}_{-1.97}$)		 &		 2.40E$+$00	 ($^{+19.61}_{-2.40}$)		&		5		 &	3.21 	&		2.01E$-$01		&		-6.42E$-$02	 ($^{+0.52}_{-0.16}$)		 &		 -3.24E$-$02		&		 -2.0 		 &		 M		&		CN		\\
1919$+$21		&		0.243$-$3.1		&		1.18E$+$01	($^{+0.6}_{-11.8}$)		&		 1.00E$-$02	($^{+1.99}_{-2.01}$)		 &		 5.00E$-$02	 ($^{+12.52}_{-0.05}$)		&		14		 &	15.70 	&		1.53E$-$01		&		2.52E$-$02	 ($^{+0.34}_{-0.21}$)		 &		 1.24E$-$02		&		-2.6 		 &		 M		&		CN		\\
1923$+$04		&		0.4$-$4.85		&		3.00E$-$01	($^{+10.10}_{-0.30}$)		&		 -4.00E$-$02	($^{+2.04}_{-1.96}$)		 &		9.20E$+$00	 ($^{+1.34}_{-9.20}$)		&		5		 &	5.44 	&		6.60E$-$02		&		-3.11E$-$03	 ($^{+0.44}_{-0.34}$)		&		 -1.51E$-$03		&		 -2.7 		 &		 CNS		&		$-$ 		\\
1937$-$26		&		0.243$-$3.1		&		1.04E$+$01	($^{+1.0}_{-10.4}$)		&		 -2.00E$-$02	($^{+2.02}_{-1.98}$)		 &		 2.60E$-$01	 ($^{+11.03}_{-0.26}$)		&		10		 &	25.70 	&		5.68E$-$04		&		-4.75E$-$02	 ($^{+0.29}_{-0.22}$)		 &		 -2.39E$-$02		&		 -0.9 		 &		 $-$ 		&		PCN		\\
2000$+$40		&		0.4$-$1.6		&		1.17E$-$01	($^{+21.48}_{-1.17{\rm E}-1}$)		&		 -1.00E$+$00	 ($^{+11.00}_{-9.00}$)		&		 2.04E$+$01	 ($^{+1.19}_{-20.41}$)		&		5		 &	0.54 	&		7.40E$-$01		 &		 -1.31E$-$02	($^{+0.27}_{-0.26}$)		&		 -2.91E$-$03		 &		 -2.2 		&		 $-$ 		&		$-$ 		 \\
2016$+$28		&		0.23$-$10.7		&		1.50E$+$01	($^{+0.6}_{-15.0}$)		&		 -2.00E$-$02	($^{+2.02}_{-1.98}$)		 &		 6.00E$-$02	 ($^{+15.47}_{-0.06}$)		&		14		 &	22.80 	&		1.89E$-$02		&		-4.85E$-$02	 ($^{+0.12}_{-0.12}$)		 &		 -2.44E$-$02		&		 -2.2 		 &		 CNS		&		CN		\\
2044$+$15		&		0.23$-$1.4		&		5.60E$-$07	($^{+17.60}_{-5.60{\rm E}-7}$)		&		 -1.02E$+$01	 ($^{+51.21}_{-40.79}$)		&		 1.66E$+$01	 ($^{+1.02}_{-16.57}$)		&		6		 &	1.94 	&		6.78E$-$01		 &		 -3.89E$-$04	($^{+0.26}_{-0.25}$)		&		 -1.09E$-$09		 &		 -1.7 		&		 CD		&		CN		 \\
2106$+$44		&		0.6$-$1.6		&		3.28E$+$01	($^{+1.2}_{-32.8}$)		&		 -2.00E$-$02	($^{+2.02}_{-1.98}$)		 &		 3.40E$-$01	 ($^{+33.75}_{-0.34}$)		&		4		 &	2.40 	&		1.21E$-$01		&		-4.82E$-$02	 ($^{+0.77}_{-0.22}$)		 &		 -2.43E$-$02		&		 -1.4 		 &		 $-$ 		&		CN		\\
2110$+$27		&		0.4$-$4.85		&		1.00E$-$01	($^{+7.30}_{-0.10}$)		&		 -2.00E$+$00	($^{+4.00}_{-0.01}$)		 &		6.40E$+$00	 ($^{+1.13}_{-6.40}$)		&		6		 &	1.51 	&		6.80E$-$01		&		-8.84E$-$02	 ($^{+0.87}_{-0.45}$)		&		 -7.25E$-$03		&		 -2.2 		 &		 CNS		&		$-$ 		\\
2111$+$46		&		0.23$-$4.85		&		7.20E$+$01	($^{+1.6}_{-72.0}$)		&		 3.00E$-$02	($^{+1.97}_{-2.03}$)		 &		 2.80E$-$01	 ($^{+72.69}_{-0.28}$)		&		10		 &	115.00 	&		0.00E$+$00		&		7.74E$-$02	 ($^{+0.10}_{-0.08}$)		 &		 3.78E$-$02		&		-2.1 		 &		 T		&		CRCN		\\
2152$-$31		&		0.23$-$1.56		&		5.42E$-$07	($^{+12.40}_{-5.42{\rm E}-7}$)		&		 -1.03E$+$01	 ($^{+51.33}_{-41.67}$)		&		 1.17E$+$01	 ($^{+0.94}_{-11.66}$)		&		6		 &	2.74 	&		5.00E$-$01		 &		 -6.03E$-$04	($^{+0.34}_{-0.34}$)		&		 -1.44E$-$09		 &		 -2.3 		&		 $-$ 		&		CN		 \\
2334$+$61		&		0.4$-$1.6		&		2.58E$+$01	($^{+1.2}_{-25.8}$)		&		 2.00E$-$02	($^{+1.98}_{-2.02}$)		 &		 2.80E$-$01	 ($^{+26.74}_{-0.28}$)		&		5		 &	0.62 	&		7.32E$-$01		&		5.06E$-$02	 ($^{+1.04}_{-0.20}$)		 &		 2.49E$-$02		&		-1.7 		 &		 $-$ 		&		$-$ 		\\

\enddata
\end{deluxetable}

%%%%%%%%%%%%%%%%%%%%%%%%%%%%%%%%%%%%%%%%%%%%%%%%%%%%%%%%%%%%%%%%%%%%%%%%%%%%%
% table 4  group C
%\usepackage{rotating}
%\begin{sideways}
\begin{deluxetable}{lllllllllllll}
\tabletypesize{\scriptsize}
\rotate
\tablewidth{0pt}
\tablecaption{Parameters for 29 group-C pulsars. \label{Tab:groupc}}
\tablehead{ \colhead{PSR B}&\colhead{Freq. (GHz)} & \colhead{A} &\colhead{$\mu$}&
\colhead{$W_{\rm 10,0} (^\circ)$ }    &\colhead{N}
& \colhead{$\chi^{2}_{\rm min}$} & \colhead{$Q$} &\colhead{$\eta$} &\colhead{$\eta^\prime$}&\colhead{$\alpha$}&
\colhead{R90} & \colhead{LM88}  }

\startdata
0138$+$59		&		0.4$-$1.72		&		1.10E$+$00	($^{+31.70}_{-1.10}$)		&		 1.22E$+$00	 ($^{+8.78}_{-11.22}$)		 &		3.07E$+$01	 ($^{+2.05}_{-30.71}$)		&		8		 &	3.28E$+$01	&		4.57E$-$06		 &	0.23 	 ($^{+0.31}_{-0.30}$)	&	0.18 	&		-1.9 		 &		 $-$ 		 &		PCN		\\
0320$+$39		&		0.23$-$4.85		&		2.35E$+$00	($^{+3.85}_{-2.03}$)		&		 1.19E$+$00	($^{+1.21}_{-0.52}$)		 &		8.62E$+$00	 ($^{+1.65}_{-3.08}$)		&		5		 &	1.68E$+$00	&		4.37E$-$01		&	1.54 	 ($^{+0.33}_{-0.27}$)	&	0.97 	&		-2.9 		 &		 CNS		 &		$-$ 		\\
0626$+$24		&		0.23$-$4.85		&		4.04E$+$00	($^{+14.56}_{-4.04}$)		&		 3.04E$-$01	 ($^{+9.70}_{-10.30}$)		 &		1.35E$+$01	 ($^{+5.16}_{-13.47}$)		&		7		 &	2.37E$+$00	&		6.78E$-$01		 &	0.21 	 ($^{+0.44}_{-0.33}$)	&	0.11 	&		-1.6 		 &		 CRS		 &		$-$ 		\\
1254$-$10		&		0.4$-$4.85		&		1.78E$+$01	($^{+1.0}_{-15.2}$)		&		 1.90E$-$01	($^{+0.83}_{-0.10}$)		 &		 3.40E$-$01	 ($^{+15.41}_{-0.34}$)		&		5		 &	2.67E$+$00	&		2.63E$-$01		&	0.59 	 ($^{+0.48}_{-0.33}$)	&	 0.26 	&		-1.8 		 &		 $-$ 		 &		$-$ 		\\
1556$-$44		&		0.631$-$3.1		&		2.00E$+$01	($^{+0.2}_{-1.2}$)		&		 3.80E$-$01	($^{+0.03}_{-0.02}$)		 &		 0.00E$+$00	 ($^{+1.36}_{0}$)		&		9		&	 2.30E$+$01	&		7.80E$-$04		&	1.58 	 ($^{+0.13}_{-0.10}$)	&	 0.60 	&		$-$ 		 &		 CRS		 &		PCN		\\
1600$-$27		&		0.4$-$1.6		&		1.62E$+$01	($^{+1.2}_{-15.8}$)		&		 1.70E$-$01	($^{+1.83}_{-0.13}$)		 &		 4.00E$-$01	 ($^{+15.95}_{-0.40}$)		&		5		 &	8.38E$-$02	&		9.59E$-$01		&	0.51 	 ($^{+3.17}_{-0.41}$)	&	 0.23 	&		-1.7 		 &		 $-$ 		 &		CN		\\
1642$-$03		&		0.243$-$4.9		&		1.06E$+$00	($^{+1.54}_{-0.72}$)		&		 1.61E$+$00	($^{+0.74}_{-0.59}$)		 &		6.67E$+$00	 ($^{+1.21}_{-1.78}$)		&		19		 &	1.06E$+$02	&		3.11E$-$15		&	1.90 	 ($^{+0.70}_{-0.54}$)	&	1.36 	&		-2.3 		 &		 CRS		 &		CRCN		\\
1732$-$07		&		0.243$-$3.1		&		1.82E$+$01	($^{+0.8}_{-17.0}$)		&		 1.00E$-$01	($^{+0.76}_{-0.06}$)		 &		 8.00E$-$02	 ($^{+17.28}_{-0.08}$)		&		10		 &	3.57E$+$01	&		8.32E$-$06		&	0.28 	 ($^{+0.20}_{-0.17}$)	&	 0.13 	&		-1.9 		 &		 $-$ 		 &		$-$ 		\\
1737$+$13		&		0.243$-$4.85		&		2.10E$+$01	($^{+2.6}_{-21.0}$)		&		 7.00E$-$02	($^{+1.93}_{-0.06}$)		 &		1.80E$+$00	 ($^{+21.42}_{-1.80}$)		&		16		 &	6.03E$+$01	&		4.71E$-$08		&	0.18 	 ($^{+0.17}_{-0.15}$)	&	0.08 	&		-1.5 		 &		 M		 &		CRCN		\\
1742$-$30		&		0.4$-$3.1		&		2.04E$+$01	($^{+1.0}_{-20.2}$)		&		 1.00E$-$01	($^{+1.9}_{-0.06}$)		 &		 2.60E$-$01	 ($^{+20.70}_{-0.26}$)		&		10		&	 2.05E$+$01	&		4.65E$-$03		&	0.28 	 ($^{+0.19}_{-0.17}$)	&	 0.13 	&		-1.6 		 &		 T		 &		$-$ 		\\
1818$-$04		&		0.6$-$4.9		&		1.16E$+$01	($^{+1.2}_{-11.6}$)		&		 1.60E$-$01	($^{+1.84}_{-2.16}$)		 &		 0.00E$+$00	 ($^{+12.90}_{0}$)		&		10		&	 7.63E$+$00	&		3.67E$-$01		&	0.49 	 ($^{+0.76}_{-0.53}$)	&	 0.22 	&		-2.4 		 &		T	 	 &		$-$ 		\\
1821$+$05		&		0.4$-$4.85		&		2.68E$+$01	($^{+3.2}_{-26.8}$)		&		 6.00E$-$02	($^{+1.94}_{-2.06}$)		 &		 2.20E$+$00	 ($^{+27.87}_{-2.20}$)		&		13		 &	2.97E$+$01	&		9.48E$-$04		&	0.15 	 ($^{+0.27}_{-0.21}$)	&	 0.07 	&		-1.7 		 &		 T		 &		CRCN		\\
1822$-$09		&		0.4$-$1.6		&		1.40E$+$01	($^{+1.2}_{-13.4}$)		&		 1.80E$-$01	($^{+1.82}_{-0.14}$)		 &		 3.00E$-$01	 ($^{+13.43}_{-0.30}$)		&		5		 &	6.61E$+$00	&		3.67E$-$02		&	0.55 	 ($^{+3.00}_{-0.43}$)	&	 0.25 	&		-1.3 		 &		 T		 &		PCN		\\
1839$+$09		&		0.4$-$4.85		&		1.44E$+$01	($^{+1.2}_{-14.4}$)		&		 4.00E$-$02	($^{+1.96}_{-2.04}$)		 &		 1.20E$-$01	 ($^{+15.60}_{-0.12}$)		&		7		 &	4.18E$+$00	&		3.82E$-$01		&	0.10 	 ($^{+0.46}_{-0.34}$)	&	 0.05 	&		-2.0 		 &		 CRS		 &		CN		\\
1851$-$14		&		0.4$-$1.6		&		1.93E$-$01	($^{+15.81}_{-1.93{\rm E}-1}$)		&		 1.54E$+$00	 ($^{+8.46}_{-11.54}$)		&		1.50E$+$01	 ($^{+1.21}_{-14.96}$)		&		6		 &	1.19E$+$01	&		7.80E$-$03		 &	 0.14 	($^{+0.44}_{-0.43}$)	&	0.12 	&		-0.8 		 &		 $-$ 		 &		$-$ 		\\
1900$+$01		&		0.6$-$4.85		&		1.04E$+$01	($^{+1.4}_{-10.4}$)		&		 5.00E$-$02	($^{+1.95}_{-2.05}$)		 &		 5.00E$-$02	 ($^{+11.79}_{-0.05}$)		&		6		 &	2.70E$+$00	&		4.40E$-$01		&	0.13 	 ($^{+0.99}_{-0.59}$)	&	 0.06 	&		-1.9 		 &		 CRS		 &		CN		\\
1907$+$10		&		0.4$-$2.65		&		8.23E$+$00	($^{+7.57}_{-8.23}$)		&		 4.41E$-$01	 ($^{+9.56}_{-10.44}$)		 &		6.05E$+$00	 ($^{+9.52}_{-6.05}$)		&		8		 &	4.80E$+$00	&		4.44E$-$01		 &	0.96 	 ($^{+20.63}_{-1.00}$)	&	0.45 	&		-2.5 		 &		 CRS		 &		CN		\\
1911$+$13		&		0.6$-$4.85		&		8.90E$-$02	($^{+15.31}_{-8.90{\rm E}-2}$)		&		 2.37E$+$00	 ($^{+9.63}_{-12.37}$)		&		1.45E$+$01	 ($^{+0.90}_{-14.53}$)		&		6		 &	4.60E$+$00	&		2.01E$-$01		 &	 0.26 	($^{+0.97}_{-1.16}$)	&	0.24 	&		-1.5 		 &		 T		 &		$-$ 		\\
1911$-$04		&		0.243$-$4.85		&		7.20E$+$00	($^{+0.80}_{-7.13}$)		&		 1.60E$-$01	 ($^{+1.84}_{-0.13}$)		 &		7.00E$-$02	 ($^{+7.50}_{-0.07}$)		&		15		 &	7.22E$+$00	&		8.43E$-$01		 &	0.49 	 ($^{+0.63}_{-0.40}$)	&	0.22 	&		-2.6 		 &		 CRS		 &		CRCN		\\
1914$+$09		&		0.6$-$4.85		&		1.72E$+$01	($^{+1.4}_{-17.2}$)		&		 1.10E$-$01	($^{+1.89}_{-2.11}$)		 &		 2.20E$-$01	 ($^{+18.46}_{-0.22}$)		&		5		 &	1.55E$+$01	&		4.38E$-$04		&	0.31 	 ($^{+0.69}_{-0.45}$)	&	 0.15 	&		-2.3 		 &		 T		 &		CN		\\
1915$+$13		&		0.43$-$4.85		&		1.56E$+$01	($^{+3.2}_{-15.6}$)		&		 5.00E$-$02	($^{+1.95}_{-2.05}$)		 &		 1.20E$+$00	 ($^{+17.61}_{-1.2}$)		&		7		 &	9.89E$-$01	&		9.11E$-$01		&	0.12 	 ($^{+0.61}_{-0.48}$)	&	 0.06 	&		-1.8 		 &		 CRS		 &		PCN		\\
1920$+$21		&		0.6$-$1.6		&		1.82E$+$01	($^{+1.2}_{-18.2}$)		&		 7.00E$-$02	($^{+1.93}_{-2.07}$)		 &		 3.00E$-$02	 ($^{+19.43}_{-0.03}$)		&		5		 &	2.45E$+$00	&		2.93E$-$01		&	0.19 	 ($^{+2.29}_{-0.46}$)	&	 0.09 	&		-2.2 		 &		 T		 &		CN		\\
1933$+$16		&		0.43$-$4.9		&		6.02E$-$01	($^{+8.00}_{-0.53}$)		&		 1.97E$+$00	($^{+1.51}_{-1.6}$)		 &		 8.39E$+$00	 ($^{+1.79}_{-8.39}$)		&		11		&	 2.79E$+$01	&		4.94E$-$04		&	1.58 	 ($^{+1.14}_{-0.88}$)	&	 1.29 	&		-1.7 		 &		 CRS		 &		CRCN		\\
1940$-$12		&		0.4$-$1.6		&		9.00E$+$00	($^{+1.00}_{-9.00}$)		&		 1.50E$-$01	($^{+1.85}_{-2.15}$)		 &		1.80E$-$01	 ($^{+9.54}_{-0.18}$)		&		5		 &	2.75E$+$00	&		2.52E$-$01		&	0.44 	 ($^{+4.71}_{-0.58}$)	&	0.20 	&		-2.4 		 &		 CNS		 &		$-$ 		\\
1943$-$29		&		0.4$-$1.4		&		9.40E$+$00	($^{+1.00}_{-9.40}$)		&		 6.00E$-$02	($^{+1.94}_{-2.06}$)		 &		3.40E$-$01	 ($^{+10.03}_{-0.34}$)		&		4		 &	2.47E$+$00	&		1.16E$-$01		&	0.16 	 ($^{+2.66}_{-0.43}$)	&	0.07 	&		-2.0 		 &		 $-$ 		 &		$-$ 		\\
1944$+$17		&		0.4$-$2.38		&		4.00E$+$01	($^{+0.8}_{-29.2}$)		&		 2.10E$-$01	($^{+0.59}_{-0.05}$)		 &		 0.00E$+$00	 ($^{+28.98}_{0}$)		&		11		&	 1.28E$+$01	&		1.17E$-$01		&	0.69 	 ($^{+0.33}_{-0.18}$)	&	 0.30 	&		-1.3 		 &		CT	 	 &		$-$ 		\\
1946$+$35		&		0.6$-$2.65		&		1.42E$+$01	($^{+1.0}_{-12.4}$)		&		 2.60E$-$01	($^{+0.92}_{-0.13}$)		 &		 7.00E$-$02	 ($^{+12.78}_{-0.07}$)		&		7		 &	2.60E$+$01	&		3.12E$-$05		&	0.91 	 ($^{+0.66}_{-0.51}$)	&	 0.38 	&		-2.4 		 &		 CRS		 &		CRCN		\\
2053$+$36		&		0.6$-$4.85		&		6.11E$-$01	($^{+14.39}_{-6.11{\rm E}-1}$)		&		 1.14E$+$00	 ($^{+8.86}_{-11.14}$)		&		1.29E$+$01	 ($^{+2.22}_{-12.89}$)		&		5		 &	1.98E$+$00	&		3.74E$-$01		 &	 0.27 	($^{+0.83}_{-1.23}$)	&	0.20 	&		-1.9 		 &		 CRS		 &		PCN		\\
2217$+$47		&		0.23$-$4.9		&		4.61E$+$00	($^{+7.59}_{-4.55}$)		&		 4.56E$-$01	($^{+2.34}_{-0.42}$)		 &		6.52E$+$00	 ($^{+5.29}_{-6.52}$)		&		10		 &	7.73E$+$00	&		3.63E$-$01		&	0.67 	 ($^{+0.85}_{-0.57}$)	&	0.35 	&		-2.6 		 &		 CRS		 &		PCN		\\

\enddata
\end{deluxetable}

\clearpage
% -------- table  pulse width list
\begin{table}
\centering
\caption{Pulsars that have been studied on the frequency dependence of pulse width or component separation in literature} \label{Tab:plswidth}
%\noalign{\smallskip}
 \begin{tabular}{llllll}
 \hline\hline
 &	 &	&	PSR B & 	&		\\
 \hline
0031$-$07	$^{4,5}$	&	0037$+$56 	$^{7}$	&	0053$+$47 	$^{7,*}$	&	0138$+$59 	$^{7}$	&	 0144$+$59 	$^{7}$	&	 0149$-$16	 $^{5}$	\\
0301$+$19 	$^{3,5-7}$	&	0329$+$54	$^{2-4,6}$	& 0402$+$61 	$^{7,\dagger}$	&	0450$-$18 	 $^{7,\dagger}$	&	 0525$+$21	 $^{2-4,6}$	&	 0559$-$05 	 $^{7}$	\\
0609$+$37 	$^{7}$	&	0611$+$22 	$^{7}$	&	0626$+$24 	$^{7,\dagger}$	&	0628$-$28 	 $^{4,5,7}$	&	0643$+$80 	$^{7}$	 &	 0751$+$32 	$^{7}$	\\
0756$-$15 	$^{7,*}$	&	0809$+$74 	$^{4,7}$	&	0818$-$13 	$^{5,7}$	&	0823$+$26	 $^{4,5}$	&   0834$+$06 	 $^{2-5,7,\dagger}$	&	 0906$-$17 	 $^{7,\dagger}$	\\
0919$+$06	$^{5}$	&	0942$-$13 	$^{7}$	&	0943$+$10	$^{5,*}$	&	0950$+$08	$^{5}$	&	 J1022$+$10	$^{7,*}$	&	 1039$-$19 	 $^{7}$	\\
1133$+$16	$^{1-6}$	&	1237$+$25	$^{2,3,5,6}$	&	1254$-$10 	$^{7}$	&	1449$-$64	 $^{4,*}$	&	1508$+$55	 $^{3}$	&	 1541$+$09 	$^{5,7}$	\\
1552$-$23 	$^{7}$	&	1604$-$00 	$^{5,7}$	&	1612$+$07 	$^{7}$	&	1620$-$09 	$^{7}$	&	 1642$-$03	$^{4,5}$	&	 1702$-$19 	$^{7}$	\\
1706$-$16	$^{4,5}$	&	1709$-$15 	$^{7,*}$	&	J1713$+$07	$^{7,*}$	&	1717$-$16 	 $^{7}$	&	1737$+$13 	$^{7}$	 &	 1738$-$08 	 $^{7}$	\\
1745$-$12 	$^{7}$	&	1749$-$28	$^{4}$	&	1750$-$24 	$^{7,*}$	&	1753$+$52 	$^{7,*}$	&	 1758$-$23 	$^{7}$	&	 1805$-$20 	 $^{7,*}$	\\
1815$-$14 	$^{7,*}$	&	1817$-$13 	$^{7,*}$	&	1818$-$04 	$^{7,\dagger}$	&	1819$-$22 	 $^{7}$	&	1820$-$11 	 $^{7,*}$	 &	 1821$+$05 	$^{5,7}$	 \\
1821$-$11 	$^{7,*}$	&	1822$-$09	$^{4}$	&	1822$-$14 	$^{7,*}$	&	1823$-$13 	$^{7,*}$	 &	1829$-$08 	$^{7,*}$	 &	 1830$-$08 	 $^{7,*}$	\\
1831$-$04 	$^{7}$	&	1839$+$09 	$^{7}$	&	1839$+$56 	$^{7}$	&	1841$-$05 	$^{7,*}$	&	 1842$+$14	$^{5}$	&	 1849$+$00 	 $^{7,*}$	\\
1855$+$02 	$^{7,*}$	&	1855$+$09 	$^{7,*}$	&	1857$-$26	$^{5,*}$	&	1859$+$07 	 $^{7,*}$	&	1900$+$01 	 $^{7}$	&	 1900$+$06 	 $^{7,*}$	\\
1905$+$39 	$^{7}$	&	1907$+$03 	$^{7}$	&	1911$+$13 	$^{7}$	&	1911$-$04 	$^{7}$	&	 1913$+$10 	$^{7,*}$	&	 1914$+$09 	 $^{7}$	\\
1914$+$13 	$^{7,*}$	&	1915$+$13 	$^{7}$	&	1916$+$14 	$^{7}$	&	1919$+$21	 $^{2-5,8,\dagger}$	&	1923$+$04 	 $^{7}$	&	 1929$+$10	$^{3}$	\\
1933$+$16	$^{4}$	&	2000$+$32 	$^{7}$	&	2000$+$40 	$^{7}$	&	2002$+$31 	$^{7}$	&	 2011$+$38 	$^{7,*}$	&	 2016$+$28 	 $^{4,5,7}$	\\
2020$+$28	$^{3,5,6}$	&	2045$-$16 	$^{2,3,5-7}$	&	2053$+$36 	$^{7}$	&	2110$+$27 	 $^{7}$	&	2111$+$46 	$^{7}$	 &	 2148$+$52 	$^{7}$	\\
2148$+$63 	$^{7}$	&	J2145$-$07	$^{7,*}$	&	2154$+$40 	$^{7}$	&	2303$+$30 	$^{7}$	&	 2310$+$42 	$^{7}$	&	 2319$+$60	 $^{3}$	\\
2315$+$21 	$^{7}$	&	2334$+$61 	$^{7}$	&			&			&			&			\\

 \hline
\end{tabular}
\tablenotetext{\space}{* Pulsars not used in this paper. }
\tablenotetext{\space}{$^\dagger$ Pulsars with the increasing component separation or pulse width identified by previous studies.}
\tablenotetext{\space}{References: 1-Craft \& Comella 1968, 2-Lyne et al. 1971, 3-Sieber et al. 1975, 4-Rankin 1983b, 5-Slee et al. 1987, 6-Thorsett 1991, 7-Kijak et al. 1998, 8-Hassall et al. 2012}
\end{table}

\clearpage

% -------- table  RFM pulsar list
\begin{table}
\centering
\caption{Pulsars that have been studied on the RFM in literature} \label{Tab:rfm}
%\noalign{\smallskip}
 \begin{tabular}{llllll}
 \hline\hline
 & & Group-A PSR B &  & &  \\
\hline
$0301+19^{1,7,8,9}$   & $0329+54^{3,5,6,8,9}$  & $0355+54^{4,5,6}$    & $0525+21^{1,4,7,8,9}$  & $0535+28^{4,*}$   & $0628-28^{7}$ \\ $0823+26^{1,2,4,5}$ & $0919+06^{1,2,4,5}$   & J$1022+1001^{6,*}$   & $1039-19^{7}$        & $1133+16^{1-9}$  & $1237+25^{8,9}$\\
$1530+27^{1}$       & $1604-00^{8,\dagger}$        & $1706-16^{3,5,6}$    & $1811+40^{7}$       & $1839-04^{7}$    & $1855+09^{6,**}$ \\
$1913+16^{1,**}$     & $1916+14^{7}$       & $1929+10^{1,5}$      & $2011+38^{7,*}$    & $2020+28^{3,5,6,8,9}$ & $2021+51^{4,5,6,7}$ \\
$2045-16^{3,4,7,8,9}$ & $2154+40^{7}$        & $2306+55^{7}$        & $2310+42^{3,6}$      & $2319+60^{3,4,6,7}$ & $2324+60^{7}$ \\
\hline
&  & Group-B PSR B  & & &  \\
\hline
 $0540+23^{1,4,5,6}$  & $0611+22^{1}$     & $0740-28^{3,4,5}$  & $0834+06^{8,\dagger}$   & $0809+74^{4}$  & $0950+08^{1,2,4}$  \\
 $1702-19^{7}$ &  J$1713+0747^{6,*}$ & $1822-09^{6}$    & $1845-01^{3,6}$  & $1919+21^{8,\dagger}$  & $2323+63^{7,*}$         \\

\hline

& &  Group-C PSR B  & & &  \\
\hline
$1642-03^{3,6}$  & $1737+13^{1,7}$ & $1915+13^{6}$ & &  \\
 \hline
\end{tabular}
\tablenotetext{\space}{ * Not used in this paper. }
\tablenotetext{\space}{ ** Millisecond pulsars, not used in this paper.}
\tablenotetext{\space}{ $^\dagger$ Regarded as none frequency dependence by Mitra \& Rankin
(2002). Except these three pulsars, all the other pulsars in this table are classified as pulsars with a negative RFM index in the references.}
\tablenotetext{\space}{References: 1-Blaskiewicz et al. 1991, 2-Phillips 1992,
3-Kramer et al. 1994, 4-Hoensbroech \& Xilouris 1997, 5-Kramer et
al. 1997, 6-Kijak \& Gil 1998, 7-Wu et al. 2002, 8-Mitra \& Rankin
2002, 9-Wang et al. 2013}
\end{table}

% Table  Student's test
\begin{table}
\centering
\caption{Result of Student's $t$-test for the mean spectral
indices of pairs of groups.}
 \label{Tab:tt}
\begin{tabular}{cccc}
\hline\hline
${\rm Parameter}$ & ${\rm A}$ & ${\rm B}$ & ${\rm C}$ \\
\hline
 $\alpha_{\rm m}$ & -1.71 & -1.92 & -1.94 \\
 $\sigma_{\alpha}$ & 0.53 & 0.48 & 0.55\\
$\eta_{\rm m}$ & -26.3 & -0.01 & 53.4\\
$\sigma_{\eta}$ & 10.9 & 0.06 & 50.3 \\
\hline
 $t-{\rm test}$& ${\rm (A, B)}$ & ${\rm (A, C)}$ & ${\rm (B, C)}$   \\
\hline
$p$    &  0.045 & 0.05 & 0.86\\
$c$    & [0.005, 0.423]  & [0, 0.47] & [-0.23, 0.28]\\
\hline
\end{tabular}
\end{table}

\end{document}